\documentclass[a4paper,fleqn,usenatbib]{mnras}

\usepackage{newtxtext,newtxmath}

\usepackage[T1]{fontenc}
\usepackage{ae,aecompl}

\usepackage{graphicx}	
\usepackage{amsmath}	
\usepackage{amssymb}	
\usepackage{aas_macros}  
\usepackage{pdflscape} 
\usepackage{subcaption} 
\title[Eclipse modelling of 15 Cataclysmic Variables]{The evolutionary status of Cataclysmic Variables: Eclipse modelling of 15 systems.}

\author[M.\ McAllister et al.]{M.\ McAllister$^{1}$, S.\,P.\ Littlefair$^{1}$, S.\,G.\ Parsons$^{1}$,  V.\,S.\ Dhillon$^{1, 2}$, T.\,R.\ Marsh$^{3}$
\newauthor B.\ T.\ G\"{a}nsicke$^{3}$, E.\ Breedt$^{4}$, C.\ Copperwheat$^{5}$, M.\ J.\ Green$^{3}$, C.\ Knigge$^{6}$ 
\newauthor D.\ I.\ Sahman$^{1}$, Martin J.\ Dyer$^{1}$, P.\ Kerry$^{1}$, R.\ P.\ Ashley$^{3}$, P.\ Irawati$^{7}$, S.\ Rattanasoon$^{7}$ \\
$^{1}$Dept of Physics and Astronomy, University of Sheffield, Sheffield, S3 7RH, UK \\
$^{2}$Instituto de Astrof\'{i}sica de Canarias (IAC), E-38200, La Laguna, Tenerife, Spain\\
$^{3}$Dept of Physics, University of Warwick, Coventry, CV4 7AL, UK\\
$^{4}$Institute of Astronomy, University of Cambridge, Madingley Road, Cambridge, CB3 0HA, UK\\
$^{5}$Astrophysics Research Institute, Liverpool John Moores University, IC2, Liverpool Science Park, L3 5RF, UK\\
$^{6}$School of Physics \& Astronomy, University of Southampton, Southampton SO17 1BJ, UK\\
$^{7}$National Astronomical Research Institute of Thailand, 191 Siriphanich Bldg., Huay Kaew Road, Chiang Mai 50200, Thailand
}
\date{Accepted XXX. Received YYY; in original form ZZZ}

\pubyear{2018}

\begin{document}
\label{firstpage}
\pagerange{\pageref{firstpage}--\pageref{lastpage}}
\maketitle

\begin{abstract}
We present measurements of the component masses in 15 Cataclysmic Variables (CVs) - 6 new estimates and 9 improved estimates.  We provide new calibrations of the relationship between superhump period excess and mass ratio, and use this relation to estimate donor star masses for 225 superhumping CVs. With an increased sample of donor masses we revisit the implications for CV evolution.  We confirm the high mass of white dwarfs in CVs, but find no trend in white dwarf mass with orbital period. We argue for a revision in the location of the orbital period minimum of CVs to $79.6 \pm 0.2$\,min, significantly shorter than previous estimates. We find that CV donors below the gap have an intrinsic scatter of only 0.005\,R$_{\odot}$ around a common evolutionary track, implying a correspondingly small variation in angular momentum loss rates. In contrast to prior studies, we find that standard CV evolutionary tracks - without additional angular momentum loss - are a reasonable fit to the donor masses just below the period gap, but that they do not reproduce the observed period minimum, or fit the donor radii below 0.1 M$_{\odot}$.
\end{abstract}




\section{Introduction}
Cataclysmic Variables (CVs) are close binary stars in which a white dwarf is accreting material from a low-mass donor star. Without angular momentum loss (AML) from the system, mass transfer could not be sustained; thus it is the AML that drives the secular evolution of CVs. The currently accepted picture of CV evolution is that CVs evolve from long to short periods under the influence of AML caused by magnetic braking. A reduction in AML due to magnetic braking is thought to arise when the donor becomes fully convective. This causes the CV to become detached, and is the cause of the dearth of CVs in the 2--3 hour orbital period range; the CV {\em period gap}. When the CV resumes mass transfer, AML is driven by gravitational radiation and the mass transfer rate is lower. The CV evolves slowly through a {\em period minimum}, which arises because the thermal timescale of the donor becomes comparable to the mass loss timescale, and the donor begins to expand in response to mass loss, which leads to a widening of the orbit.

This long-standing picture has survived for over 35 years \citep{rappaport82, rappaport83} despite the fact that it struggles to explain the observed value of the period minimum \citep{gaensicke09}, the scarcity of known post-period-minimum systems \citep{hernandez18} and the average high white dwarf mass in CVs \citep{zorotovic11}. Modifications to the standard model exist that can potentially explain some of these issues. The orbital period minimum problem can be solved with an additional source of AML for short period systems \citep{patterson98, knigge11}, and AML in nova outbursts may cause CVs with low-mass white dwarfs to be unstable, explaining the high average white dwarf mass \citep{schreiber16, nelemans16}. However, it remains to be seen if those modifications can correctly describe the observed properties of known CVs. In particular, the mass and radius of the donor star is a sensitive probe of the secular evolution. This is because the radius of the donor star in a CV can be inflated from the main-sequence value, by some amount that depends upon the mass-loss history of the donor. In particular, the donor radius is more likely to track the long-term average mass loss rate than other physical properties of the CV such as the accretion light or the effective temperature of the accreting white dwarf \cite[see][and references within]{knigge11}.

One of the best methods of measuring
donor masses and radii is to model the primary eclipse. During primary eclipse, the white dwarf and accretion disc are occulted, along with the bright spot, located where the accretion stream impacts the outer rim of the disc. The path of the gas stream is determined by the mass ratio, and so the detailed shape of the primary eclipse contains enough information to derive extremely precise masses that are consistent with conventional spectroscopic methods \cite[see][for example]{tulloch09, copperwheat10, savoury12}. The photometric method has the advantage that it does not rely on detection of the light from the donor star, which is often invisible given the much brighter white dwarf and accretion disc, particularly for CVs with shorter orbital periods. It does however require high quality lightcurves of the eclipses, which occur on timescales of minutes. With this in mind, our group has been acquiring high quality lightcurves of eclipsing CVs with the high time-resolution instruments ULTRACAM \citep{dhillon07} and ULTRASPEC \citep{dhillon14}. Here we present the analysis of 15 systems, and review the evolutionary status of CV systems in light of the results. 

\subsection{Systems selected for eclipse modelling}
\label{sec:addsys}

The 15 systems modelled in this paper are listed in Table~\ref{table:ephem}. CTCV 1300, DV UMa, SDSS 1152, SDSS 1501 have existing mass determination from eclipse modelling of ULTRACAM data \citep{savoury11}, whilst Z Cha, OY Car, IY UMa, GY Cnc and SDSS 1006 have existing mass determinations in the literature from various methods \citep{wood86,wadehorne88,woodhorne90, thorstensen00, steeghs03, southworth09, copperwheat12}. The existing mass determinations have large associated errors, and we re-analyse them here  in the light of new data, and an updated modelling approach \cite[see][for details]{mcallister17a}. The remaining 6 systems have no existing donor mass estimates, and were chosen from the eclipsing CVs observed with ULTRACAM/ULTRASPEC to date; the primary reason for their selection was an eclipse shape suitable for modelling, with visible white dwarf and bright spot eclipses.

\section{Observations and Data Reduction}
The observations in this paper span a range of dates from May 2003 to Feb 2017. All data were taken with the triple-band fast camera ULTRACAM, or the single-band fast camera ULTRASPEC. ULTRACAM data were taken on three telescopes; the 4.2-m William Herschel Telescope (WHT) situated at the Roque de los Muchachos Observatory on La Palma, Spain, the 8.2-m Very Large Telescope (VLT) at Paranal, Chile, and the 3.5-m New Technology Telescope (NTT) located at La Silla, Chile. All ULTRASPEC data were taken using the 2.4-m Thai National Telescope (TNT), located on Doi Inthanon in Thailand. All observations were obtained using the Sloan Digital Sky Survey (SDSS) filter set, with the exception of some of the ULTRASPEC observations, which use the KG5 filter. This filter is described in detail in \cite{hardy17}; it is a broadband filter encompassing the SDSS $u'$, $g'$ and $r'$ passbands. For a full journal of observations, see Table~\ref{table:obsj}. 

Data reduction was carried out using the ULTRACAM pipeline reduction software \citep[see][]{dhillon07}. One or more nearby, photometrically stable comparison stars were used to correct for transparency variations during observations. If the comparison stars have tabulated SDSS magnitudes, we used these to transform the photometry into the $u'$\,$g'$\,$r'$\,$i'$\,$z'$ standard system \citep{smith02}, otherwise observations of standard stars from the nearest photometric night were used. Photometry was corrected for extinction using the median extinction coefficients for each observatory, as derived from long duration time-series taken with ULTRACAM and ULTRASPEC.  

\section{Methods}

\subsection{Orbital Ephemerides}
\label{subsec:ephem}

\begin{table*}
\setlength{\tabcolsep}{3pt}
\begin{tabular}{p{3.8cm}cccllccc}
\hline
Object & Right & Declination & Out-of eclipse Mag. & $T_{0}$ & $P_{\mathrm{orb}}$ & $N_{\mathrm{ecl}}$ & Add. Ecl. \\
& Ascension && ($g'$) & (MJD) & (d) &  & Times \\ \hline
CTCV J1300$-$3052  & 13 00 29.05 & $-$30 52 57.1 & 18.6 & 54262.099166(18)$^{h}$ & 0.0889406998(17) & 4 & 1 \\ 
DV UMa & 09 46 36.65 & +44 46 45.1 & 18.7 & 52782.973948(10)$^{h}$ & 0.0858526308(7) & 4 & 2,3,4 \\ 
SDSS J115207.00+404947.8 & 11 52 07.01 & +40 49 48.0 & 19.5 & 55204.101279(6)$^{h}$ & 0.0677497026(3) &  7 & 5 \\
SDSS J150137.22+550123.4 & 15 01 37.24 & +55 01 23.5 & 19.0 & 56178.870444(8)$^{h}$ & 0.05684126603(21) &  12  & -- \\
CSS080623 J140454$-$102702 & 14 04 53.97 & $-$10 27 02.3 & 19.5 & 55329.234631(13)$^{h}$ & 0.059578971(3) &  10 &  -- \\
CSS110113 J043112$-$031452 & 04 31 12.45 & $-$03 14 51.6 & 19.5 & 55942.014642(15)$^{h}$ & 0.0660508707(18) & 12 & -- \\
GY Cnc & 09 09 50.55 & +18 49 47.5 & 16.7 & 55938.263734(22)$^{b}$ & 0.175442399(6) &  12 & -- \\
IY UMa& 10 43 56.73 & +58 07 31.9 & 17.1 & 56746.6395010(9)$^{h}$ & 0.07390892818(21) & 10 &  8 \\
OY Car & 10 06 22.07 & $-$70 14 04.6 & 15.6 & 55353.996477(3)$^{h}$ & 0.06312092545(24) &  7 & -- \\
SDSS J090103.94+480911.0 & 09 01 03.94 & +48 09 11.0 & 19.5 & 55942.116358(8)$^{h}$ & 0.0778805321(5) & 10 & 9 \\
SDSS J100658.40+233724.4 & 10 06 58.42 & +23 37 24.6 & 18.6 & 56682.72973(5)$^{h}$ & 0.185913107(13) & 11 & 7,10 \\
SSS130413 J094551$-$194402 & 09 45 51.00 & $-$19 44 00.8 & 16.7 & 56683.673971(12)$^{h}$ & 0.0657692903(12) & 17 & 6 \\
SSS100615 J200331$-$284941 & 20 03 31.27 & -28 49 41.3 & 19.6 & 56873.023625(5)$^{h}$ & 0.0587045(4) & 3 & -- \\
V713 Cep & 20 46 38.70 & +60 38 02.8 & 18.5 & 56176.936402(7)$^{h}$ & 0.0854185080(12) & 15 & 11 \\
Z Cha & 08 07 27.75 & $-$76 32 00.7 & 15.6 & 53498.011471(4)$^{h}$ & 0.0744992631(3) & 14 & -- \\
\hline
\end{tabular}
\caption[Ephemerides]{\label{table:ephem}Ephemerides for the CVs modelled in this paper. $T_{0}$ is the mid-eclipse time of cycle 0, $P_{\rm orb}$  is the orbital period, while $N_{\mathrm{ecl}}$ is the total number of eclipses obtained. References for additional eclipse times: 
(1) \cite{tappert04}, (2) \cite{howell88}, (3) \cite{patterson00}, (4) \cite{nogami01}, (5) \cite{southworth10}, 
(6) \cite{thorstensen16}, (7) Woudt (priv. comm.),  (8) Coppejans (priv. comm.), (9) \cite{dillon08}, 
(10) \cite{southworth07},  (11) Bours (priv. comm.).}
\vspace{0.15cm}
$^{h}$Heliocentric times in HMJD(UTC), $^{b}$Barycentric times in BMJD(TDB).
\end{table*}

Updated orbital ephemerides for the CVs in this paper were calculated, and are shown in Table~\ref{table:ephem}.
Mid-eclipse times were determined by averaging the time of white dwarf ingress and egress, as determined by locating the minima and maxima of a smoothed lightcurve derivative. Mid eclipse times were corrected to the Solar System Heliocentre or Barycentre using \textsc{astropy} \citep{astropy18}. The correction used was decided upon a system-to-system basis, and depended on previous mid-eclipse times and ephemerides in the literature. Heliocentric times are recorded in Coordinated Universal Time (UTC), Barycentric times in Barycentric Dynamical Time (TDB). Mid-eclipse times for each individual eclipse observed are presented in Table~\ref{table:obsj}.

\subsection{Eclipse light-curve modelling}
\label{subsec:modelling}
The model used to fit the eclipse light curve is described by \cite{savoury11}. The important assumptions in the model are that the bright spot lies on the ballistic trajectory from the donor, the white dwarf follows a theoretical mass-radius relation and that the white dwarf is unobscured. The model has recently received two major improvements, as outlined in \cite{mcallister17a}. The model now has the ability to fit multiple lightcurves simultaneously whilst sharing parameters that do not change; such as the mass ratio $q$, the white dwarf eclipse width $\Delta \Phi$ and the white dwarf radius, scaled by the binary separation $R_{1}/a$. In addition, the model now has a statistical treatment of flickering using Gaussian Processes (GPs) that makes the uncertainty estimates for these parameter robust in the presence of flickering. For each system we either fit all the individual eclipses, or averaged several eclipses in the same filter. Averaging eclipses can ease convergence of the model, by reducing the number of free parameters, but it is not suitable when the lightcurve features change between eclipses, for example due to a changing accretion disc radius. 

Eclipse averaging was used for six systems: CSS080623, CSS110113, DV UMa, SDSS 0901, SDSS 1152, SSS100615. All systems have multiple eclipse light curves observed close together in time (e.g.\ during the same observing run), and contain only low amplitude flickering. When selecting eclipses for the construction of each average eclipse, great care was taken to exclude any eclipses with differing disc radius/flux and/or bright spot shape/flux changes. Firstly, only eclipses obtained during the same observing run were considered for each average eclipse. Secondly, before averaging, all eclipses were phase-folded and overlaid, with any differing eclipses removed from consideration. An average eclipse was created for each available wavelength band, typically $u'g'r'$ or $u'g'i'$. As both CSS080623 and SDSS 0901 have multiple eclipses from two separate observing runs, two average eclipses in each wavelength band were created. For the remaining nine systems, we did not average lightcurves prior to fitting. 

In general, the majority of eclipses showing a clear bright spot ingress feature were selected for modelling. However, for systems with many high signal-to-noise eclipses containing very clear bright spot eclipse features (e.g.\ OY Car and Z Cha), only six were selected. In these cases, the inclusion of additional eclipses had an insignificant effect on the system parameter values and errors, and did not justify the resulting increased model complexity and computational time. This approach was also taken with SSS130413 and V713 Cep, two systems with moderately clear bright spot features. 

\section{Results}

\subsection{Simultaneous Eclipse Light Curve Modelling}
\label{subsec:lcmod}
For each of the 15, the chosen eclipses were fit with the CV eclipse model, with GPs used to model the flickering component. The binary model contains two possible versions of the bright spot \cite[see][for details]{savoury11}. A more complex bright spot model was used for all but three systems (SDSS 1501, SSS100615, V713 Cep). The simple bright spot was used in these three systems due to each containing a weak bright spot component in their eclipse light curves. The typical phase range of the eclipse light curves modelled was $-$0.10 to 0.15, however an extended phase range was used for a number of systems. The phase range was increased for systems with a prominent bright spot (e.g.\ CTCV 1300, GY Cnc, SDSS 1006) in addition to SDSS 1501 (tenuous bright spot component) and V713 Cep (combination of heavy flickering post-eclipse and significant disc contribution).

Posterior probability distributions of all parameters in the binary model were estimated using a Markov Chain Monte Carlo (MCMC) approach. The full results of all eclipse fits are shown in  Figures~\ref{fig:modfit_css080623}--\ref{fig:modfit_zcha}. Figure~\ref{fig:modelfits_addsys} shows an example $g'$-band eclipse light curve fit for each system. In addition to the most probable fit of the eclipse model (blue line), a blue band is plotted that covers $1\sigma$ from the mean of a random sample (size 1000) of the MCMC chain. The grey points represent the actual eclipse light curves, while the black points are the result of subtracting the GP's posterior mean (itself shown, $\pm1\sigma$, by the red band covering the residuals below each plot). Also plotted are the separate components of the eclipse model: white dwarf (purple), bright spot (red), accretion disc (yellow) and donor (green).

\begin{figure*}
\begin{center}
\includegraphics[width=1.5\columnwidth,trim=0 30 0 20]{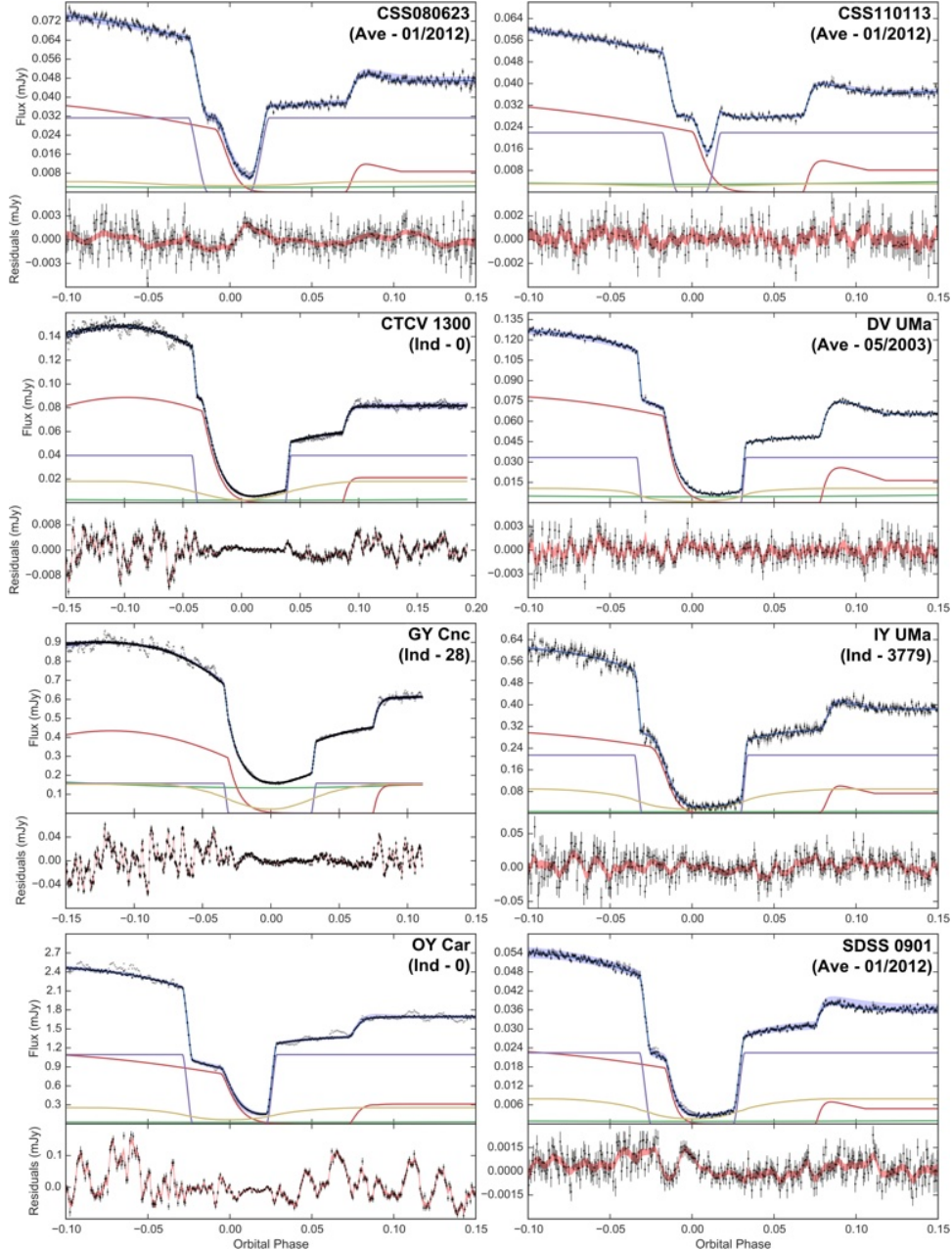}
\caption[Eclipse model fits to $g$-band light curves of 15 CVs.]{\label{fig:modelfits_addsys}Eclipse model fits to $g$-band light curves of 15 CVs. The lightcurves are shown in grey points; see Section~\ref{subsec:lcmod} for full details of what is plotted. The system name is displayed in the top-right corner of each plot, along with whether the eclipse is an individual (Ind) or average (Ave) eclipse. For individual eclipses the cycle number is shown, while for average eclipses the month and year of the eclipses is shown. See Appendix~\ref{app:modfits} for a complete set of eclipse plots.}
\end{center}
\end{figure*}
\begin{figure*}\ContinuedFloat
\begin{center}
\includegraphics[width=1.5\columnwidth,trim=0 60 0 20]{gband_fits_2.pdf}
\caption[\textit{Continued.}]{\textit{Continued.}}
\end{center}
\end{figure*}

\subsection{System Parameters}
\label{subsec:syspars_addsys}
Once the parameters of the binary model are estimated, the system parameters can be found. A full discussion can be found in \cite{mcallister17a}. In brief, this involves an iterative procedure where the white dwarf spectral energy distribution (SED) -- measured from the eclipse depth of the white dwarf -- is fit by white dwarf atmosphere models \citep{bergeron95}. This yields estimates of the white dwarf temperature $T_{1}$ and distance $d$. The values of $q$, $\Delta \Phi$ and $R_{1}/a$ from the binary model, combined with Kepler's third law and a temperature-corrected mass-radius relationship for the white dwarf, are used to calculate the posterior probability distribution functions (PDFs) of the system parameters:
\begin{itemize}
\item mass ratio, $q$;
\item white dwarf mass, $M_{1}$;
\item white dwarf radius, $R_{1}$;
\item white dwarf $\log g$;
\item donor mass, $M_{2}$;
\item donor radius, $R_{2}$;
\item binary separation, $a$;
\item inclination, $i$.
\end{itemize}
 
System parameter values (see Table~\ref{table:syspars_addsys}) were then obtained from the peak of each posterior PDF, with errors from the 67\% confidence level. The results of the white dwarf SED fits are shown in Figure~\ref{fig:fluxes_addsys}, and the resulting $T_{1}$ and $d$ values for each system are also displayed in Table~\ref{table:syspars_addsys}. Note that the white dwarf flux fitting was not carried out for either IY UMa or SDSS 1006, due to the lack of $u'$-band eclipses in their eclipse model fits. Thankfully, precise measurements of $T_{1}$ for both IY UMa\footnote{IY UMa entered outburst between the observations of \cite{pala17} and this work, so this $T_{1}$ measurement may be slightly lower than $T_{1}$ of the white dwarf in the eclipse light curves.} ($18000\pm1000\,\mathrm{K}$) and SDSS 1006 ($16000\pm1000\,\mathrm{K}$) from spectral fitting are given in \cite{pala17}.

\begin{figure}
\begin{center}
\includegraphics[width=1.0\columnwidth,trim=0 30 0 30]{wdatmos_fits_1.pdf}
\caption[Fits to white dwarf atmosphere predictions.]{\label{fig:fluxes_addsys}White dwarf fluxes for 13 CVs, showing the white dwarf fluxes from the eclipse model fits (blue) and white dwarf atmosphere predictions (red), at wavelengths corresponding to $u'$ (355.7\,nm), $g'$ (482.5\,nm), \textit{KG5} (507.5\,nm), $r'$ (626.1\,nm) and $i'$ (767.2\,nm) filters. The name of each system is displayed in the top-right corner of each plot.}
\end{center}
\end{figure}

\begin{figure}\ContinuedFloat
\begin{center}
\includegraphics[width=1.0\columnwidth,trim=0 30 0 30]{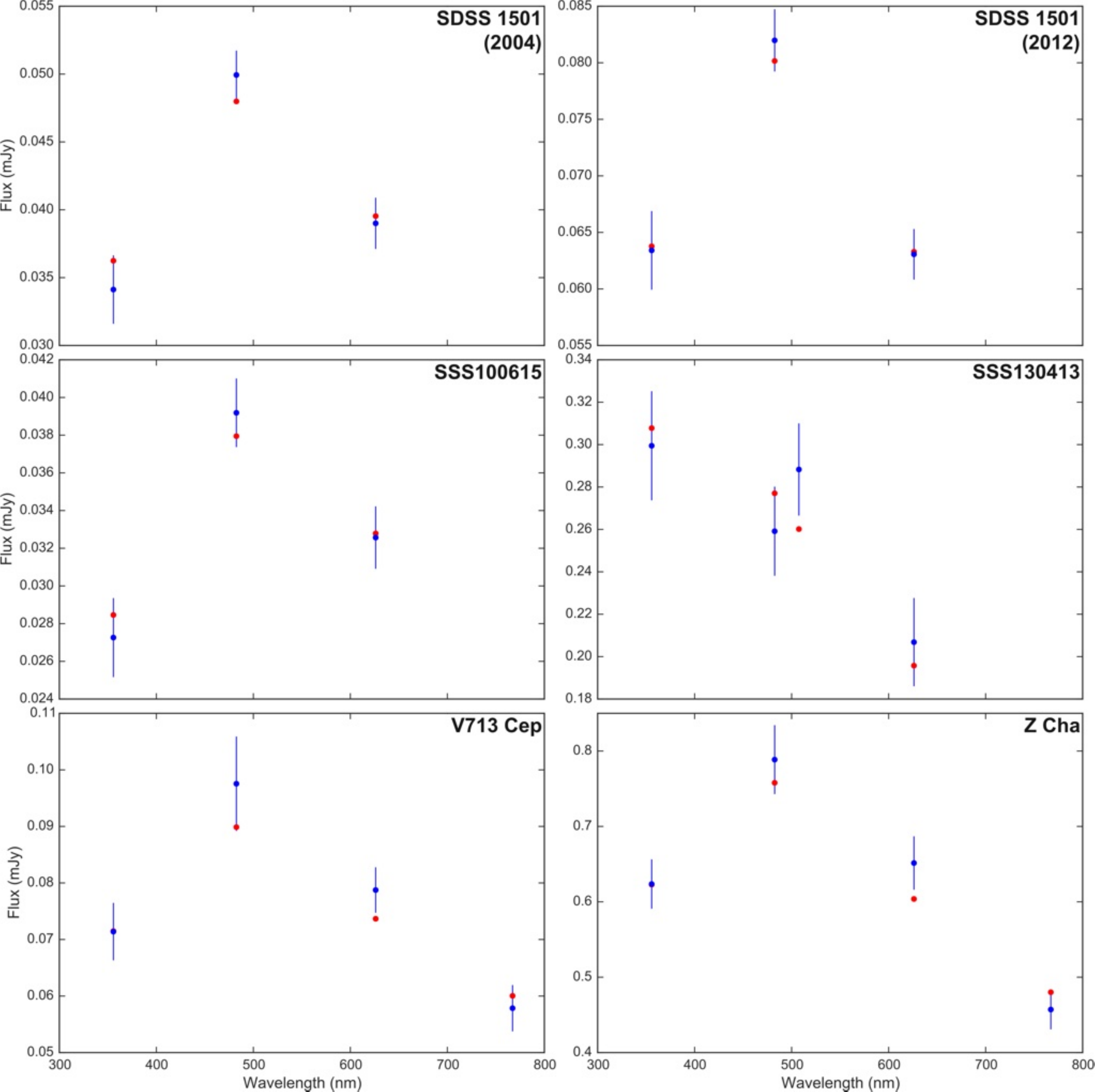}
\caption[\textit{Continued.}]{\textit{Continued.}}
\end{center}
\end{figure}

As a sanity check on our white dwarf atmosphere fitting, we can compare the derived distances with the parallaxes found in Gaia DR2 \citep{lindegren18}. We naively converted our distances to parallaxes, and compared to the parallaxes in Gaia DR2. The results are perfectly consistent with Gaussian statistics; the parallaxes of all but 4 out of 15 CVs agree within 1 standard deviation, whilst the most discrepant CV (GY Cnc) has a 2$\sigma$ discrepancy between our derived distance and the Gaia DR2 parallax. This gives us confidence on our distance estimates and also their uncertainties. 

In section~\ref{sec:discussion} we discuss the implications of the measured system parameters for CV evolution. However, before then, we discuss some remarkable aspects of the data for two individual systems.

\subsection{White Dwarf Flux and Orbital Period Variations in SDSS 1501\:}
\label{subsec:var_sdss1501}

The eclipses of SDSS 1501 are white dwarf dominated, but some show faint bright spot features. There are a total of 15 available ULTRACAM eclipses of SDSS 1501, obtained during observing runs in 2004 (one eclipse), 2006 (eight), 2010 (two) and 2012 (one) (see Table~\ref{table:obsj} for further details). However, only the single eclipses from 2004 and 2012 show signs of a bright spot eclipse, so both\footnote{Each ULTRACAM eclipse is in three bands ($u'g'r'$), giving six individual eclipses for modelling.} were chosen for simultaneous eclipse modelling described above.

It became apparent that there was an appreciable increase in white dwarf flux across all three ($u'g'r'$) bands between the 2004 and 2012 eclipses. For this reason, model atmosphere fitting to the white dwarf fluxes was carried out separately for each eclipse, as shown in Figure~\ref{fig:fluxes_addsys}. The resulting $d$ and $T_{1}$ for each eclipse are shown in Table~\ref{table:syspars_addsys}. The white dwarf in 2012 appears marginally hotter, but note the $1.6\sigma$ discrepancy in $d$, which should of course remain constant. 
Both distances are formally consistent with a formal inference of the distance from Gaia DR2, including a weak distance prior \citep{bailer-jones18}. However, the most likely distance from Gaia DR2 is 340\,pc; favouring the 2012 distance estimate. The white dwarf flux fitting was repeated for both eclipses, but this time with $d$ held fixed at 360\,pc. This now gives $T_{1}\mathrm{(2004)} = 12100\,\pm\,300\,\mathrm{K}$ and $T_{1}\mathrm{(2012)} = 15800\,\pm\,300\,\mathrm{K}$, a much larger increase of 3700\,K.

Such a large discrepancy in $T_{1}$ indicates that the white dwarf in SDSS 1501 underwent a period of enhanced accretion between 2004 and 2012, most likely a superoutburst. According to vsnet-alert~12169\footnote{http://ooruri.kusastro.kyoto-u.ac.jp/mailarchive/vsnet-alert/12169}, the superoutburst occurred in Sep 2010, with the observer claiming to have observed SDSS 1501 in outburst in addition to obtaining part of a superhump. Unfortunately, there is not enough coverage of this outburst to determine a superhump period.

\begin{figure}
\begin{center}
\includegraphics[width=1.0\columnwidth,trim=0 10 0 10]{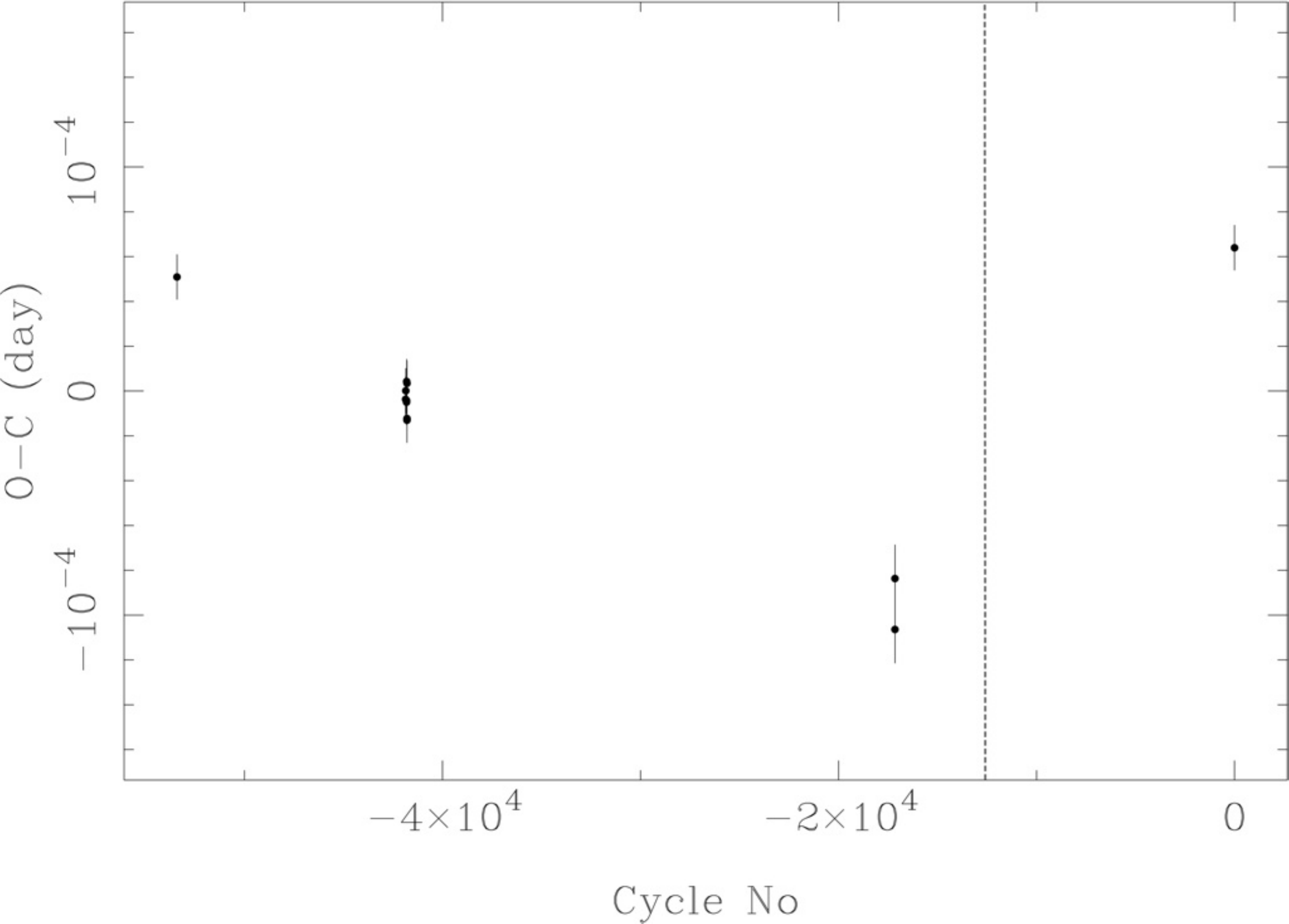}
\caption[$\mathrm{O}-\mathrm{C}$ diagram for all 12 ULTRACAM eclipses of SDSS 1501.]{\label{fig:sdss1501_pchanges}$\mathrm{O}-\mathrm{C}$ diagram for all 12 available ULTRACAM eclipses of SDSS 1501, spanning $\sim$\,8\,yrs. The vertical dashed line corresponds to Sep 2010, when SDSS 1501 was reportedly observed in superoutburst. The $y$-axis covers $\pm15\,\mathrm{s}$.}
\end{center}
\end{figure}

In addition to the white dwarf flux variations, SDSS~1501 also exhibits small orbital period variations. The white dwarf-dominated SDSS 1501 eclipses enable very precise mid-eclipse times to be obtained. We show the mid-eclipse times -- after the subtraction of a linear ephemeris -- in Figure~\ref{fig:sdss1501_pchanges}. The orbital period of SDSS 1501 appears to depart from linearity by approximately $\pm7\,\mathrm{s}$ over the $\sim$\,8\,yr ULTRACAM observational baseline. Such variations are not uncommon in CVs, and are thought to be caused by a magnetically-driven process within the donor. However, they are not observed in CVs with donors of spectral type later than M6 \citep{bours16}, due to magnetic activity in the donor decreasing with later spectral types. SDSS 1501's donor mass obtained through eclipse modelling is substellar ($0.061\pm0.004\,\mathrm{M}_{\odot}$), strongly indicating a spectral type later than M6, and so the observation of period variations is surprising.

A logical deduction from looking at Figure~\ref{fig:sdss1501_pchanges} is that the superoutburst from Sep 2010 (dashed line) may have caused the observed change in orbital period, as the ephemeris appears approximately linear up until this point. In this scenario, the 2012 eclipse occurs $\sim$\,21\,s later than expected, implying an increase in SDSS 1501's orbital period of 0.0016\,s ($\Delta P_{\rm{orb}}/P_{\rm{orb}}=3.2\times10^{-7}$). It is not clear how the superoutburst could have caused such a large change in the orbital period. If some fraction of the disc mass was ejected during superoutburst, we would expect $\Delta P_{\rm orb}/P_{\rm{orb}} = 2M_{\rm ej}/(M_1 + M_2)$, where $M_{\rm ej}$ is the mass ejected. This implies ejected masses of $10^{-7}\, \mathrm{M}_{\odot}$, and disc masses in excess of this. A period change might be induced by a change in the quadropole moment of the white dwarf and disc, due to the disc draining onto the white dwarf. In this case, \cite{applegate92} gives 

$$\Delta P_{\rm{orb}}/P_{\rm{orb}} \approx -\frac{9\Delta Q}{M_1 a^2},$$

where $\Delta Q$ is the change in quadropole moment. We can obtain an order-of-magnitude estimate for $\Delta Q$ if we approximate the disc as a ring of mass $M_{\rm d}$ and radius $a/3$, and assume that during superoutburst the disc completely drains onto the white dwarf, giving $\Delta Q \approx - M_{\rm d} a^2 / 9$. Therefore $\Delta P_{\rm{orb}}/P_{\rm{orb}} \approx M_{\rm d}/M_1$, again implying disc masses of order $10^{-7}\, \mathrm{M}_{\odot}$. With SDSS 1501's system parameters known (Table~\ref{table:syspars_addsys}), the pre-outburst white dwarf temperature can be used to determine a medium-term average mass transfer rate for SDSS 1501 \citep{townsleybildsten03,townsleybildsten04,townsleygaensicke09} of $\dot{M}=9.3\times10^{-11}\,\mathrm{M}_{\odot}\,\mathrm{yr}^{-1}$. Period minimum systems are observed to have superoutburst cycles of order 20--30\,yrs. Therefore the required disc masses are unrealistic, and the Sep 2010 superoutburst is not (at least not fully) responsible for the period variations exhibited by SDSS 1501. Another possible cause of the period variations is the presence of a third body within the system, however additional precise mid-eclipse timings are required in order to investigate this further. 

\subsection{Observed Low State of V713 Cep}
\label{subsec:ls_v713cep}

\begin{figure}
\begin{center}
\includegraphics[width=0.8\columnwidth,trim=40 10 40 10]{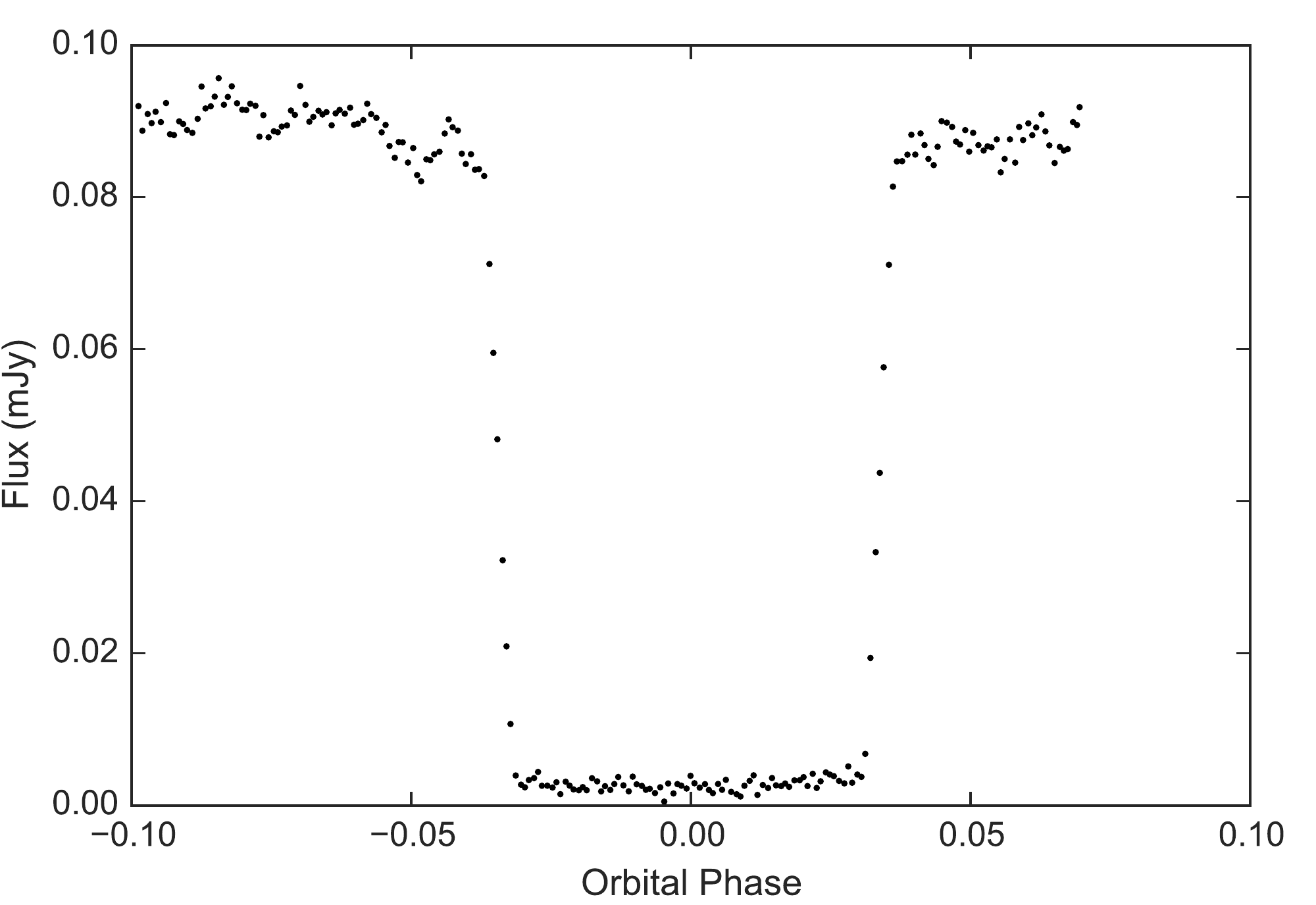}
\caption[$g'$-band eclipse light curve of V713 Cep during a low state.]{\label{fig:v713cep_low}$g'$-band eclipse light curve (24 Jun 2015, cycle no.\ 11955) of V713 Cep during a low state.}
\end{center}
\end{figure}

The ULTRACAM/ULTRASPEC data archive contains a total of 15 V713 Cep eclipses, with two ULTRACAM eclipses (cycle nos.\ 11 [$u'g'r'$] and 3655 [$u'g'i'$]) showing clear bright-spot features suitable for eclipse modelling. A feature of these two eclipses is a notable disc contribution (see Figure~\ref{fig:modfit_v713cep}), which is seen in all other V713 Cep eclipses in the archive, with the exception of one. The ULTRACAM $u'g'r'$ eclipse of 24 Jun 2015 (cycle no.\ 11955, $g'$-band eclipse shown in Figure~\ref{fig:v713cep_low}) contains no obvious signs of either a disc or bright spot eclipse, and at first glance resembles an eclipse of a detached, non-accreting binary. However, on closer inspection there are signs of flickering outside of white dwarf eclipse, as well as a very slight curvature inside eclipse. These two features are both evidence for the presence of an -- albeit considerably diminished -- accretion disc. A dwindling accretion disc and no sign of a bright spot indicates that the secondary has stopped supplying the disc with material and the system is in what is known as a `low state'.

Low states are relatively common phenomena for both magnetic CVs and a subgroup of novalike (NL) CVs called VY Scl stars, however they appear to be very rare (and unexpected) for DNe below the period gap. In fact, there is only one other documented occurrence in the literature -- an extended ($>$\,2\,yrs) low state of IR Com \citep{mansergaensicke14}. Given the rarity of low states in DNe, it is notable that IR Com and V713 Cep have similar orbital periods, just at the lower edge of the period gap. With only one eclipse of V713 Cep obtained during its low state, it is not known exactly how long this low state lasted. An upper limit of 403 days can be estimated based on the timings of other ULTRACAM eclipses, and therefore it was significantly shorter than the low state of IR Com.

\renewcommand{\arraystretch}{1.5}

\begin{table*}
\begin{center}
\setlength{\tabcolsep}{3.5pt}
\begin{tabular}{lcccc}
\hline
Parameter & \textbf{CSS080623} & \textbf{CSS110113} & \textbf{CTCV 1300} & \textbf{DV UMa} \\ \hline
$q$ & $0.114\pm0.005$ & $0.105\pm0.006$ & $0.233\pm0.004$ & $0.172\,^{+0.002}_{-0.007}$ \\
$M_{1}$ ($\mathrm{M}_{\odot}$) & $0.710\pm0.019$ & $1.00\,^{+0.04}_{-0.01}$ & $0.717\pm0.017$ & $1.09\pm0.03$ \\
$R_{1}$ ($\mathrm{R}_{\odot}$) & $0.0117\,^{+0.0001}_{-0.0004}$ & $0.0080\pm0.0003$ & $0.01133\pm0.00021$ & $0.0072\pm0.0004$ \\
$M_{2}$ ($\mathrm{M}_{\odot}$) & $0.081\pm0.005$ & $0.105\pm0.007$ & $0.166\,^{+0.006}_{-0.003}$ & $0.187\,^{+0.003}_{-0.012}$ \\
$R_{2}$ ($\mathrm{R}_{\odot}$) & $0.1275\pm0.0024$ & $0.149\pm0.003$ & $0.2111\,^{+0.0025}_{-0.0014}$ & $0.215\,^{+0.001}_{-0.005}$ \\
$a$ ($\mathrm{R}_{\odot}$) & $0.593\pm0.005$ & $0.711\,^{+0.009}_{-0.003}$ & $0.805\pm0.007$ & $0.889\,^{+0.006}_{-0.012}$ \\
$K_{1}$ (km\,s$^{-1}$) & $50.8\pm2.3$ & $51.1\,^{+2.9}_{-2.4}$ & $86.4\pm1.4$ & $76.1\,^{+0.9}_{-2.9}$ \\
$K_{2}$ (km\,s$^{-1}$) & $449\,^{+1}_{-6}$ & $487\pm3$ & $371\pm3$ & $444\pm4$ \\
$i$ $(^{\circ})$ & $80.76\pm0.19$ & $79.94\pm0.19$ & $86.9\,^{+0.5}_{-0.2}$ & $83.29\,^{+0.29}_{-0.10}$ \\
log\,$g$ & $8.15\,^{+0.01}_{-0.04}$ & $8.63\pm0.03$ & $8.186\pm0.019$ & $8.77\pm0.04$ \\ \hline
$T_{1}$ (K) & $15500\pm1700$ & $14500\pm2200$ & $11000\pm1000$ & $17400\pm1900$ \\
$d$ (pc) & $550\pm60$ & $430\pm60$ & $340\pm40$ & $380\pm40$ \\ \hline \hline
Parameter & \textbf{GY Cnc} & \textbf{IY UMa} & \textbf{OY Car} & \textbf{SDSS 0901} \\ \hline
$q$ & $0.448\,^{+0.014}_{-0.021}$ & $0.146\,^{+0.009}_{-0.001}$ & $0.1065\,^{+0.0009}_{-0.0029}$ & $0.182\,^{+0.009}_{-0.004}$ \\
$M_{1}$ ($\mathrm{M}_{\odot}$) & $0.881\pm0.016$ & $0.955\,^{+0.013}_{-0.028}$ & $0.882\,^{+0.011}_{-0.015}$ & $0.752\,^{+0.024}_{-0.018}$ \\
$R_{1}$ ($\mathrm{R}_{\odot}$) & $0.00976\,^{+0.00021}_{-0.00018}$ & $0.0087\,^{+0.0003}_{-0.0001}$ & $0.00957\,^{+0.00018}_{-0.00012}$ & $0.01105\,^{+0.00022}_{-0.00029}$ \\
$M_{2}$ ($\mathrm{M}_{\odot}$) & $0.394\,^{+0.016}_{-0.022}$ & $0.141\pm0.007$ & $0.093\,^{+0.004}_{-0.001}$ & $0.138\pm0.007$ \\
$R_{2}$ ($\mathrm{R}_{\odot}$) & $0.446\,^{+0.006}_{-0.009}$ & $0.1770\pm0.0028$ & $0.1388\,^{+0.0018}_{-0.0003}$ & $0.182\pm0.003$ \\
$a$ ($\mathrm{R}_{\odot}$) & $1.429\pm0.012$ & $0.765\,^{+0.004}_{-0.009}$ & $0.662\pm0.003$ & $0.739\pm0.007$ \\
$K_{1}$ (km\,s$^{-1}$) & $125\pm4$ & $66\,^{+4}_{-1}$ & $50.4\pm0.9$ & $73\pm3$ \\
$K_{2}$ (km\,s$^{-1}$) & $278.0\pm2.4$ & $453\pm3$ & $475.9\pm2.1$ & $401\pm3$ \\
$i$ $(^{\circ})$ & $77.06\,^{+0.29}_{-0.18}$ & $84.9\,^{+0.1}_{-0.5}$ & $83.27\,^{+0.10}_{-0.13}$ & $81.4\,^{+0.1}_{-0.3}$ \\ 
log\,$g$ & $8.40\pm0.019$ & $8.54\pm0.03$ & $8.422\,^{+0.017}_{-0.013}$ & $8.228\,^{+0.022}_{-0.025}$ \\ \hline
$T_{1}$ (K) & $25900\pm2300$ & -- & $18600\,^{+2800}_{-1600}$ & $14900\pm2000$ \\
$d$ (pc) & $320\pm30$ & -- & $90\pm5$ & $600\pm70$ \\ \hline
\end{tabular}
\caption[System parameters for 15 eclipsing systems.]{\label{table:syspars_addsys}System parameters for the 15 eclipsing systems analysed in this paper.}
\end{center}
\end{table*}

\begin{table*}\ContinuedFloat
\begin{center}
\setlength{\tabcolsep}{3.5pt}
\begin{tabular}{lcccc}
\hline
Parameter & \textbf{SDSS 1006} & \textbf{SDSS 1152} & \textbf{SDSS 1501} & \textbf{SSS100615} \\ \hline
$q$ & $0.46\pm0.03$ & $0.153\,^{+0.015}_{-0.011}$ & $0.084\pm0.004$ & $0.095\pm0.004$ \\
$M_{1}$ ($\mathrm{M}_{\odot}$) & $0.82\pm0.11$ & $0.62\pm0.04$ & $0.723\,^{+0.017}_{-0.013}$ & $0.88\pm0.03$ \\
$R_{1}$ ($\mathrm{R}_{\odot}$) & $0.0102\pm0.0013$ & $0.0129\pm0.0006$ & $0.01142\,^{+0.00016}_{-0.00022}$ & $0.0095\pm0.0003$ \\
$M_{2}$ ($\mathrm{M}_{\odot}$) & $0.37\pm0.06$ & $0.094\,^{+0.016}_{-0.009}$ & $0.061\pm0.004$ & $0.083\pm0.005$ \\
$R_{2}$ ($\mathrm{R}_{\odot}$) & $0.457\,^{+0.022}_{-0.026}$ & $0.147\pm0.006$ & $0.1129\,^{+0.0025}_{-0.0016}$ & $0.1276\,^{+0.0028}_{-0.0024}$ \\
$a$ ($\mathrm{R}_{\odot}$) & $1.46\pm0.07$ & $0.627\pm0.014$ & $0.574\pm0.004$ & $0.628\pm0.007$ \\
$K_{1}$ (km\,s$^{-1}$) & $124\pm9$ & $62\pm5$ & $39.5\,^{+2.2}_{-1.3}$ & $46.5\,^{+2.2}_{-1.7}$ \\
$K_{2}$ (km\,s$^{-1}$) & $270\pm13$ & $402\pm7$ & $468\pm3$ & $493\pm5$ \\
$i$ $(^{\circ})$ & $83.1\,^{+1.2}_{-0.7}$ & $82.6\pm0.5$ & $83.89\,^{+0.20}_{-0.27}$ & $85.1\pm0.3$ \\
log\,$g$ & $8.33\pm0.13$ & $8.01\pm0.05$ & $8.182\,^{+0.016}_{-0.019}$ & $8.43\pm0.03$ \\ \hline
$T_{1}$ (K) & -- & $15900\pm2000$ & $^{13400\,\pm\,1100\,(2004)}_{14900\,\pm\,1000\,(2012)}$ & $13600\pm1500$ \\
$d$ (pc) & -- & $610\pm80$ & $^{400\,\pm\,30\,(2004)}_{338\,\pm\,21\,(2012)}$ & $350\pm30$ \\ \hline \noalign{\vskip 0.7mm} \cline{1-4}
Parameter & \textbf{SSS130413} & \textbf{V713 Cep} & \textbf{Z Cha} &  \\ \cline{1-4}
$q$ & $0.169\,^{+0.011}_{-0.006}$ & $0.246\,^{+0.006}_{-0.014}$ & $0.189\pm0.004$ \\
$M_{1}$ ($\mathrm{M}_{\odot}$) & $0.84\pm0.03$ & $0.703\,^{+0.012}_{-0.015}$ & $0.803\pm0.014$ \\
$R_{1}$ ($\mathrm{R}_{\odot}$) & $0.0102\,^{+0.0006}_{-0.0002}$ & $0.01173\,^{+0.00020}_{-0.00015}$ & $0.01046\pm0.00017$ \\
$M_{2}$ ($\mathrm{M}_{\odot}$) & $0.140\,^{+0.012}_{-0.008}$ & $0.176\,^{+0.007}_{-0.018}$ & $0.152\pm0.005$ \\
$R_{2}$ ($\mathrm{R}_{\odot}$) & $0.163\pm0.004$ & $0.208\,^{+0.002}_{-0.005}$ & $0.1820\pm0.0020$ \\
$a$ ($\mathrm{R}_{\odot}$) & $0.680\,^{+0.007}_{-0.011}$ & $0.781\pm0.006$ & $0.734\pm0.005$ \\
$K_{1}$ (km\,s$^{-1}$) & $75\pm4$ & $91\,^{+2}_{-5}$ & $78.4\,^{+1.4}_{-1.8}$ \\
$K_{2}$ (km\,s$^{-1}$) & $443\,^{+3}_{-7}$ & $367.6\,^{+2.6}_{-2.3}$ & $413.2\,^{+2.5}_{-2.0}$ \\
$i$ $(^{\circ})$ & $82.5\pm0.3$ & $81.7\pm0.3$ & $80.44\pm0.11$ \\
log\,$g$ & $8.35\pm0.04$ & $8.147\,^{+0.017}_{-0.014}$ & $8.304\pm0.016$ \\ \cline{1-4}
$T_{1}$ (K) & $24000\pm3000$ & $17000\,^{+6000}_{-3000}$ & $16300\pm1400$ \\
$d$ (pc) & $240\pm40$ & $320\pm30$ & $103\pm6$ \\ \cline{1-4}
\end{tabular}
\caption[\textit{Continued.}]{\textit{Continued.}}
\end{center}
\end{table*}

\renewcommand{\arraystretch}{1.0}
 
\section{Discussion}
\label{sec:discussion}
With the new and revised system parameters obtained in this work, we now discuss what impact these results may have on the current understanding of CVs and their evolution. In what follows, we combine the parameters presented here with a compilation of reliable parameters for 46 CVs from the literature. This compilation is presented in Table~\ref{table:supp_sys}.

It has been shown that there is a significant discrepancy between the mean white dwarf mass in the field and that within CVs. \cite{zorotovic11} obtained a mean CV white dwarf mass of $0.82\pm0.03\,\mathrm{M}_{\odot}$, and an intrinsic scatter of white dwarf masses of $\sigma=0.15\,\mathrm{M}_{\odot}$. With the updated sample of CV masses now available, we can revise the mean white dwarf mass in CVs, following the procedure outlined in Appendix B of \cite{knigge06}, to $0.81\pm0.02\,\mathrm{M}_{\odot}$ ($\sigma=0.13\,\mathrm{M}_{\odot}$), entirely consistent with Zorotovic et al's value.

One way to explain the presence of high white dwarf masses in CVs is through white dwarf mass growth through steady accretion across the lifetime of a CV. Since CVs evolve to shorter orbital periods over their lives, this requires the observation of higher white dwarf masses in systems with lower orbital periods. To test this, $\langle{M_{1}}\rangle$ was re-calculated for 31 systems below the period gap ($P_{\mathrm{orb}}\sim2.15\,\mathrm{hrs}$), giving $\langle{M_{1}}\rangle\mathrm{(below\:gap)} = 0.81\pm0.02\,\mathrm{M}_{\odot}$ ($\sigma=0.10\,\mathrm{M}_{\odot}$), and for 16 systems above the gap ($P_{\mathrm{orb}}\sim3.18\,\mathrm{hrs}$), giving $\langle{M_{1}}\rangle\mathrm{(below\:gap)} = 0.82\pm0.02\,\mathrm{M}_{\odot}$ ($\sigma=0.10\,\mathrm{M}_{\odot}$). We therefore see no evidence for white dwarf mass growth in CVs. While white dwarf mass growth in CVs appears doubtful, further precise white dwarf masses from systems at long period ($>$\,3\,hrs) are required before it can be entirely dismissed.

\subsection{Testing the Validity of the Empirical CAML Model}
\label{subsec:ecaml_test}

An alternative explanation for the high white dwarf mass in CVs was proposed by \cite{schreiber16}. The authors put forward an empirical consequential angular momentum loss (eCAML) model, which produces a  dynamical stability limit on $q$, causing systems with low-mass white dwarfs to become unstable to mass transfer. These systems consequently merge, removing them from the CV population. The eCAML model is attractive as it can simultaneously explain the low observed space density of CVs \citep{belloni18} and the origin of isolated low-mass white dwarfs \citep{zorotovic17}.

\begin{figure}
\begin{center}
\includegraphics[width=1.0\columnwidth,trim=0 5 0 10]{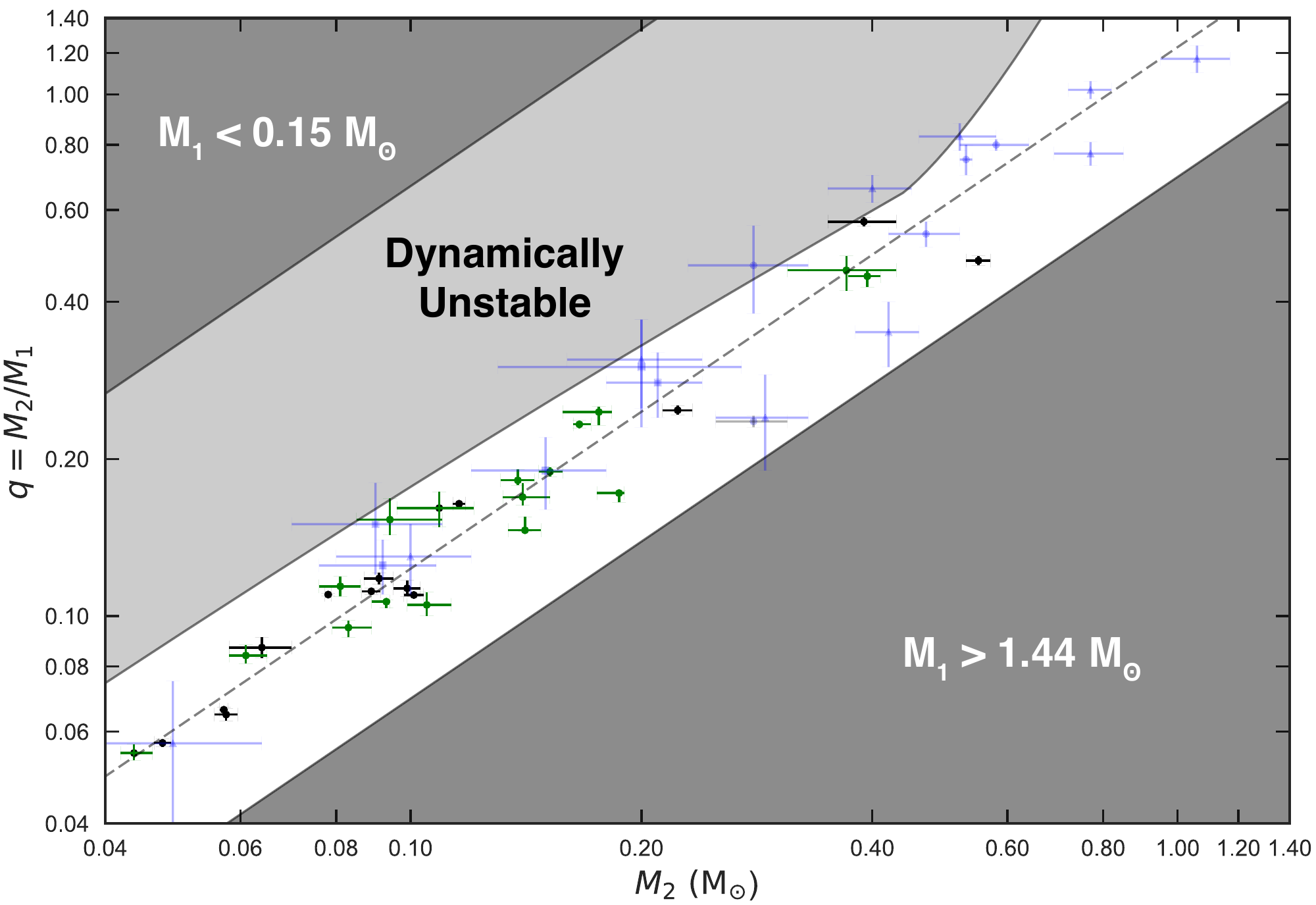}
\caption[$q$ vs $M_{2}$ plot for CVs.]{\label{fig:ecaml}$q$ vs $M_{2}$ plot for CVs. The grey regions are theoretically prohibited due to constraints put on $M_{1}$. The dark grey regions cover unrealistically low white dwarf masses ($\lesssim$\,0.15\,M$_{\odot}$) and masses greater than the Chandrasekhar mass limit (1.44\,M$_{\odot}$), while the light grey region is forbidden by the empirical consequential angular momentum loss (eCAML) model of \cite{schreiber16}. The dashed grey line represents the mean value of $M_{1}$ from this work. The green and black points represent masses obtained from eclipse modelling of ULTRACAM/ULTRASPEC data, either from this work (green) or otherwise (black). The faint blue points represent measured CV masses from other methods: eclipse modelling of other data (circles), contact phase timing (squares), and radial velocity (triangles).}
\end{center}
\end{figure}

The top-left plot of Figure 2 in \cite{schreiber16} was updated to take into account the results of this work (Figure~\ref{fig:ecaml}). This plot is in $M_{2}$ vs $q$ parameter space, with regions (grey) that are theoretically prohibited due to constraints put on $M_{1}$. The dark grey prohibited region in the bottom right of Figure~\ref{fig:ecaml} is an upper mass limit on $M_{1}$, resulting from the Chandrasekhar mass limit of a white dwarf (1.44\,M$_{\odot}$). The light grey prohibited region is a lower mass limit on $M_{1}$ and is a consequence of the dynamical stability limit on $q$ supplied by the eCAML model. Also plotted in Figure~\ref{fig:ecaml} are systems with measured $M_{2}$ and $q$, either from this work (green points) or elsewhere (black/blue points; see Table~\ref{table:supp_sys}). These systems with measured system parameters provide a test of the eCAML model, as all should lie within the valid region (white). Any systems lying inside the prohibited dynamically unstable region would compromise the credibility of the model.

All systems modelled in this work lie comfortably within the valid region of Figure~\ref{fig:ecaml}, along with the vast majority of other systems. Two appear to (just) violate the dynamical instability constraint, namely SDSS 0756+0858 \citep{tovmassian14} and DQ Her \citep{horne93}, however both systems could feasibly be stable under the eCAML model after taking into account their uncertainties. This outcome offers support to the validity of the eCAML model as a solution to the CV white dwarf mass problem, however a much larger sample of systems with precise system parameters is necessary in order to provide a more stringent test of the model.

\subsection{Reviewing the Properties of the Period Spike}
\label{sec:pps}
The period spike is a feature of the orbital period distribution which is expected to occur as systems ``pile-up'' near the orbital period minimum due to the long evolutionary timescale. It was finally observed by \cite{gaensicke09} through analysing the orbital period distribution of newly identified CVs from SDSS \citep{york00}. These systems were all identified spectroscopically (e.g.\ \citealt{szkody02}), and therefore not affected by the same biases/limitations as systems discovered through other means, e.g.\ DN outbursts and X-ray emission (see \citealt{gaensicke09} for more details). Spectroscopic identification, coupled with a survey depth of $g'\sim19.5$, gives this particular sample the ability to provide the closest representation of the true orbital period distribution of CVs to date, a claim supported by the emergence of the long predicted-but-elusive period spike at the period minimum. \cite{gaensicke09} produced estimates for the location ($82.4\pm0.7\,\mathrm{min}$) and width (FWHM = 5.7\,min) of the period spike. Eight years on, the sample has increased and more $P_{\mathrm{orb}}$ measurements have become available, enabling the orbital period distribution -- and in particular the properties of the period spike -- to be reviewed.

The \cite{gaensicke09} sample consisted of 49 spectroscopically identified SDSS CVs below the period gap ($P_{\mathrm{orb}}\lesssim129$\,min; \citealt{knigge06}) with precise $P_{\mathrm{orb}}$ measurements (errors $<30$\,s). Precise $P_{\mathrm{orb}}$ measurements for an additional 23 systems (and updated measurements for a handful from the original sample) have since become available, increasing the sample to 72 systems. Of the new systems, six are eclipsing systems with observations using ULTRACAM/ULTRASPEC, 10 are from \cite{thorstensen15,thorstensen17} and the remaining seven are from the \cite{ritterkolb03} catalogue (v7.24; see references within). All systems were discovered by the SDSS (e.g.\ \citealt{szkody11}) except two, PHL 1445 and  CSS110113, which were discovered by the 6dF Galaxy Survey (6dFGS; \citealt{jones04}).

\begin{figure}
\begin{center}
\includegraphics[width=1.0\columnwidth,trim=10 10 0 20]{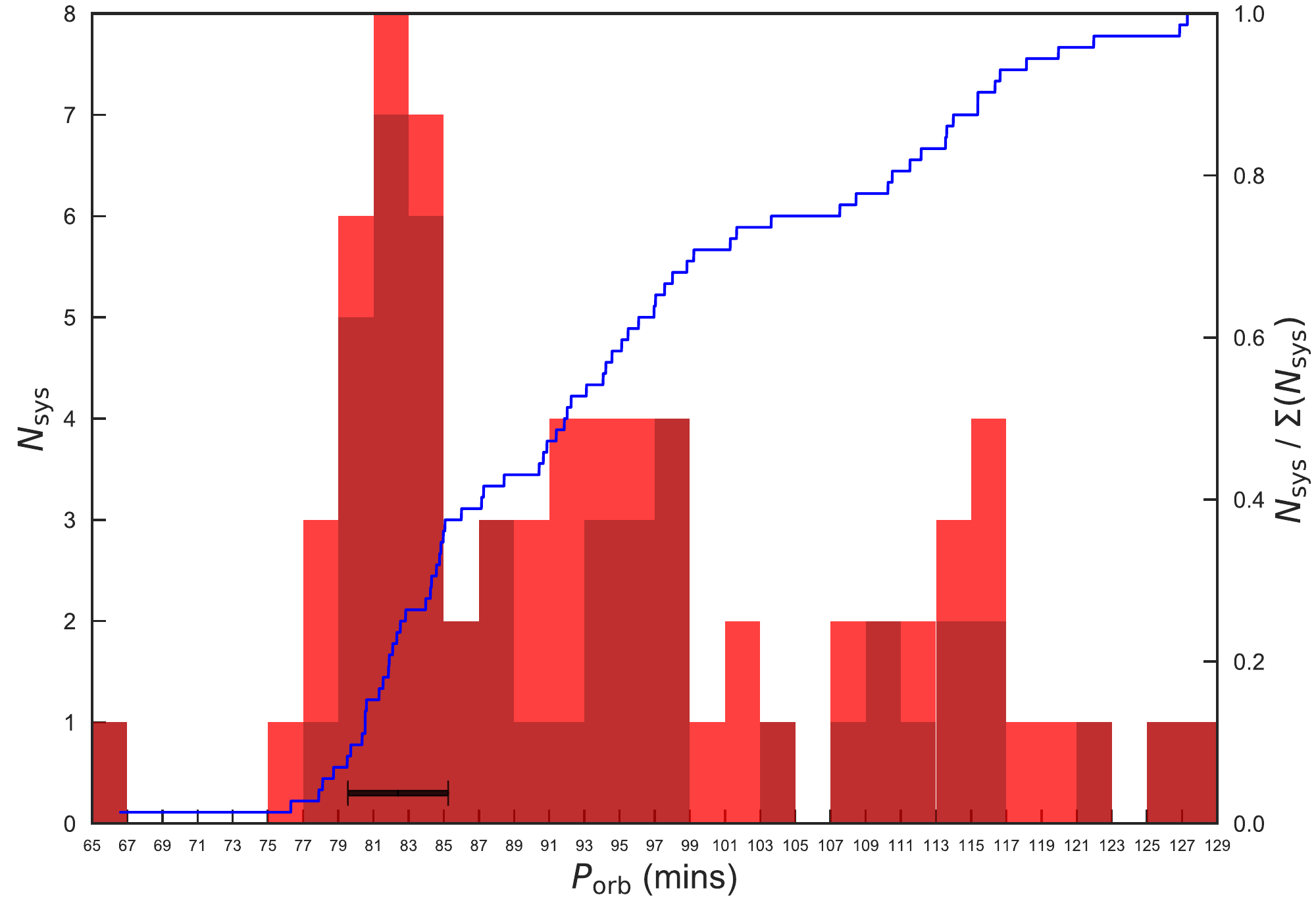}
\caption[Updated orbital period distribution for spectroscopically identified CVs below the period gap.]{\label{fig:porb_hist}Histogram (red) and cumulative plot (blue) for 72 spectroscopically identified (from SDSS and 6dFGS) CVs below the period gap with precise $P_{\mathrm{orb}}$ measurements (sub-30\,s errors). For comparison, the sample of \cite{gaensicke09} is also shown (dark red histogram), in addition to the position and FWHM of the period spike estimated in the same study (black bar).}
\end{center}
\end{figure}

Figure~\ref{fig:porb_hist} shows the orbital period distribution of all 72 spectroscopically identified CVs in the form of both a histogram (red) and cumulative plot (blue). As with the \cite{gaensicke09} sample (dark red histogram), the new sample shows a clear accumulation of systems centred around $\sim$\,82\,min, which is clearly identifiable as the period spike. Estimating $P_{\mathrm{spike}}$ involved the fitting of a Gaussian distribution to the orbital period distribution between 77 and 87\,min. An estimate of $P_{\mathrm{spike}}=82.7\pm0.4$\,min ($\sigma=2.35$\,min, $\mathrm{FWHM}=5.53$\,min) was obtained, which is largely unchanged from the \cite{gaensicke09} sample. This is not surprising, as the majority ($\sim$\,75\%) of additional systems have $P_{\mathrm{orb}}>\,89$\,min, and therefore do not belong to the period spike. 

We note here that there is a hint of bi-modality in the period distribution of systems {\em below} the period gap, with a dearth of systems with orbital periods around 88 minutes. A Hartigan dip test \citep{hartigan85} reveals that this is not statistically significant.

\subsection{Updating the Calibration of the superhump period excess-mass ratio Relation}
\label{sec:updated_epq}

During superoutburst the accretion disc is driven into an elliptical state by resonances between the donor star and material within the disc. Tidal interactions between the elliptical disc and the donor lead to periodic fluctuations in the elliptical, precessing, disc known as superhumps. The disc precesses at a slow rate, with a period ($P_{\mathrm{prec}}$) significantly longer than $P_{\mathrm{orb}}$. These two periods therefore both contribute to the formation of the superhump period ($P_{\mathrm{sh}}$), which is simply the `beat period' of $P_{\mathrm{prec}}$ and $P_{\mathrm{orb}}$ \citep{hellier01}:
\begin{equation}
\frac{1}{P_{\mathrm{sh}}} = \frac{1}{P_{\mathrm{orb}}} - \frac{1}{P_{\mathrm{prec}}}.
\label{eq:p_sh}
\end{equation}
$P_{\mathrm{sh}}$ is therefore usually a few percent longer than $P_{\mathrm{orb}}$, but does not stay constant throughout the superoutburst. In fact, a superoutburst can be split up into three distinct stages (A, B and C), with sharp transitions observed between each stage. Stage A represents the start of the superoutburst, with a long, stable $P_{\mathrm{sh}}$. Stage B is the middle part of the superoutburst, with a shorter, unstable $P_{\mathrm{sh}}$. The final stage (C) exhibits the shortest $P_{\mathrm{sh}}$, which is stable once again \citep{olech03,kato09}. The general trend of decreasing $P_{\mathrm{sh}}$ across the superoutburst hints at an increasing $P_{\mathrm{prec}}$ (from equation~\ref{eq:p_sh}) and therefore a dwindling disc radius \citep{murray00}.

The superhump excess ($\epsilon$) is defined as $\epsilon = \frac{P_{\mathrm{sh}} - P_{\mathrm{orb}}}{P_{\mathrm{orb}}}$, and is directly related to the mass ratio, $q$. A calibration of this relationship \cite[e.g][]{patterson05, knigge06} allows estimates of mass ratios for all superhumping systems. From this current work and the work of others (e.g.\ \citealt{savoury11}), new potential calibration systems have emerged, in addition to revised $q$ values for existing calibration systems. Revised superhump periods have also been measured, courtesy of the SU UMa-type DNe survey of \cite{kato09,kato10,kato12,kato13,kato14a,kato14b,kato15,kato16,kato17}. With all of these new measurements becoming available since the work of \cite{knigge06}, it is appropriate to update the calibration of the $\epsilon$($q$) relation.

\begin{table*}
\setlength{\tabcolsep}{4pt}
\begin{center}
\begin{tabular}{lcccl}
\hline
System & $P_{\mathrm{orb}}$\,(d) & $P_{\mathrm{sh}}^{\mathrm{B}}$\,(d) & $P_{\mathrm{sh}}^{\mathrm{C}}$\,(d) & Ref.(s) \\ \hline
SDSS 1507 & 0.04625828(4) & 0.046825(4) & -- & 1,2 \\
SSS100615$^{*}$ & 0.0587045(4)$^{\S}$ & 0.05972(9) & -- & 3 \\
SDSS 1502 & 0.05890961(5) & 0.060463(13) & 0.060145(19) & 1,4 \\
SDSS 0903 & 0.059073543(9) & 0.06036(5) & 0.06007(5) & 1,4 \\
ASASSN-14ag & 0.060310665(9)$^{\S}$ & 0.06206(6) & -- & 5 \\
XZ Eri & 0.061159491(5) & 0.062807(18) & 0.06265(12) & 1,6 \\
SDSS 1227 & 0.062959041(7) & 0.064604(29) & 0.06440(5) & 1,7 \\
OY Car$^{*}$ & 0.06312092545(24)$^{\S}$ & 0.064653(28) & 0.06444(5) & 8\\
SSS130413$^{*}$ & 0.0657692903(12)$^{\S}$ & -- & 0.06751(24) & 5 \\
CSS110113$^{*}$ & 0.0660508707(18)$^{\S}$ & 0.067583(26) & 0.06731(4) & 7 \\
SDSS 1152$^{*}$ & 0.0677497026(3)$^{\S}$ & 0.07036(4) & 0.069914(19) & 8\\
OU Vir & 0.072706113(5) & 0.074912(17) & -- & 1,6 \\
IY UMa$^{*}$ & 0.07390892818(21)$^{\S}$ & 0.076210(25) & 0.075729(19) & 4 \\
Z Cha$^{*}$ & 0.0744992631(3)$^{\S}$ & 0.07736(8) & 0.076948(23) & 5 \\
SDSS 0901$^{*}$ & 0.0778805321(5)$^{\S}$ & 0.08109(5) & 0.08072(10) & 9 \\
DV UMa$^{*}$ & 0.0858526308(7)$^{\S}$ & 0.08880(3) & 0.08841(3) & 6 \\
SDSS 1702 & 0.10008209(9) & 0.10507(8) & -- & 1,6 \\ 
\hline
WZ Sge & 0.0566878460(3) & 0.057204(5) & -- & 6,10 \\
V2051 Oph & 0.06242785751(8)$^{\S}$ & 0.06471(9) & 0.06414(4) & 5,11 \\
HT Cas & 0.0736471745(5)$^{\S}$ & 0.076333(5) & 0.075886(5) & 3,12 \\
V4140 Sgr & 0.0614296779(9) & 0.06351(4) & 0.06309(7) & 6,11 \\
\hline
V348 Pup & 0.101838931(14) & 0.108567(2)$^{\dagger}$ & -- & 13 \\
V603 Aql & 0.13820103(8) & 0.14686(7)$^{\dagger}$ & -- & 14,15\\
DW UMa & 0.136606499(3) & 0.14539(13)$^{\dagger}$ & -- & 16,17 \\
UU Aqr & 0.1638049430 & 0.17510(18)$^{\dagger}$ & -- & 18,19 \\
\hline
\end{tabular}
\caption[Orbital and superhump periods of the systems used to calibrate the $\epsilon(q)$ relation.]{\label{table:sh_calib}Orbital ($P_{\mathrm{orb}}$) and superhump ($P_{\mathrm{sh}}$) periods of the systems used to calibrate the $\epsilon(q)$ relation. The majority of systems are SU UMa-type DNe, however the bottom four are CNe/NLs. $P_{\mathrm{sh}}^{\mathrm{B}}$ and $P_{\mathrm{sh}}^{\mathrm{C}}$ are the periods for stage B and C superhumps, respectively. See Tables~\ref{table:syspars_addsys} and~\ref{table:supp_sys} for $q$ values. References: (1) \cite{savoury11}, (2) \cite{patterson17}, (3) \cite{kato16}, (4) \cite{kato10}, (5) \cite{kato15}, (6) \cite{kato09}, (7) \cite{kato12}, (8) \cite{kato17}, (9) \cite{kato13}, (10) \cite{patterson98}, (11) \cite{baptista03}, (12) \cite{horne91}, (13) \cite{rolfe00}, (14) \cite{peters06}, (15) \cite{patterson97}, (16) \cite{araujobetancor03}, (17) \cite{patterson02}, (18) \cite{baptistabortoletto08}, (19) \cite{patterson05}. 
\vspace{0.15cm}\\

$^{*}$Updated $q$ value produced in this work (Table~\ref{table:syspars_addsys}), $^{\dagger}$Superhump period from permanent superhumps, $^{\S}P_{\mathrm{orb}}$ from this work}
\end{center}
\end{table*}

Table~\ref{table:sh_calib} contains all of the calibrating systems currently available\footnote{The calibration system KV UMa used by \cite{knigge06} was not included on the basis of it being a low-mass X-ray binary, rather than a CV.}, along with their orbital and superhump periods (and references). The two superhump period columns, $P_{\mathrm{sh}}^{\mathrm{B}}$ and $P_{\mathrm{sh}}^{\mathrm{C}}$, represent the superhump periods during stage B and stage C of superoutburst, respectively. All but the final four systems in Table~\ref{table:sh_calib} are SU UMa-type DNe that undergo superoutbursts. The other four systems are either Classical Novae or Novalikes that display permanent superhumps, and it is assumed these superhump periods resemble those of $P_{\mathrm{sh}}^{\mathrm{B}}$ for SU UMa-type DNe.

\begin{figure}
\begin{center}
\includegraphics[width=1.0\columnwidth,trim=0 0 0 0]{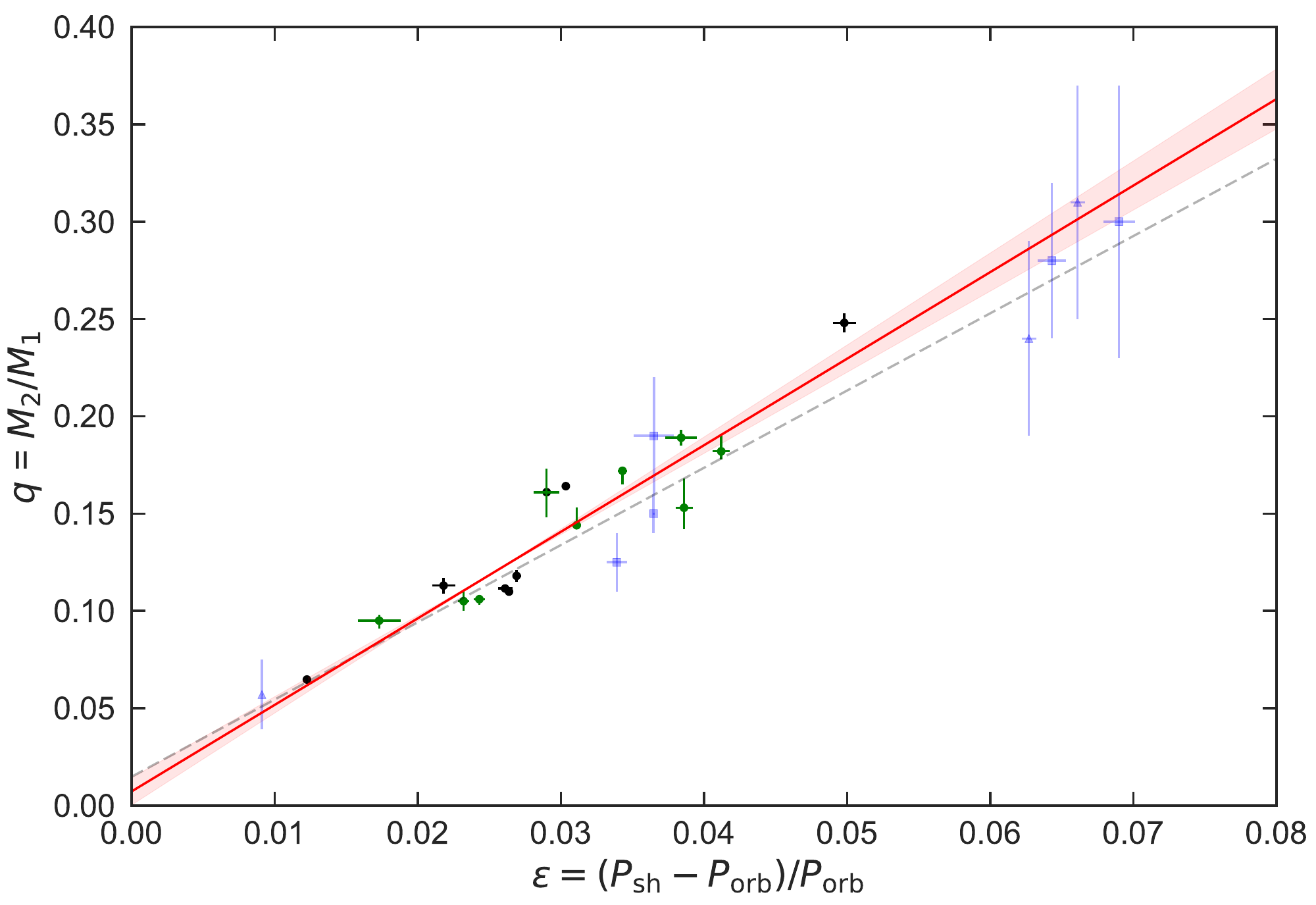}
\includegraphics[width=1.0\columnwidth,trim=0 0 0 0]{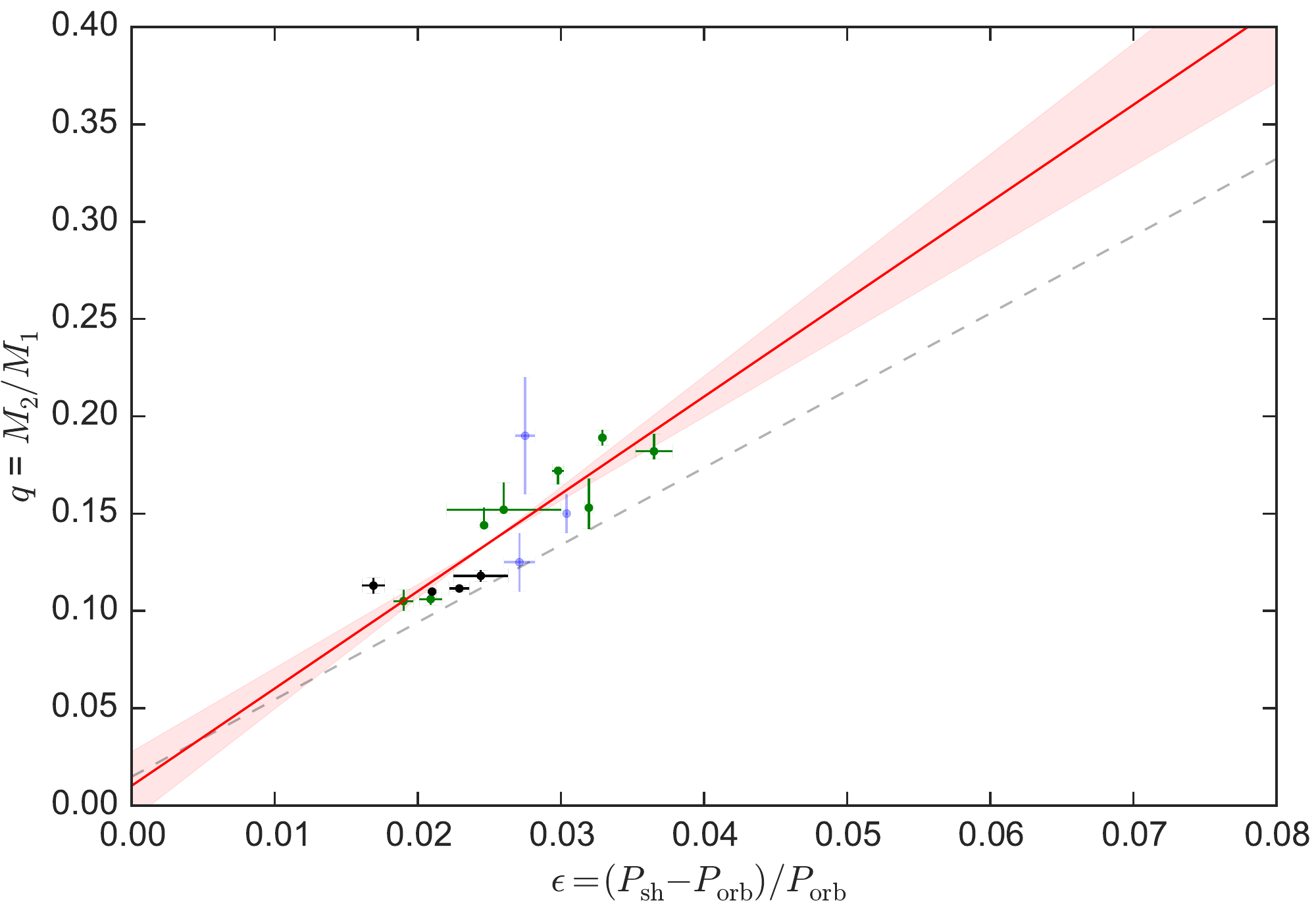}
\caption[Updated calibration of the $\epsilon(q)$ relation.]{\label{fig:epsilon-q_b}Measured $\epsilon_{\mathrm{B}}$ and $q$ values of superhumping and eclipsing CVs, with the same data point colour/shape scheme as Figure~\ref{fig:ecaml}. The dashed grey line shows the existing linear calibration of the $\epsilon(q)$ relation for superhumping CVs from \cite{knigge06}, while the red line shows an updated calibration from this work. The red shaded region represents 1$\sigma$ errors. The top plot shows the relationship for stage B superhumps, the bottom plot that for stage C superhumps.}
\end{center}
\end{figure}

For each system in Table~\ref{table:sh_calib}, the superhump period excess was calculated for stage B ($\epsilon_{\mathrm{B}}$) and stage C ($\epsilon_{\mathrm{C}}$) depending on $P_{\mathrm{sh}}^{\mathrm{B}}/P_{\mathrm{sh}}^{\mathrm{C}}$ availability. Figure~\ref{fig:epsilon-q_b} shows $\epsilon_{\mathrm{B}}$ plotted against $q$ for the 24 calibration systems from Table~\ref{table:sh_calib} with available $P_{\mathrm{sh}}^{\mathrm{B}}$ measurements. The dashed grey line shows the existing calibration from \cite{knigge06}, while the red line represents the following, updated linear calibration:
\begin{equation}
q(\epsilon_{\mathrm{B}}) = (0.118 \pm 0.003) + (4.45 \pm 0.28) \times (\epsilon_{\mathrm{B}}-0.025).
\label{eq:epsilon_q_b}
\end{equation}
This updated calibration was obtained through the same $\chi^{2}$ minimisation technique employed by \cite{knigge06} (see Appendix A of reference), and has an intrinsic dispersion ($\sigma$) of 0.012. While there is good coverage for systems with $0.1<q<0.2$, more calibration systems with $q$ outside this range are required in order to further constrain the gradient. For example, due to its position in Figure~\ref{fig:epsilon-q_b}, SDSS 1702 ($q\approx0.25$) has a rather large influence on the gradient, so therefore more systems with precisely measured values of $q$ greater than 0.2 are highly coveted. Unfortunately, this includes period gap systems, which are rare, and systems above the gap, for which precise measurements of $q$ are hard to obtain. It is clear from Figure~\ref{fig:epsilon-q_b} that the new calibration has a steeper gradient that the existing one from \cite{knigge06}. A possible reason for this is the variation in measurement of $P_{\mathrm{sh}}$ between \cite{patterson05} and \cite{kato09}; the sources of $P_{\mathrm{sh}}$ for both the existing and new calibration, respectively. \cite{patterson05} measures $P_{\mathrm{sh}}$ from `common' superhumps, which typically cover stage B, but can also cover only a fraction of this stage or spread into stages A and C.

The same treatment was given to the 15 calibration systems in Table~\ref{table:sh_calib} with available $P_{\mathrm{sh}}^{\mathrm{C}}$ measurements, producing the following linear relation (with $\sigma=0.012$ again inferred):
\begin{equation}
q(\epsilon_{\mathrm{C}}) = (0.135 \pm 0.004) + (5.0 \pm 0.7) \times (\epsilon_{\mathrm{C}}-0.025).
\label{eq:epsilon_q_c}
\end{equation}

This relation is also shown in Figure~\ref{fig:epsilon-q_b}.

\subsection{Donor Masses and Radii of Superhumping CVs} 
\label{sec:donors_super}
Given our updating of the superhump-mass ratio relations above, we revisit the analysis of donor star properties
in \cite{knigge06} and \cite{knigge11}. Firstly, $P_{\mathrm{sh}}$ values for all SU UMa-type DNe in the \cite{patterson05} sample (70 systems) were replaced by $P_{\mathrm{sh}}^{\mathrm{B}}$ measurements from the SU UMa-type DNe survey of \cite{kato09,kato10,kato12,kato13,kato14a,kato14b,kato15,kato16,kato17}. For a number of systems, $P_{\mathrm{orb}}$ was also updated, either from measurements made by Kato et al., or  additional studies (see references within Kato et al.). Values of $\epsilon_{\mathrm{B}}$ were obtained from $P_{\mathrm{sh}}^{\mathrm{B}}$ and $P_{\mathrm{orb}}$, then subsequently converted into $q$ via the newly calibrated $\epsilon_{\mathrm{B}}(q)$ relation (equation~\ref{eq:epsilon_q_b}). Equation~\ref{eq:epsilon_q_b} was also used to determine $q$ for the eight systems displaying permanent superhumps. Assuming a constant white dwarf mass of $\langle{M_{1}}\rangle=0.81\,\mathrm{M}_{\odot}$, donor mass estimates were obtained for all systems in the superhumper sample.

As the donor fills its Roche lobe, the \cite{eggleton83} approximation for the volume-averaged Roche lobe size, combined with Kepler's 3rd law, can be used to obtain estimates for donor radii from $q$, $M_{2}$ and $P_{\mathrm{orb}}$:
\begin{equation}
\frac{R_{2}}{\mathrm{R}_{\odot}} = 0.2478\left(\frac{M_{2}}{\mathrm{M}_{\odot}}\right)^{1/3}P_{\mathrm{orb}}^{2/3}\left[\frac{q^{1/3}(1+q)^{1/3}}{0.6q^{2/3}+\ln(1+q^{1/3})}\right],
\label{eq:r2_sh}
\end{equation}
where $P_{\mathrm{orb}}$ is in units of hrs. The \cite{eggleton83} approximation for the volume-averaged Roche lobe size is the same one used to determine $R_{2}$ for systems that have been eclipse modelled, establishing consistency between the superhumping and eclipsing samples. It is important to note that \cite{knigge11} use a more complex, accurate approximation for the volume-averaged size of the Roche lobe based on the results of \cite{sirotkinkim09}, which represents the donor as a polytrope, rather than a point source. However, the advantage of using the \cite{sirotkinkim09} approximation is small, with only a $\sim$\,1\% difference between the two approximations (Figure 3 of \citealt{knigge11}).

In addition to the 78 superhumper sample from \cite{patterson05}, Kato et al. provide $P_{\mathrm{sh}}^{\mathrm{B,C}}$ and $P_{\mathrm{orb}}$ values for a further 147 systems. These systems were given the same treatment as the \cite{patterson05} sample (outlined above). A handful of systems only have available $P_{\mathrm{sh}}^{\mathrm{C}}$ values, in which case equation~\ref{eq:epsilon_q_c} was used. This brings the total number of superhumping systems with inferred donor properties to 225.

\subsection{Updating the Semi-Empirical Mass-Radius Relation for CV Donor Stars}
\label{subsec:updated_semrr}

With donor masses and radii for 15 eclipsing systems in this work, a further 31 (mostly) eclipsing systems from the literature (see Table~\ref{table:supp_sys}) and 225 superhumpers, it is possible to update the mass-radius relation for CV donor stars from \cite{knigge06} and \cite{knigge11}. The same fitting procedure used by \cite{knigge06} was followed to update the mass-radius relation. Assumptions for some parameters in this model are required, since they are not well-constrained by the donor masses and radii. Assumptions for the donor mass within the period gap ($M_{\mathrm{conv}}$), and the upper and lower ($P_{\mathrm{gap,+}}$, $P_{\mathrm{gap,-}}$) bounds of the period gap from \cite{knigge11} remained unchanged. We do adopt a smaller value for $P_{\mathrm{bounce}}$ (called $P_{\mathrm{min}}$ in \cite{knigge11}). $P_{\mathrm{bounce}}$ is the orbital period where the pre-bounce and post-bounce power-law relationships intersect.
\cite{knigge11} used the location of the period spike from \cite{gaensicke09} for $P_{\mathrm{bounce}}$. However, real systems do not reach this orbital period, because the smooth track followed by real systems near period minimum is not well represented by two power laws. PHL 1445 \citep{mcallister15}  is expected to be close to the absolute minimum period for main sequence CVs, and so its orbital period of 76.3\,min is used for $P_{\mathrm{bounce}}$ here. The value of $M_{\mathrm{bounce}}$ shown above was determined from the optimal short-period fit.

\begin{equation*}
\begin{array}{ll}
M_{\mathrm{bounce}} = 0.063\,^{+0.005}_{-0.002}\,\mathrm{M}_{\odot}, & \quad P_{\mathrm{bounce}} = 76.3\pm1.0\,\mathrm{min} \\
M_{\mathrm{conv}} = 0.20\pm0.02\,\mathrm{M}_{\odot}, & \quad P_{\mathrm{gap,-}} = 2.15\pm0.03\,\mathrm{hrs}, \\
M_{\mathrm{evol}} \simeq 0.6\mathrm{-}0.8\,\mathrm{M}_{\odot}, &  \quad P_{\mathrm{gap,+}} = 3.18\pm0.04\,\mathrm{hrs}.\\
\end{array}
\label{eq:bpl_vals_new}
\end{equation*}

\begin{figure*}
\begin{center}
\includegraphics[width=1.8\columnwidth,trim=5 10 5 25]{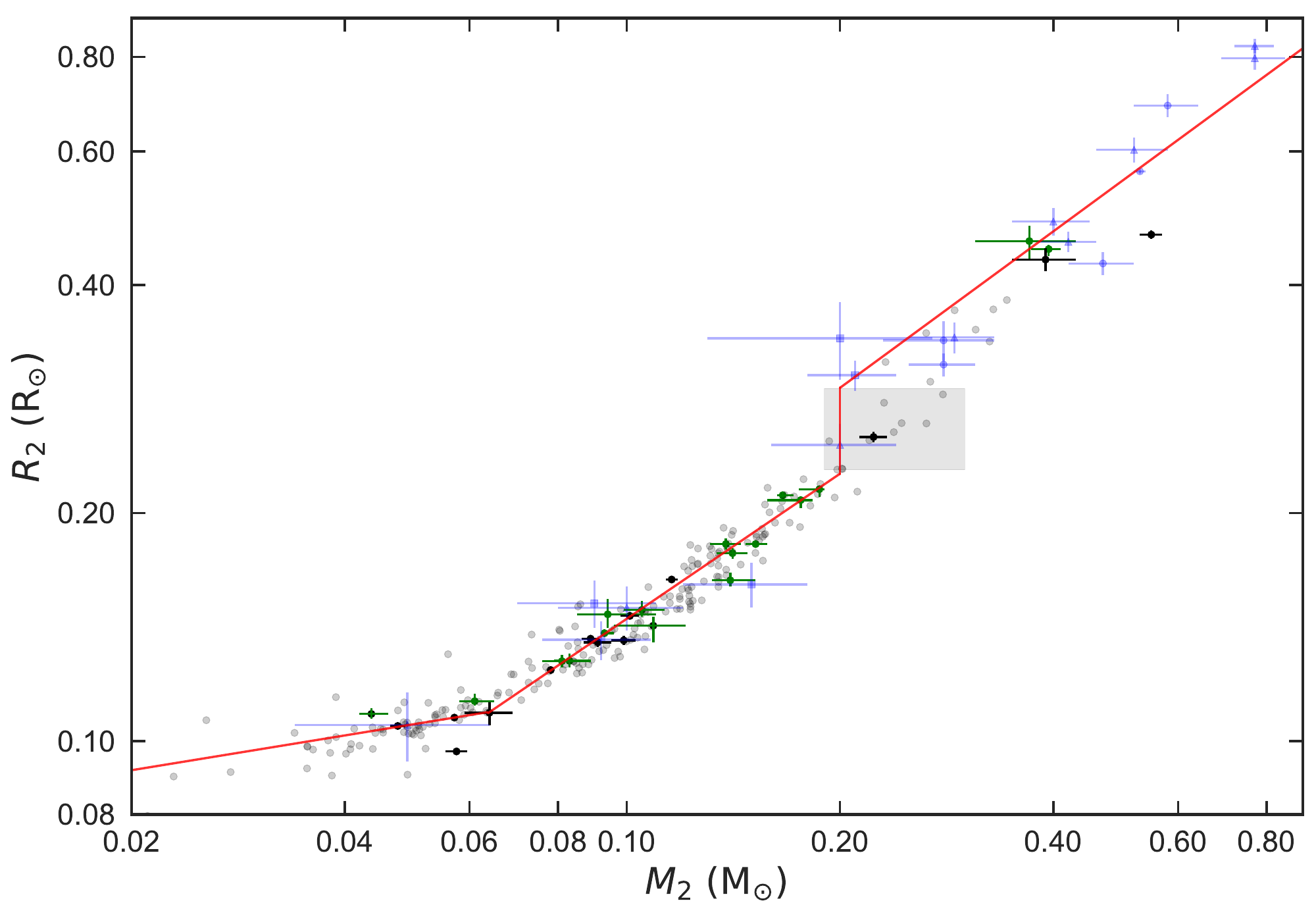}
\caption[Existing and updated mass-radius relations for CV donors.]{\label{fig:r2vsm2_bpl}Measured CV donor masses ($M_{2}$) and radii ($R_{2}$). The data point colour/shape scheme is the same as in Figure~\ref{fig:ecaml}, but with additional superhumping systems (grey points), for which error bars have been omitted for clarity. The red line is the semi-empirical mass-radius relation from this work. The grey shaded region contains systems assumed to lie within the period gap, and are therefore not included in the updated broken-power-law fit.}
\end{center}
\end{figure*}

The donor masses and radii for all but 12 systems were included in the fits. The majority of these systems were excluded due to being period gap systems (see grey box in bottom plot of Figure~\ref{fig:r2vsm2_bpl}), while SDSS 1507 (outlying black data point in period bouncer regime) was excluded as it is known to be a Galactic halo object \citep{patterson08,uthas11}. The results from the three power law fits are shown in Figure~\ref{fig:r2vsm2_bpl}, and take the following form:
\begin{equation*}
\frac{R_{2}}{\mathrm{R}_{\odot}} = \left\{
	\begin{array}{ll}
	0.109\pm0.003\left(\frac{M_{2}}{M_{\mathrm{bounce}}}\right)^{0.152\pm0.018} &  M_{2} < M_{\mathrm{bounce}}\\[8pt]
	0.225\pm0.008\left(\frac{M_{2}}{M_{\mathrm{conv}}}\right)^{0.636\pm0.012} &  M_{\mathrm{bounce}} < M_{2} < M_{\mathrm{conv}}\\[8pt]
	0.293\pm0.010\left(\frac{M_{2}}{M_{\mathrm{conv}}}\right)^{0.69\pm0.05} &  M_{\mathrm{conv}} < M_{2} < M_{\mathrm{evol}}.\\
	\end{array}
\right.
\label{eq:bpl_new}
\end{equation*}
Comparing these results with \cite{knigge11}, there is little change in the exponents of the mass-radius relation in both the long- and short-period regimes. One notable difference, however, is the amount of intrinsic scatter, $\sigma_{\mathrm{int}}$, required for the short-period systems, reduced from approximately $0.02\,\mathrm{R}_{\odot}$ to $0.005\,\mathrm{R}_{\odot}$. The small scatter provides strong evidence for a very tight evolutionary path followed by non-evolved CV donors, implying little spread in AML loss rates for CVs with the same component masses. The scatter within the long-period regime, at $0.04\,\mathrm{R}_{\odot}$, is almost a factor of 10 larger than that at short periods. Figure~\ref{fig:r2vsm2_bpl} shows two outlying long-period systems with $R_{2}\simeq0.40\,\mathrm{R}_{\odot}$, namely IP Peg \citep{copperwheat10} and HS 0220+0603 \citep{rodriguezgil15}. The donors within these two systems are undersized for their masses, and may even be in thermal equilibrium, which is unexpected for a CV donor. It is possible that both IP Peg and HS 0220+0603 have donors in thermal equilibrium due to recently starting mass transfer.

The mass-radius relation for period-bouncers has changed significantly. The new power law exponent of $0.152\pm0.018$ is much smaller than that of \cite{knigge11}, a consequence of using lower values for both $M_{\mathrm{bounce}}$ and $P_{\mathrm{bounce}}$, in addition to the inclusion of many more period-bouncers in the new donor sample, which enables a better constraint of the power law in this regime. There has been a long-standing issue with the number of confirmed period-bounce CVs, which has always been much lower than the predicted 40--70\% \citep{goliaschnelson15, kolb93}. Whilst the sample of donor masses collected here is far from homogeneous, and the presence of large numbers of superhumping systems introduces complicated selection effects, we note here that 30\% of our sample has a donor mass below 0.063\,$M_{\odot}$ and are therefore likely to be period-bouncers. 

\subsection{Comparison to theoretical CV evolution tracks}
\label{sec:evotracks_revisited}

\begin{figure*}
\begin{center}
\includegraphics[width=1.8\columnwidth,trim=0 0 0 0]{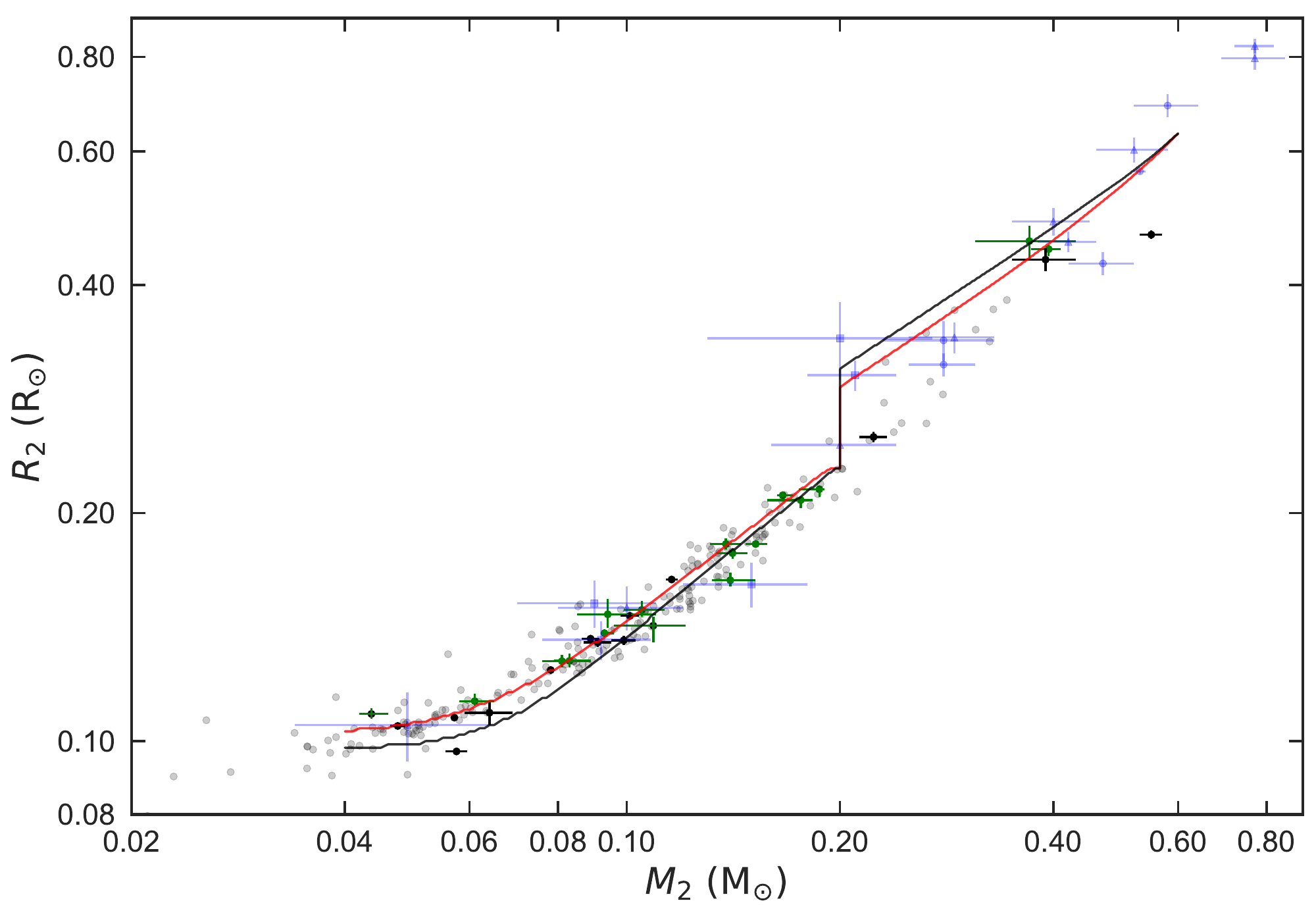}
\caption[Donor-Based CV Evolution tracks -- $R_{2}$-$M_{2}$ plane.]{\label{fig:r2vsm2_evo}Measured CV donor masses ($M_{2}$) and radii ($R_{2}$). The data point colour/shape scheme is the same as in Figure~\ref{fig:r2vsm2_bpl}. The red and black lines represent the best-fit ($f_{\mathrm{GR}}=2.47\pm0.22$, $f_{\mathrm{MB}}=0.66\pm0.05$) and `standard' ($f_{\mathrm{GR}}=f_{\mathrm{MB}}=1$) evolutionary tracks from \cite{knigge11}, respectively.}
\end{center}
\end{figure*}

\begin{figure*}
\begin{center}
\includegraphics[width=1.8\columnwidth,trim=0 0 0 0]{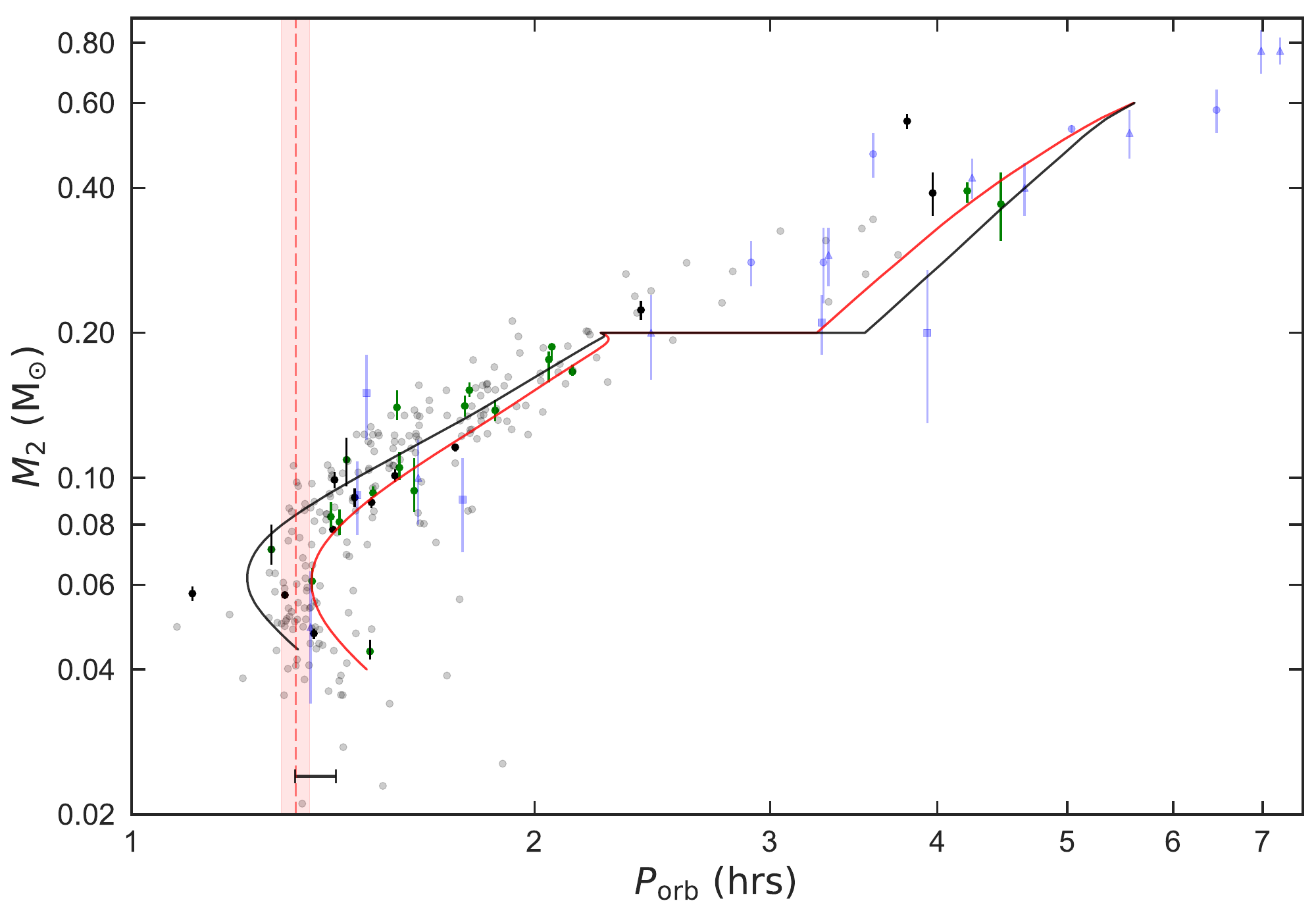}
\caption[Donor-Based CV Evolution tracks -- $M_{2}$-$P_{\mathrm{orb}}$ plane.]{\label{fig:m2vsp}Measured CV donor masses ($M_{2}$) as a function of orbital period ($P_{\mathrm{orb}}$). The data point colour/shape scheme is the same as in Figure~\ref{fig:r2vsm2_bpl}. The red and black lines represent the best-fit ($f_{\mathrm{GR}}=2.47\pm0.22$, $f_{\mathrm{MB}}=0.66\pm0.05$) and `standard' ($f_{\mathrm{GR}}=f_{\mathrm{MB}}=1$) evolutionary tracks from \cite{knigge11}, respectively. The vertical dashed red line and shaded region is an estimate of the true $P_{\mathrm{min}}$ based on fitting a Gaussian distribution to the orbital periods in the range 76--82 mins.}
\end{center}
\end{figure*}

In addition to a broken-power-law mass-radius relation for CV donors, \cite{knigge11} present a theoretical evolutionary track, produced with the aim of quantifying the secular mass transfer rate in CVs.  The track which best reproduces their donor sample requires reduced magnetic braking above the gap ($f_{\mathrm{MB}}=0.66\pm0.05$), but additional angular momentum loss below the gap ($f_{\mathrm{GR}}=2.47\pm0.22$). The donor sample presented in this work is shown in the $\textit{M}_{2}$-$\textit{R}_{2}$ and $\textit{P}_{\mathrm{orb}}$-$\textit{M}_{2}$ planes 
in Figures~\ref{fig:r2vsm2_evo} and~\ref{fig:m2vsp}, respectively.  Also shown is the `best fit' track from \cite{knigge11}, and the `standard' track ($f_{\mathrm{GR}}=f_{\mathrm{MB}}=1$).
It is clear from these figures that the best-fit evolutionary track from \cite{knigge11} under predicts the donor mass at orbital periods below the period gap, and has a period minimum that is longer than that observed. This again implies that less additional angular momentum loss is needed below the period gap than suggested by \cite{knigge11}. In contrast, we find that the `standard' track provides a better fit to the donor sample immediately below the gap, where the donor mass is in the range 0.10--0.20\,M$_{\odot}$. This is most apparent in Figure~\ref{fig:m2vsp}. Although the standard track is a good fit to systems immediately below the gap, it diverges from the donor sequence at lower masses, and predicts a period minimum shorter than the observed value. Therefore, the donor properties in CVs appears to argue for an additional source of AML that is small compared to gravitational radiation just below the period gap, but becomes more significant at shorter orbital periods and/or donor masses.

The eCAML model of \cite{schreiber16} might provide something similar to the behaviour required. All models of CV evolution require a term $\nu$, which expresses the AML which arises as a consequence of mass transfer. In the standard model, it is assumed that the mass lost from the white dwarf during nova eruptions carries with it the specific angular momentum of the white dwarf, leading to $\nu = M_{2}^{2} / (M_{1}M)$, where $M$ is the total mass of the system. In the eCAML model, an alternative form of $\nu \sim 0.35/M_{1}$ is proposed. We used equation~1 from \cite{knigge11} to roughly estimate the mass loss rates under the eCAML model at key points in the evolution of the donor. Just below the period gap, we take $M_{1}=0.82$\,$M_{\odot}$, $M_{2}=0.15$\,$M_{\odot}$ and we assume the donor is roughly in thermal equilibrium, so the mass-radius index is $\xi = 0.8$. This implies that in the eCAML model, mass loss rates just after the period gap are only around 35\% higher than the `standard', $f_{\mathrm{GR}}=1$, model. For systems near the period minimum, we take $M_{1}=0.82$\,$M_{\odot}$, $M_{2}=0.065$\,$M_{\odot}$ and $\xi = -1/3$, which suggests mass loss rates around 9 times higher than the $f_{\mathrm{GR}}=1$ case. Therefore, the eCAML model provides a mass loss law which is qualitatively similar to the one implied by CV donor properties. However, it is worth bearing in mind that the $\nu \sim 0.35/M_{1}$ prescription is not physically motivated. \cite{schreiber16} suggest that angular momentum loss during nova outbursts might produce a similar behaviour, but the frequency of nova outbursts will drop as the accretion rate falls. Therefore eCAML may be less important for CVs near the period minimum than implied above. It will require a physically plausible model of CV evolution, including AML during nova outbursts, to determine if such a model can reproduce both the high white dwarf mass in CVs and the $\textit{P}_{\mathrm{orb}}$-$\textit{M}_{2}$ locus of the donor stars.

Finally, we note that our results introduce a tension between the donor masses and radii, and the temperatures of white dwarfs in CVs. As described in \cite{townsleygaensicke09}, compressional heating of the white dwarfs due to accretion sets the equilibrium temperature of the white dwarf in a CV. The observed white dwarf temperature thus depends upon the accretion rate, averaged over the  thermal timescale of the non-degenerate layer on the white dwarf surface \citep{townsleybildsten03}. The best study of white dwarf temperatures in CVs to date is \cite{pala17}, who show that the white dwarfs in CVs below the period gap imply AML rates approximately twice that implied by $f_{\mathrm{GR}}=1$. As discussed extensively in Section~4 of \cite{knigge11}, one plausible explanation for the discrepancy is the presence of mass transfer rate fluctuations, coupled with the fact that the white dwarf temperature reflects the mass transfer rate averaged over much shorter timescales than the donor star radius. However, this would presumably lead to white dwarf temperatures scattered around the expected values; whereas they are systematically warmer than expected.

\subsection{The Period Minimum}
\label{subsec:pmin_2}

It is apparent from Figure~\ref{fig:m2vsp} that the current donor sample contains a sufficiently large number of systems at the shortest orbital periods to finally begin to reveal the locus of CVs evolving through the period minimum. The period minimum of the current donor sample covers an approximate period range of 76--82\,min (1.27--1.37\,hrs). Fitting a Gaussian distribution to the donor sample within this period range returned the following estimates for both the period minimum ($P_{\mathrm{min}}=79.6\pm0.2$\,min) and its width ($\mathrm{FWHM}=4.0\,\mathrm{min}$). These estimates for $P_{\mathrm{min}}$ and its width are shown by the red vertical dashed line and shaded area within Figure~\ref{fig:m2vsp}.

It was briefly mentioned in Section~\ref{subsec:updated_semrr} that the observed location of the period minimum appears to be slightly lower than the value $P_{\mathrm{min}}=81.8\pm0.9$\,min predicted by the best-fit track of \cite{knigge11}. The new measurement of $P_{\mathrm{min}}$ from the donor sample confirms this, with the two $P_{\mathrm{min}}$ estimates differing by approximately 2.4$\sigma$. A lower value of $P_{\mathrm{min}}$ than the existing estimate of \cite{knigge11} was previously hinted at by \cite{mcallister15}. Figure~\ref{fig:m2vsp} shows that with the new estimate for $P_{\mathrm{min}}$, PHL~1445 ($P_{\mathrm{orb}}=1.27$\,hrs) and SDSS~1433 ($P_{\mathrm{orb}}=1.30$\,hrs) are no longer troublesome outliers.

\section{Conclusions}

We present new measurements of the system parameters for 15 eclipsing CVs, six of which are published for the first time. We also compile a list of reliable system parameter determinations from the literature. We use these measurements to refine the calibration of the relationship between superhump period excess and mass ratio; allowing us to estimate the donor properties of 225 CVs showing superhump phenomena. This provides an extensive sample of CVs with known system parameters which we can use to test models of CV evolution. 

We confirm the high average white dwarf mass in CVs, but we find no evidence for a trend in white dwarf mass with orbital period. Contrary to previous studies, we find that the donor properties of CVs immediately below the period gap are consistent with the standard model, in which AML due to magnetic braking is small compared to gravitational radiation. We do, however, find that CVs at shorter orbital periods and lower masses still require an additional source of AML. We argue that the eCAML model of \cite{schreiber16} predicts an AML law that is qualitatively similar to this behaviour. We find that, for systems below the period gap, donor radii at a given orbital period show a very small intrinsic scatter of only 0.005\,R$_{\odot}$, suggesting that most CVs below the gap follow a common evolutionary path. We estimate a value for the orbital period minimum of $79.6\pm0.2$\,min, shorter than previously estimated by \cite{knigge11}.

The CVs with donor properties estimated from superhumps show a sizeable fraction of systems which appear to have evolved past the period minimum. As a result, 30\% of our sample appear to be post-period minimum systems. This hints that post-period minimum systems may be as common as models predict, but the superhump sample is strongly biassed towards low mass ratios. The advent of Gaia means that detailed follow up of a relatively complete volume-limited sample may resolve this question in the near future.

\section*{Acknowledgements}
VSD, SPL, ULTRACAM and ULTRASPEC are supported by the STFC grant ST/R000964/1. This work has made use of data obtained at the Thai National Observatory on Doi Inthanon, operated by NARIT. VSD, TRM and SPL acknowledge the support of the Royal Society and the Leverhulme Trust for the operation of ULTRASPEC at the TNT. The WHT is operated on the island of La Palma by the Isaac Newton Group of Telescopes in the Spanish Observatorio del Roque de los Muchachos of the Instituto de Astrof{\'i}sica de Canarias. This work is based on observations collected at the European Organisation for Astronomical Research in the Southern Hemisphere.





\bibliographystyle{mnras}
\bibliography{Thesis.bib}


\appendix

\section{Fits to eclipse lightcurves}
\label{app:modfits}

The following figures show the full sets of eclipse model fits for the 15 additional systems from Section~\ref{subsec:lcmod}.

\textheight = 712pt

\begin{figure}
\begin{center}
\includegraphics[width=1.0\columnwidth,trim=10 40 10 80]{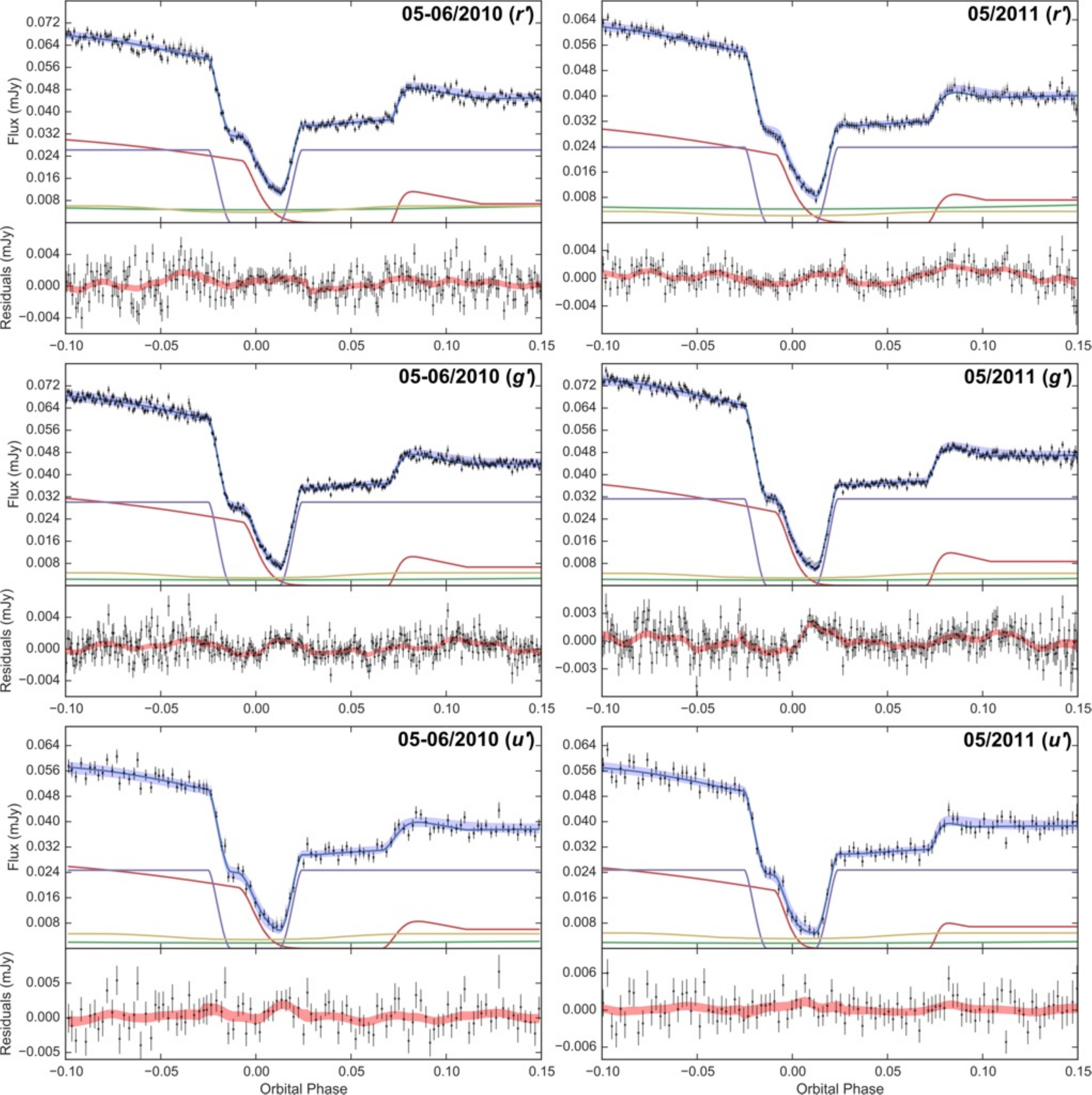}
\caption[Simultaneous eclipse model fit to six average CSS080623 eclipse light curves.]{\label{fig:modfit_css080623}Simultaneous eclipse model fit to six average CSS080623 eclipse light curves. See Section~\ref{subsec:lcmod} for full details of what is plotted. Displayed in the top-right corner of each average eclipse plot is the date(s) and wavelength band each of the constituent eclipses were observed in.}
\end{center}
\end{figure}

\begin{figure}
\begin{center}
\includegraphics[width=1.0\columnwidth,trim=10 40 10 80]{ctcv1300.pdf}
\caption[Simultaneous eclipse model fit to five CTCV 1300 eclipse light curves.]{\label{fig:modfit_ctcv1300}Simultaneous eclipse model fit to five CTCV 1300 eclipse light curves. See Section~\ref{subsec:lcmod} for full details of what is plotted. Displayed in the top-right corner of each eclipse plot is the cycle number of the eclipse and the wavelength band it was observed in.}
\end{center}
\end{figure}

\begin{figure}
\begin{center}
\includegraphics[width=1.0\columnwidth,trim=0 40 0 150]{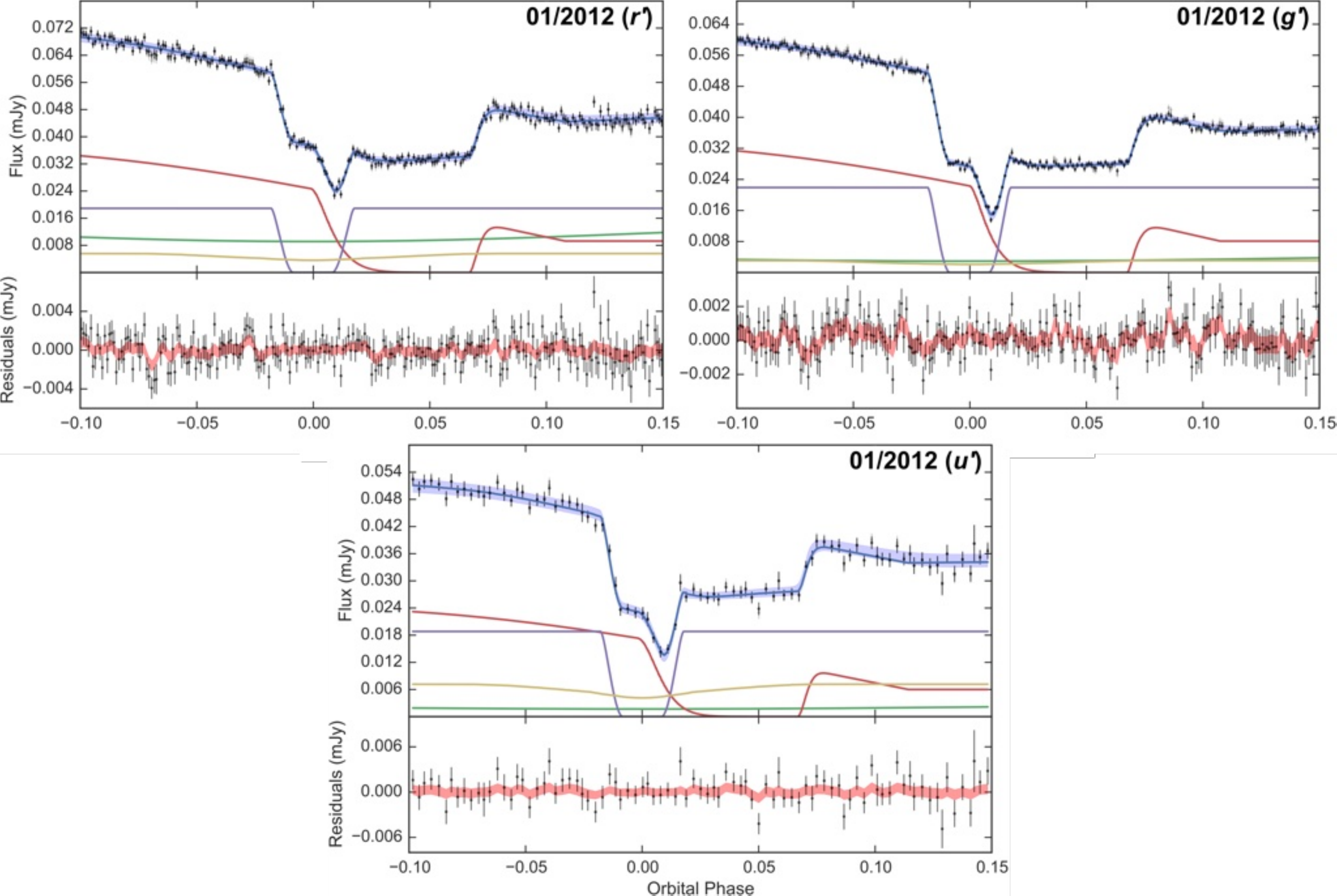}
\caption[Simultaneous eclipse model fit to three average CSS110113 eclipse light curves.]{\label{fig:modfit_css110113}Simultaneous eclipse model fit to three average CSS110113 eclipse light curves. See Section~\ref{subsec:lcmod} for full details of what is plotted. Displayed in the top-right corner of each average eclipse plot is the date and wavelength band each of the constituent eclipses were observed in.}
\end{center}
\end{figure}

\begin{figure}
\begin{center}
\includegraphics[width=1.0\columnwidth,trim=0 40 0 150]{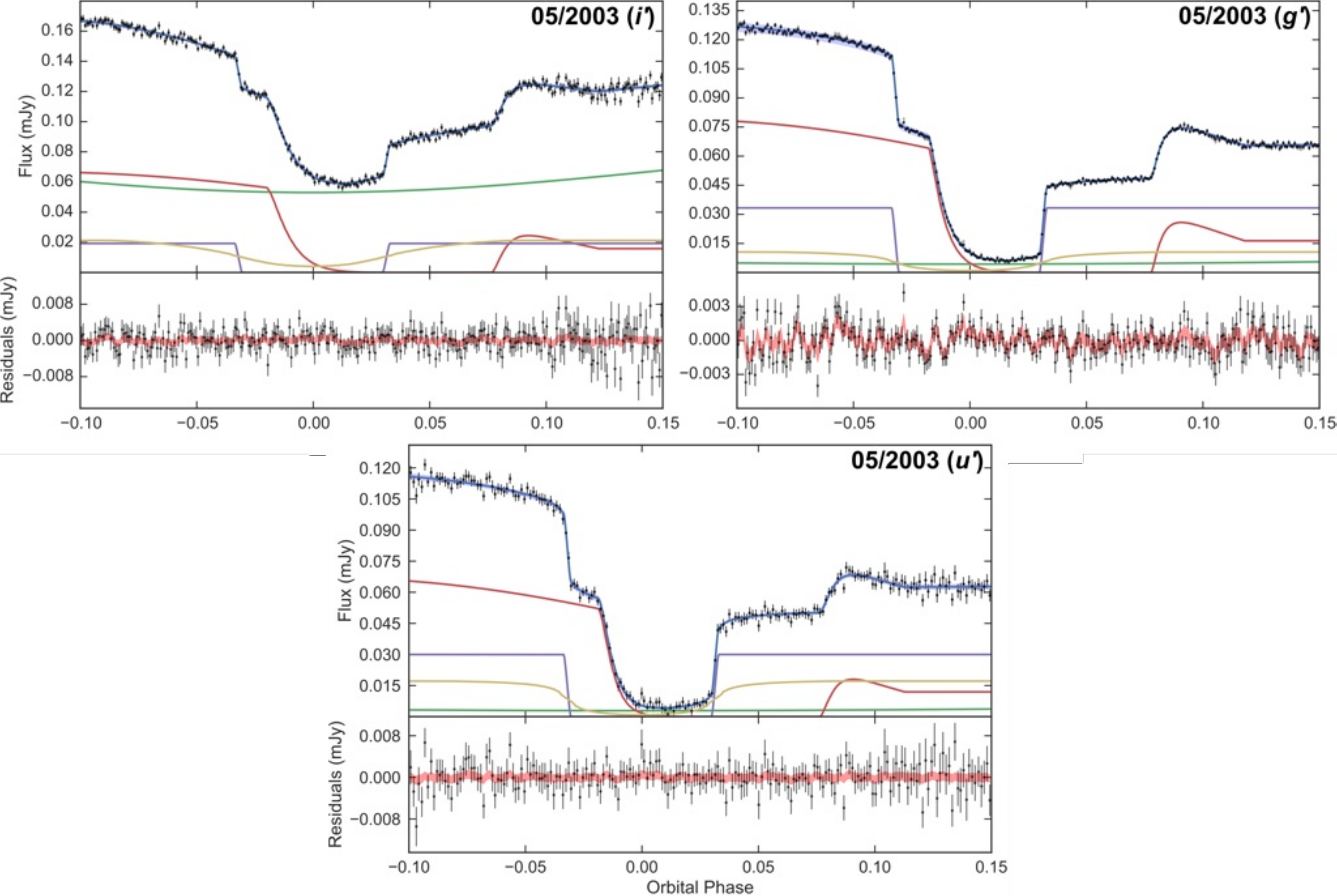}
\caption[Simultaneous eclipse model fit to three average DV UMa eclipse light curves.]{\label{fig:modfit_dvuma}Simultaneous eclipse model fit to three average DV UMa eclipse light curves. See Section~\ref{subsec:lcmod} for full details of what is plotted. Displayed in the top-right corner of each average eclipse plot is the date and wavelength band each of the constituent eclipses were observed in.}
\end{center}
\end{figure}

\begin{figure}
\begin{center}
\includegraphics[width=1.0\columnwidth,trim=10 40 10 80]{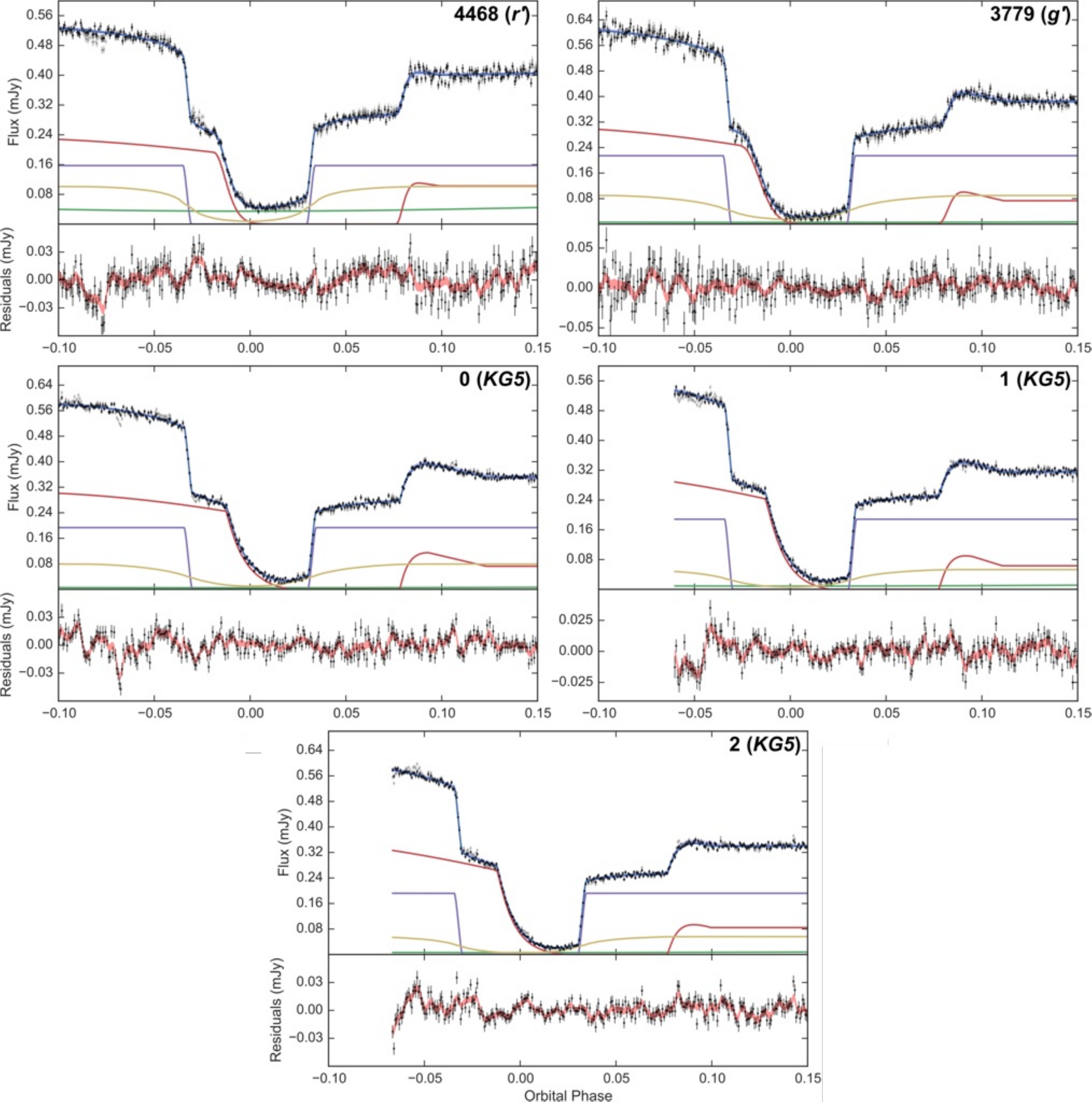}
\caption[Simultaneous eclipse model fit to five IY UMa eclipse light curves.]{\label{fig:modfit_iyuma}Simultaneous eclipse model fit to five IY UMa eclipse light curves. See Section~\ref{subsec:lcmod} for full details of what is plotted. Displayed in the top-right corner of each eclipse plot is the cycle number of the eclipse and the wavelength band it was observed in.}
\end{center}
\end{figure}

\begin{figure}
\begin{center}
\includegraphics[width=1.0\columnwidth,trim=10 40 10 80]{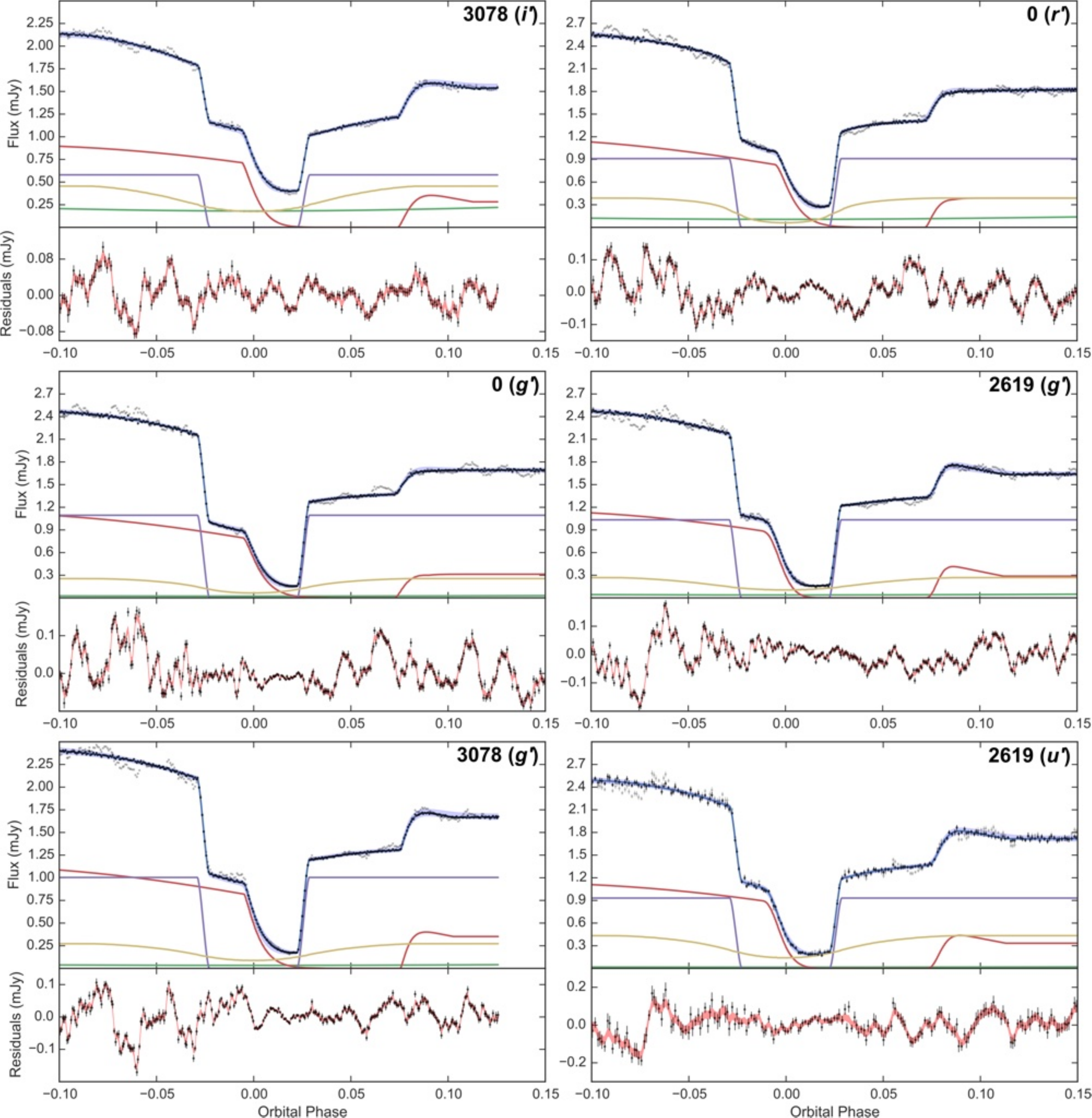}
\caption[Simultaneous eclipse model fit to six OY Car eclipse light curves.]{\label{fig:modfit_oycar}Simultaneous eclipse model fit to six OY Car eclipse light curves. See Section~\ref{subsec:lcmod} for full details of what is plotted. Displayed in the top-right corner of each eclipse plot is the cycle number. of the eclipse and the wavelength band it was observed in.}
\end{center}
\end{figure}

\begin{figure}
\begin{center}
\includegraphics[width=1.0\columnwidth,trim=10 40 10 80]{sdss0901.pdf}
\caption[Simultaneous eclipse model fit to five average SDSS 0901 eclipse light curves.]{\label{fig:modfit_sdss0901}Simultaneous eclipse model fit to five average SDSS 0901 eclipse light curves. See Section~\ref{subsec:lcmod} for full details of what is plotted. Displayed in the top-right corner of each average eclipse plot is the date and wavelength band each of the constituent eclipses were observed in.}
\end{center}
\end{figure}

\begin{figure}
\begin{center}
\includegraphics[width=1.0\columnwidth,trim=0 40 0 150]{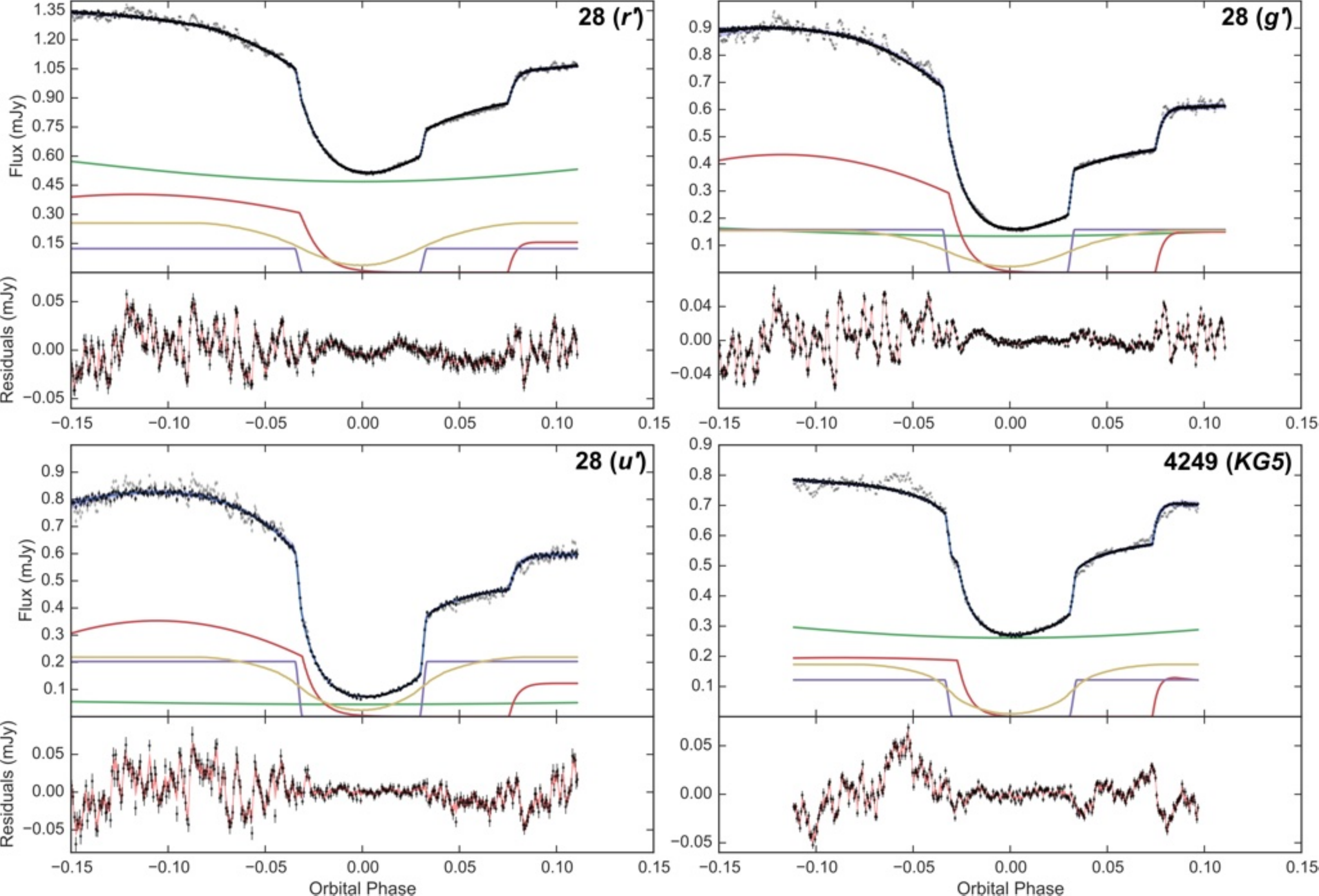}
\caption[Simultaneous eclipse model fit to four GY Cnc eclipse light curves.]{\label{fig:modfit_gycnc}Simultaneous eclipse model fit to four GY Cnc eclipse light curves. See Section~\ref{subsec:lcmod} for full details of what is plotted. Displayed in the top-right corner of each eclipse plot is the cycle number of the eclipse and the wavelength band it was observed in.}
\end{center}
\end{figure}

\begin{figure}
\begin{center}
\includegraphics[width=1.0\columnwidth,trim=0 40 0 150]{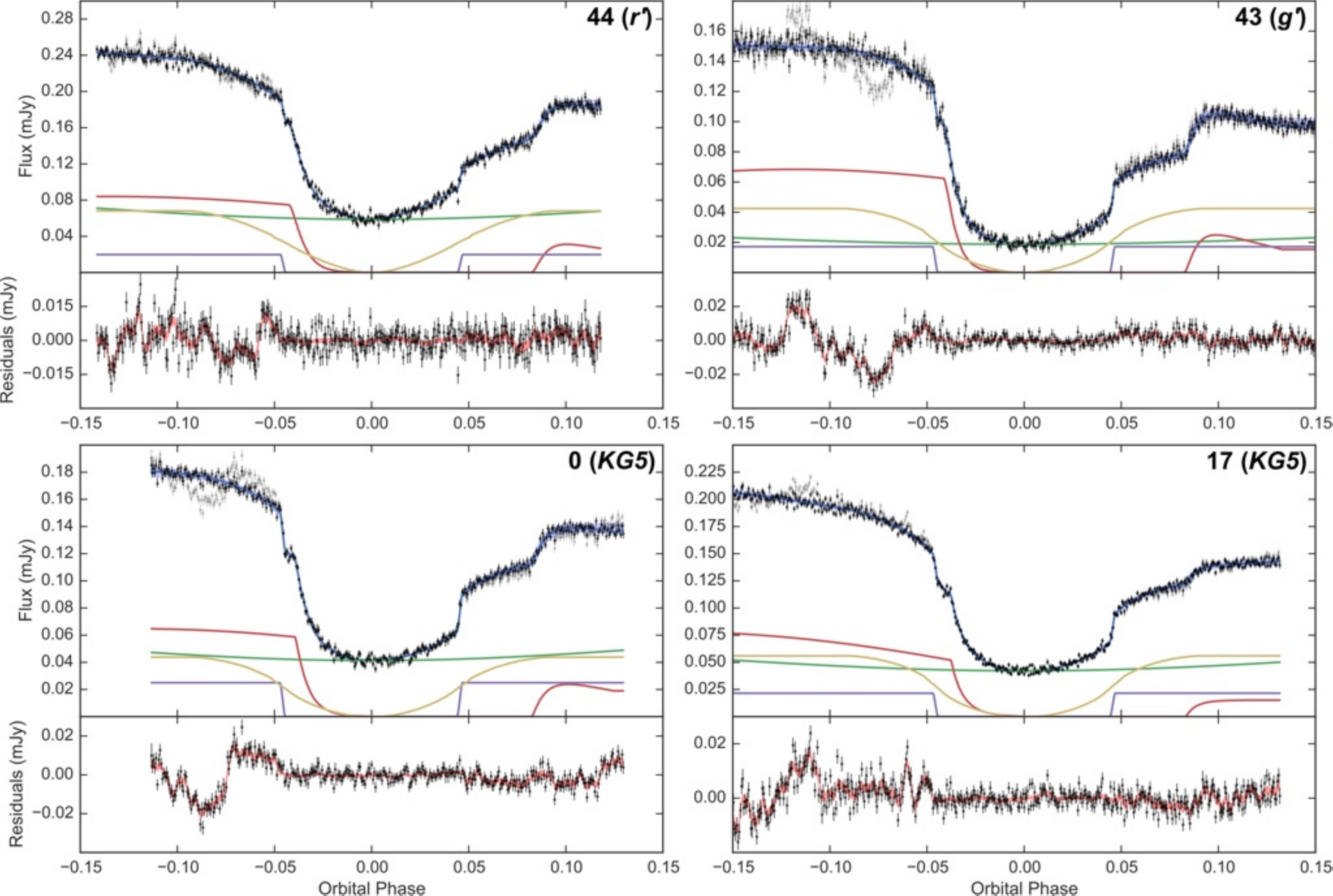}
\caption[Simultaneous eclipse model fit to four SDSS 1006 eclipse light curves.]{\label{fig:modfit_sdss1006}Simultaneous eclipse model fit to four SDSS 1006 eclipse light curves. See Section~\ref{subsec:lcmod} for full details of what is plotted. Displayed in the top-right corner of each eclipse plot is the cycle number of the eclipse and the wavelength band it was observed in.}
\end{center}
\end{figure}

\begin{figure}
\begin{center}
\includegraphics[width=1.0\columnwidth,trim=0 40 0 150]{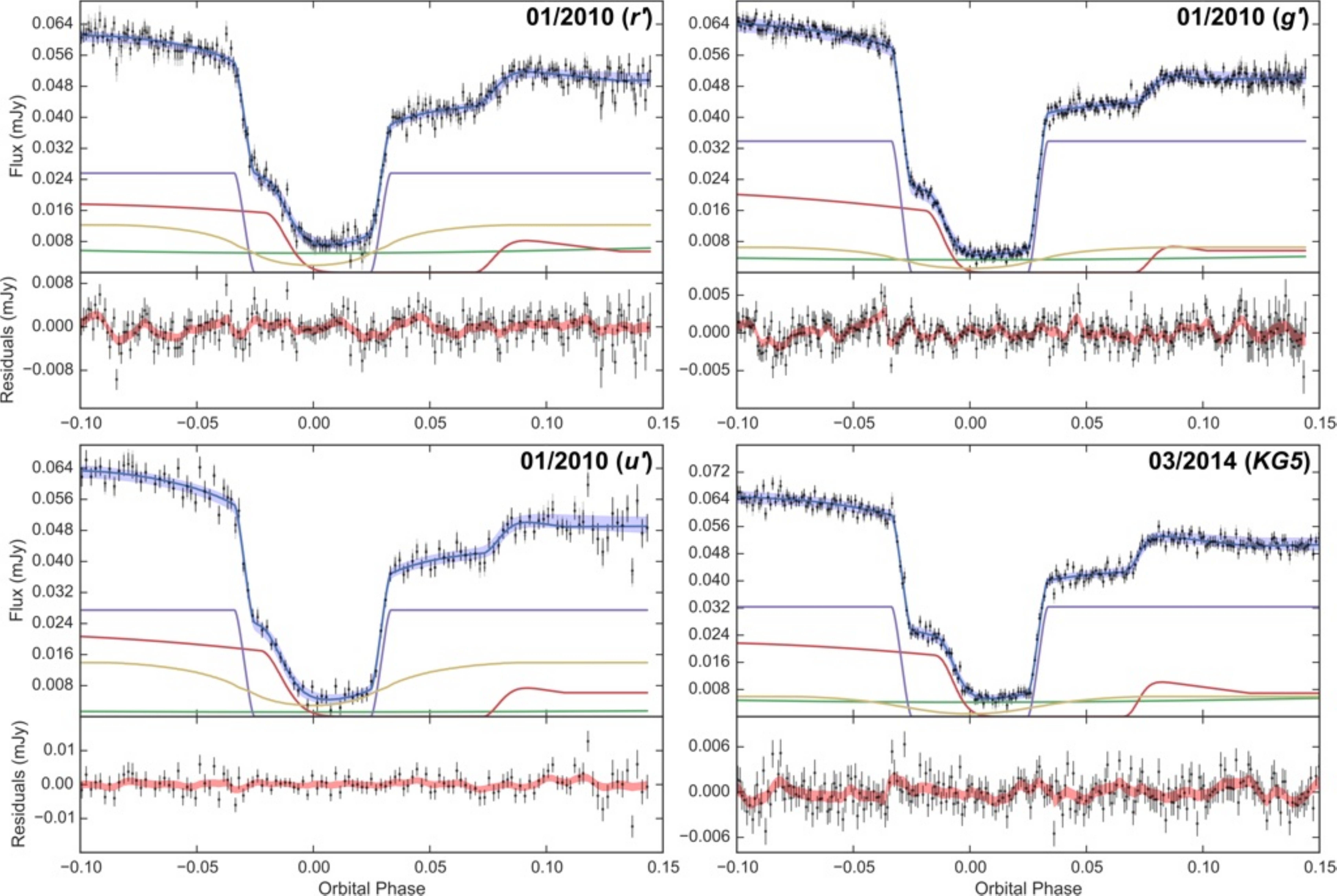}
\caption[Simultaneous eclipse model fit to four average SDSS 1152 eclipse light curves.]{\label{fig:modfit_sdss1152}Simultaneous eclipse model fit to four average SDSS 1152 eclipse light curves. See Section~\ref{subsec:lcmod} for full details of what is plotted. Displayed in the top-right corner of each average eclipse plot is the date and wavelength band each of the constituent eclipses were observed in.}
\end{center}
\end{figure}

\begin{figure}
\begin{center}
\includegraphics[width=1.0\columnwidth,trim=0 40 0 150]{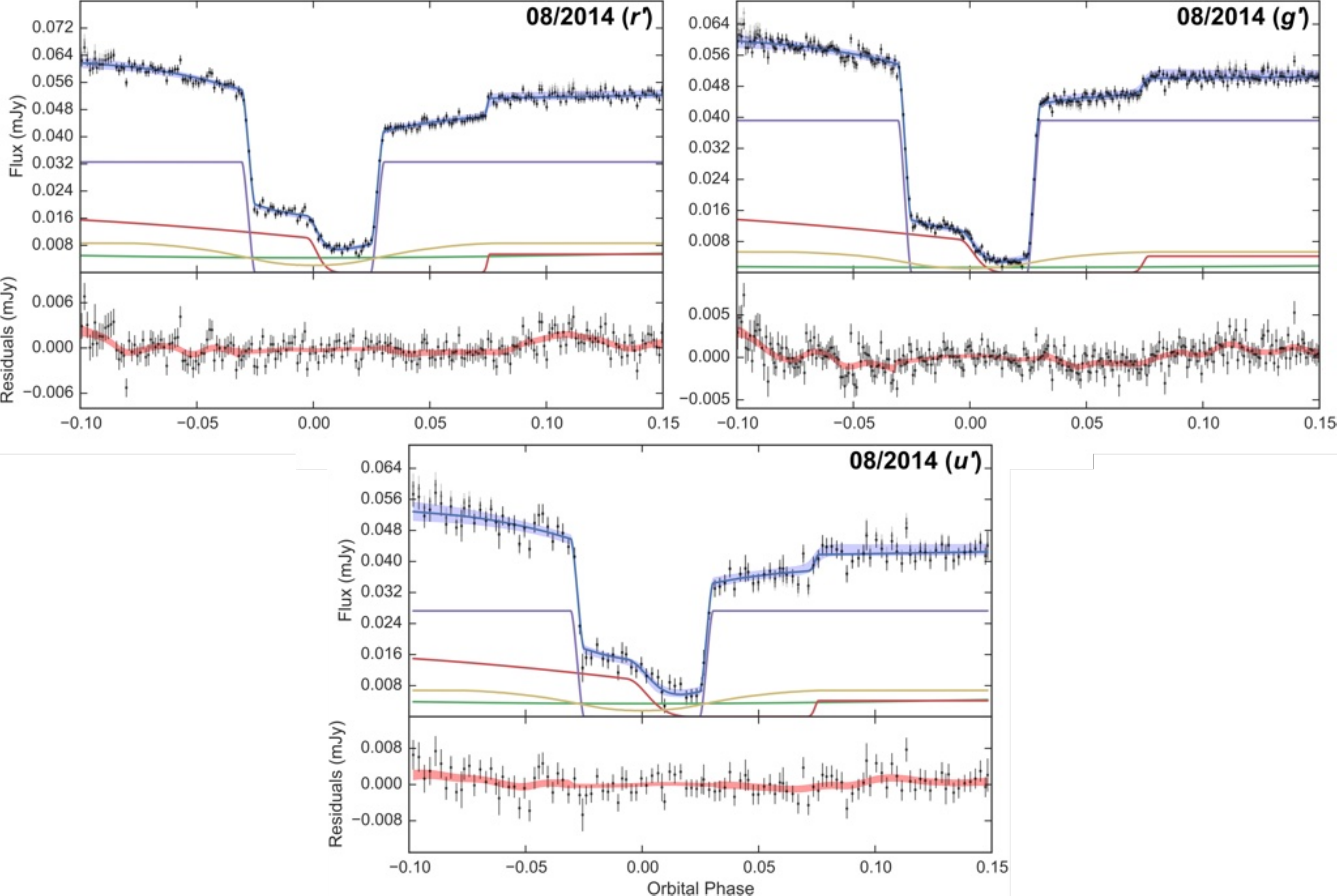}
\caption[Simultaneous eclipse model fit to three average SSS100615 eclipse light curves.]{\label{fig:modfit_sss100615}Simultaneous eclipse model fit to three average SSS100615 eclipse light curves. See Section~\ref{subsec:lcmod} for full details of what is plotted. Displayed in the top-right corner of each average eclipse plot is the date and wavelength band each of the constituent eclipses were observed in.}
\end{center}
\end{figure}

\begin{figure}
\begin{center}
\includegraphics[width=1.0\columnwidth,trim=10 40 10 80]{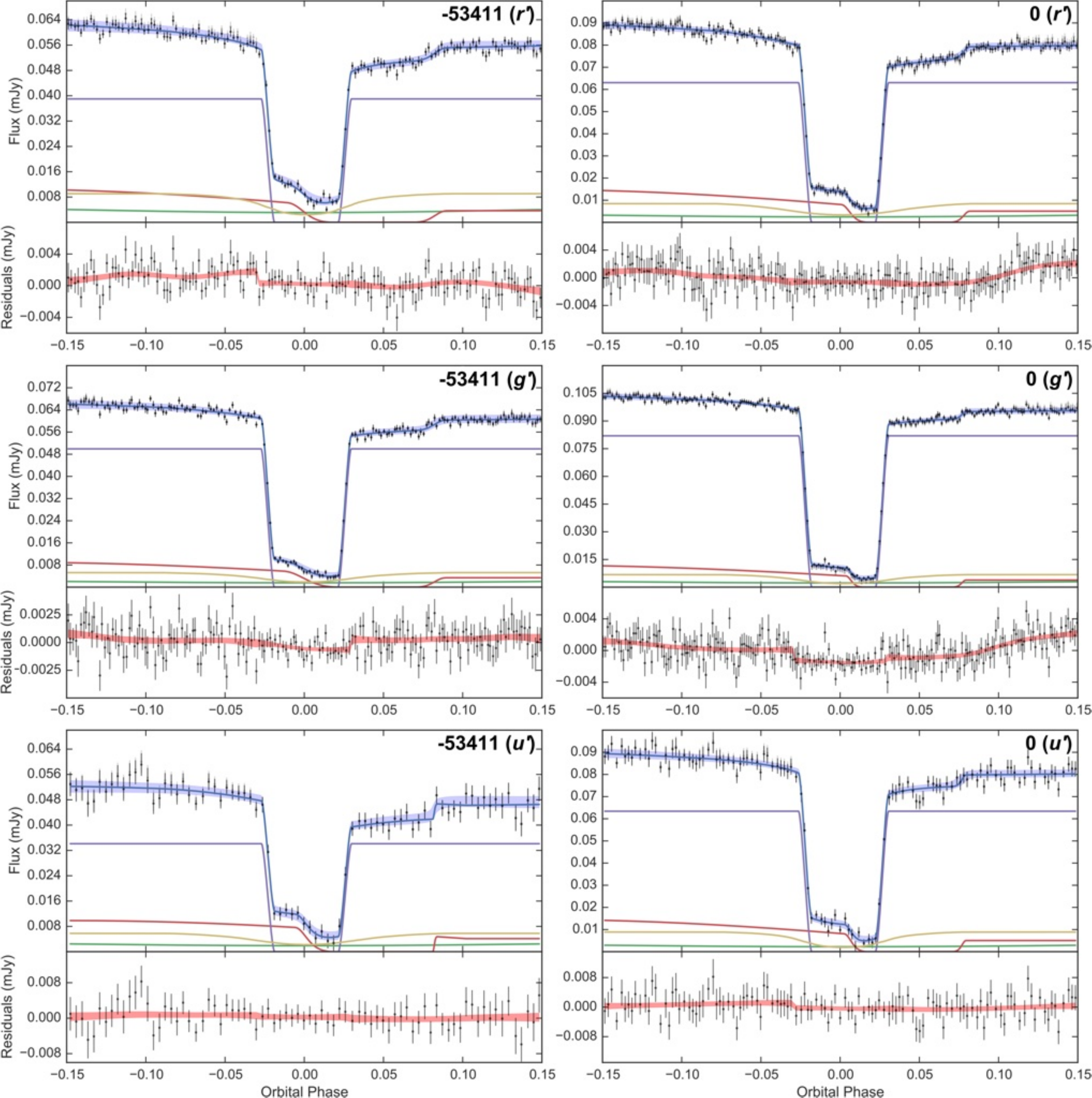}
\caption[Simultaneous eclipse model fit to six SDSS 1501 eclipse light curves.]{\label{fig:modfit_sdss1501}Simultaneous eclipse model fit to six SDSS 1501 eclipse light curves. See Section~\ref{subsec:lcmod} for full details of what is plotted. Displayed in the top-right corner of each eclipse plot is the cycle number of the eclipse and the wavelength band it was observed in.}
\end{center}
\end{figure}

\begin{figure}
\begin{center}
\includegraphics[width=1.0\columnwidth,trim=10 40 10 80]{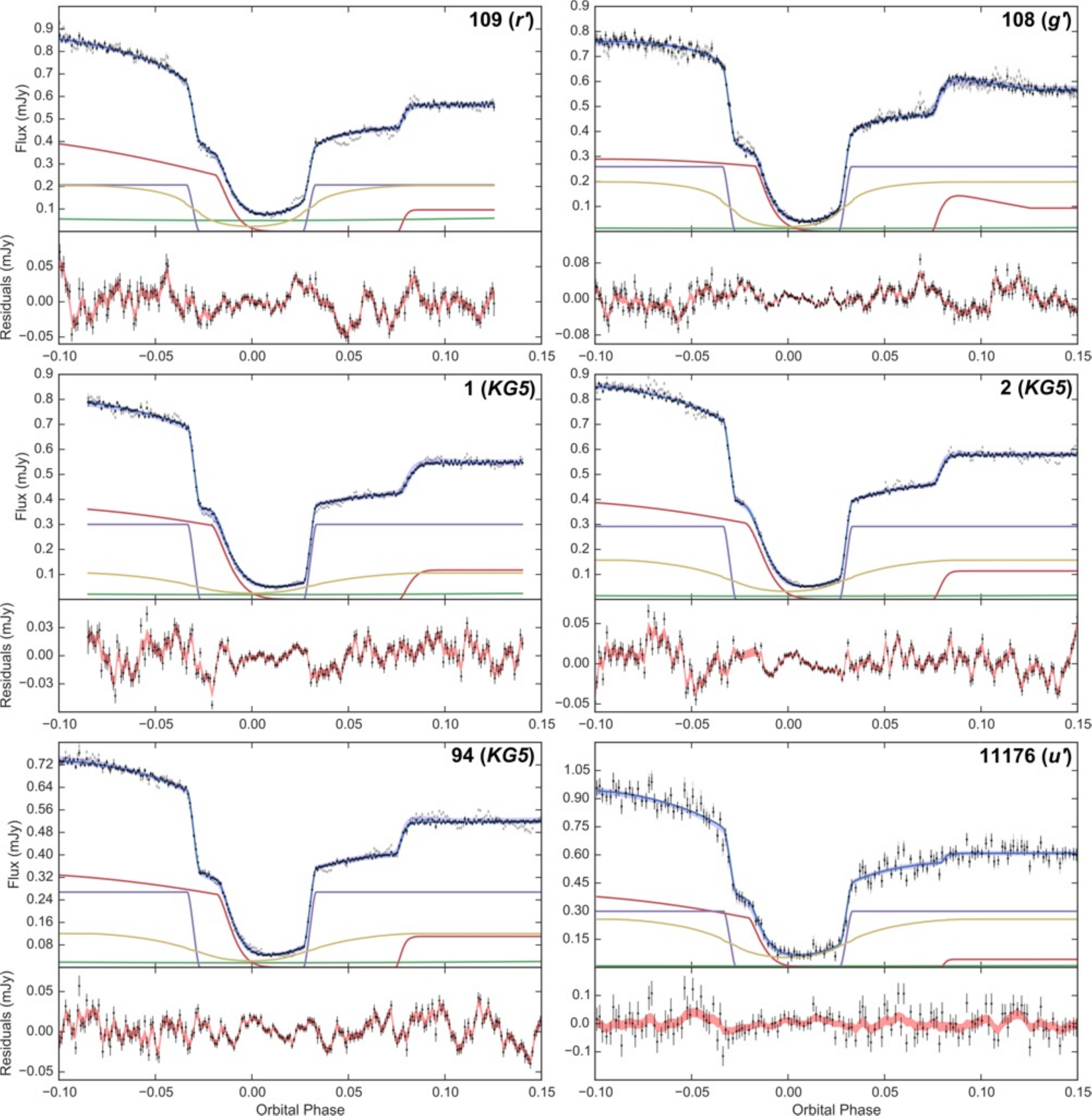}
\caption[Simultaneous eclipse model fit to six SSS130413 eclipse light curves.]{\label{fig:modfit_sss130413}Simultaneous eclipse model fit to six SSS130413 eclipse light curves. See Section~\ref{subsec:lcmod} for full details of what is plotted. Displayed in the top-right corner of each eclipse plot is the cycle number of the eclipse and the wavelength band it was observed in.}
\end{center}
\end{figure}

\begin{figure}
\begin{center}
\includegraphics[width=1.0\columnwidth,trim=10 40 10 80]{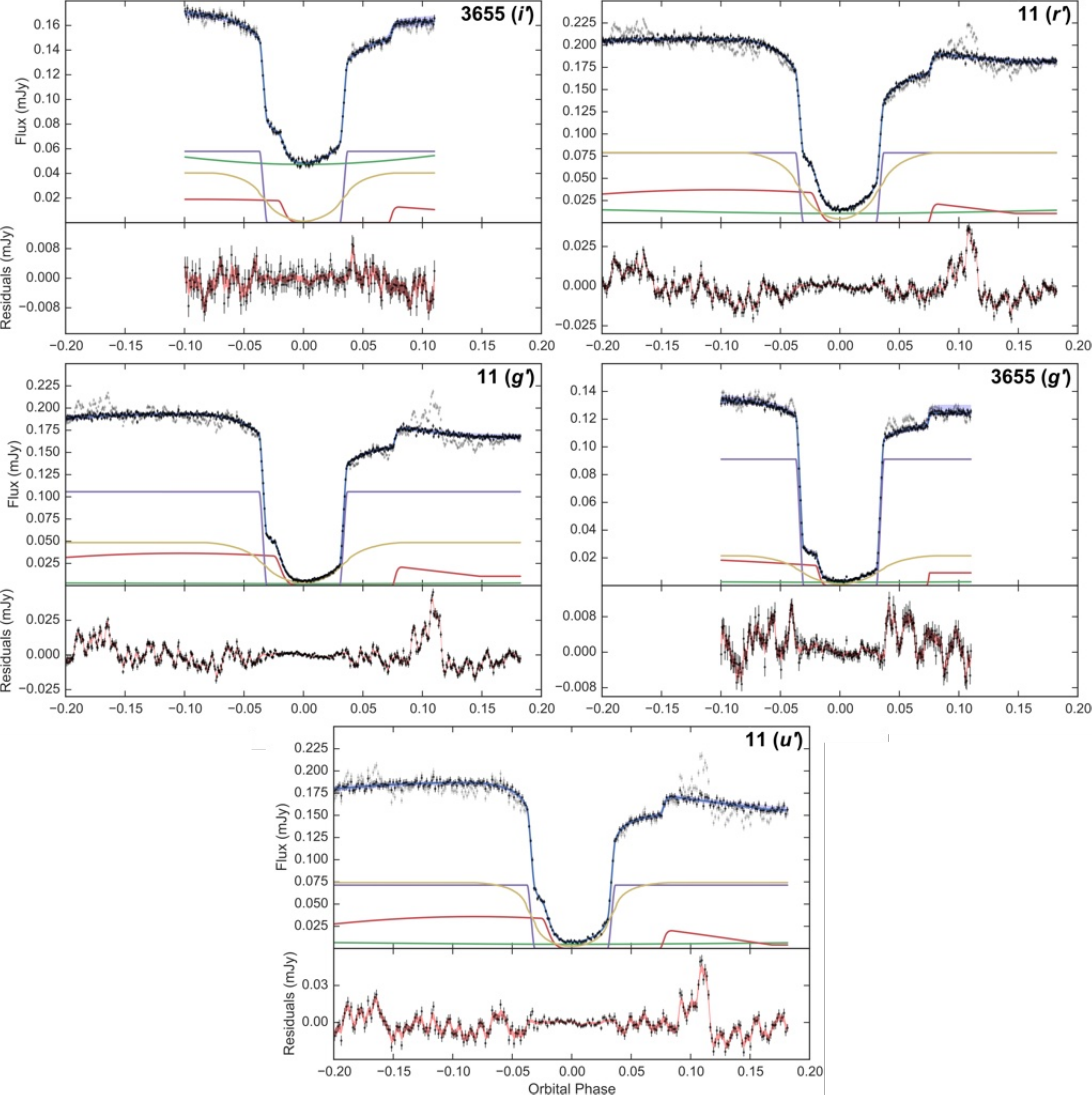}
\caption[Simultaneous eclipse model fit to five V713 Cep eclipse light curves.]{\label{fig:modfit_v713cep}Simultaneous eclipse model fit to five V713 Cep eclipse light curves. See Section~\ref{subsec:lcmod} for full details of what is plotted. Displayed in the top-right corner of each eclipse plot is the cycle number of the eclipse and the wavelength band it was observed in.}
\end{center}
\end{figure}

\begin{figure}
\begin{center}
\includegraphics[width=1.0\columnwidth,trim=10 40 10 80]{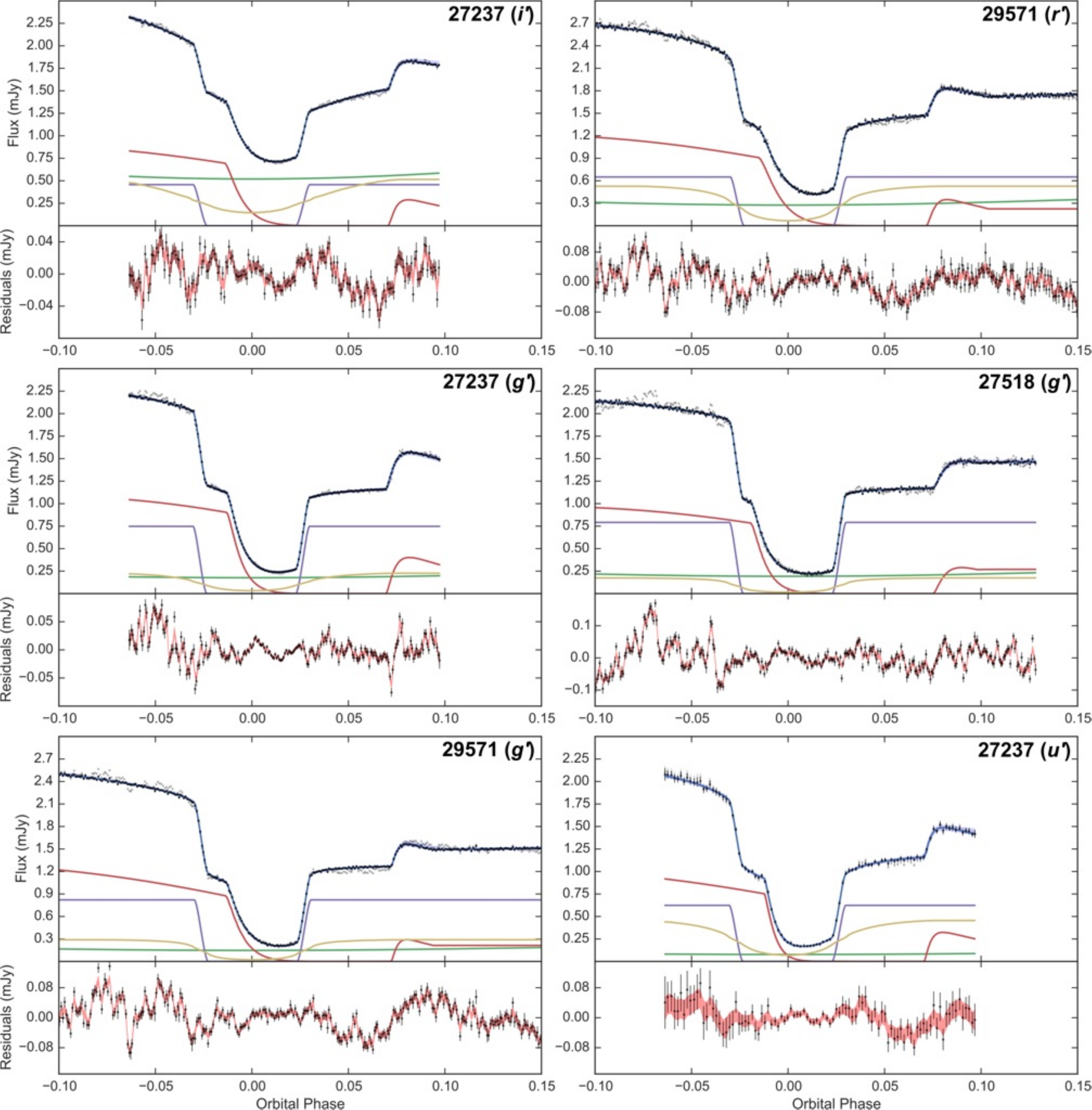}
\caption[Simultaneous eclipse model fit to six Z Cha eclipse light curves.]{\label{fig:modfit_zcha}Simultaneous eclipse model fit to six Z Cha eclipse light curves. See Section~\ref{subsec:lcmod} for full details of what is plotted. Displayed in the top-right corner of each eclipse plot is the cycle number of the eclipse and the wavelength band it was observed in.}
\end{center}
\end{figure}

\section{System Parameters for Supplementary Systems}
\label{app:supp_sys}
The following table includes reliably determined system parameters for CVs from the literature. 

\renewcommand{\arraystretch}{1.2}

\begin{table*}
\setlength{\tabcolsep}{1.6pt}
\footnotesize
\begin{tabular}{lccccccl}
\hline
System & $P_{\mathrm{orb}}$ & $q$ & $M_{1}$ & $M_{2}$ & $R_{2}$ & Method & Ref.\\ 
& (d) & & $(\mathrm{M}_{\odot})$ & $(\mathrm{M}_{\odot})$ & $(\mathrm{R}_{\odot})$ & & \\ \hline
SDSS J1433+1011 & 0.054240679(2) & 0.0661(7) & 0.865(5) & 0.0571(7) & 0.1074(4) & EM(U) & 1 \\
SDSS J1507+5230 & 0.04625828(4) & 0.0647(18) & 0.892(8) & 0.0575(20) & 0.0969(11) & EM(U) & 1 \\
SDSS J1035+0551 & 0.0570067(2) & 0.0571(10) & 0.835(9) & 0.0475(12) & 0.1047(8) & EM(U) & 1 \\
CTCV J2354$-$4700 & 0.065550270(1) & 0.1097(8) & 0.935(31) & 0.101(3) & 0.1463(16) & EM(U) & 1 \\
SDSS J1152+4049$^{*}$ & 0.0677497026(3)$^{\S}$ & 0.155(6) & 0.560(28) & 0.087(6) & 0.142(3) & EM(U) & 1 \\
SDSS J0903+3300 & 0.059073543(9) & 0.113(4) & 0.872(11) & 0.099(4) & 0.1358(20) & EM(U) & 1 \\
SDSS J1227+5139 & 0.062959041(7) & 0.1115(16) & 0.796(18) & 0.0889(25) & 0.1365(13) & EM(U) & 1 \\
XZ Eri & 0.061159491(5) & 0.118(3) & 0.769(17) & 0.091(4) & 0.1350(18) & EM(U) & 1 \\
SDSS J1502+3334 & 0.05890961(5) & 0.1099(7) & 0.709(4) & 0.0781(8) & 0.1241(3) & EM(U) & 1 \\
SDSS J1501+5501$^{*}$ & 0.05684126603(21)$^{\S}$ & 0.101(10) & 0.767(27) & 0.077(10) & 0.122(5) & EM(U) & 1 \\
CTCV J1300$-$3052$^{*}$ & 0.0889406998(17)$^{\S}$ & 0.240(21) & 0.736(14) & 0.177(21) & 0.215(8) & EM(U) & 1 \\
OU Vir & 0.072706113(5) & 0.1641(13) & 0.703(12) & 0.1157(22) & 0.1634(10) & EM(U) & 1 \\
DV UMa$^{*}$ & 0.0858526308(7)$^{\S}$ & 0.1778(22) & 1.098(24) & 0.196(5) & 0.2176(18) & EM(U) & 1 \\
SDSS J1702+3229 & 0.10008209(9) & 0.248(5) & 0.91(3) & 0.223(10) & 0.252(4) & EM(U) & 1 \\
PHL 1445 & 0.0529848884(13) & 0.087(6) & 0.73(3) & 0.064(5) & 0.109(4) & EM(U) & 2 \\
SDSS J1057+2759 & 0.0627919557(6) & 0.0546(20) & 0.800(15) & 0.0436(20) & 0.1086(17) & EM(U) & 3\\
ASASSN-14ag & 0.060310665(9) & 0.149(16) & 0.63(4) & 0.093(13) & 0.135(7) & EM(U) & 4 \\
KIS J1927+4447 & 0.165308(5) & 0.570(11) & 0.69(7) & 0.39(4) & 0.432(15) & EM(U) & 5,6 \\
IP Peg & 0.1582061029(3) & 0.48(1) & 1.16(2) & 0.55(2) & 0.466(6) & EM(U) & 7 \\
EX Dra & 0.20993718(2) & 0.75(5) & 0.71(4) & 0.53(1) & 0.565(4) & EM & 8 \\
SDSS J1006+2337$^{*}$ & 0.185913107(13)$^{\S}$ & 0.51(8) & 0.78(12) & 0.40(10) & 0.47(4) & EM & 9 \\
CSS111003 (Te 11) & 0.120971471(9) & 0.236(6) & 1.18(11) & 0.28(3) & 0.314(11) & EM & 10 \\
HS 0220+0603 & 0.14920775(8) & 0.54(3) & 0.87(9) & 0.47(5) & 0.427(15) & EM & 11 \\
1RXS J0644+3344 & 0.26937431(22) & 0.80(2) & 0.73(7) & 0.58(6) & 0.690(24) & EM & 12,13 \\
SDSS J0756+0858& 0.1369745(4) & 0.47(9) & 0.60(12) & 0.28(5) & 0.338(20) & EM & 14 \\
\hline
\end{tabular}
\caption[System parameters for supplementary systems included in section~\ref{sec:discussion}.]{\label{table:supp_sys}System parameters for supplementary systems included in section~\ref{sec:discussion} (Figures~\ref{fig:ecaml} --\ref{fig:m2vsp}). The second-to-last column indicates the method used to obtain system parameters: EM\,$-$\,eclipse modelling (U\,$-$\,using ULTRACAM/ULTRASPEC data), CPT\,$-$\,contact phase timing, RV\,$-$\,radial velocity, GR\,$-$\,gravitational redshift, SM\,$-$\,spectrophotometric modelling. For consistency, all $R_{2}$ values were calculated using equation~\ref{eq:r2_sh} (ensuring all systems follow the same period-density relation). References: (1) \cite{savoury11}, 
(2) \cite{mcallister15}, (3) \cite{mcallister17a}, (4) \cite{mcallister17b} (5) \cite{scaringi13}, (6) \cite{littlefair14}, (7) \cite{copperwheat10}, (8) \cite{shafter03}, (9) \cite{southworth09}, (10) \cite{miszalski16}, (11) \cite{rodriguezgil15}, (12) \cite{sing07}, (13) \cite{hernandez17}, (14) \cite{tovmassian14}, (15) \cite{steeghs03}, (16) \cite{horne91}, (17) \cite{woodhorne90}, (18) \cite{littlefair08}, (19) \cite{baptista03}, (20) \cite{borgesbaptista05}, (21) \cite{araujobetancor03}, (22) \cite{patterson05}, (23) \cite{baptistabortoletto08}, (24) \cite{baptista94}, (25) \cite{thorstensen00}, (26) \cite{wadehorne88}, (27) \cite{echevarria16}, (28) \cite{arnold76}, (29) \cite{echevarria07}, (30) \cite{horne93}, (31) \cite{thoroughgood05}, (32) \cite{rolfe00}, (33) \cite{rodriguezgil01}, (34) \cite{peters06}, (35) \cite{arenas00}, (36) \cite{robinson74}, (37) \cite{welsh07}, (38) \cite{thoroughgood04}, (39) \cite{patterson98}, (40) \cite{steeghs01}, (41) \cite{steeghs07}, (42) \cite{vanamerongen87}, (43) \cite{smith06}, (44) \cite{szkody77}, (45) \cite{gaensicke06a}.
\vspace{0.15cm}\\
$^{*}$Updated system parameters produced in this work (Table~\ref{table:syspars_addsys}), $^{\S}P_{\mathrm{orb}}$ from this work}
\end{table*}

\begin{table*}\ContinuedFloat
\setlength{\tabcolsep}{1.6pt}
\footnotesize
\begin{tabular}{lccccccl}
\hline
System & $P_{\mathrm{orb}}$ & $q$ & $M_{1}$ & $M_{2}$ & $R_{2}$ & Method & Ref.\\ 
& (d) & & $(\mathrm{M}_{\odot})$ & $(\mathrm{M}_{\odot})$ & $(\mathrm{R}_{\odot})$ & & \\ \hline
IY UMa$^{*}$ & 0.07390892818(21)$^{\S}$ & 0.125(8) & 0.79(4) & 0.10(1) & 0.158(5) & CPT & 15 \\
HT Cas & 0.0736471745(5) & 0.15(3) & 0.61(4) & 0.09(2) & 0.152(11) & CPT & 16 \\
OY Car$^{*}$ & 0.06312092545(24)$^{\S}$ & 0.102(3) & 0.84(4) & 0.086(5) & 0.1354(26) & CPT & 17,18 \\
V2051 Oph & 0.06242785751(8) & 0.19(3) & 0.78(6) & 0.15(3) & 0.161(11) & CPT & 19 \\
V4140 Sgr & 0.0614296779(9) & 0.125(15) & 0.73(8) & 0.092(16) & 0.136(8) & CPT & 19,20 \\
DW UMa & 0.136606499(3) & 0.28(4) & 0.73(3) & 0.21(3) & 0.304(14) & CPT & 21,22 \\
UU Aqr & 0.1638049430 & 0.30(7) & 0.67(14) & 0.20(7) & 0.34(4) & CPT & 23,24 \\
GY Cnc$^{*}$ & 0.175442399(6)$^{\S}$ & 0.41(4) & 0.82(14) & 0.33(7) & 0.42(3) & RV & 25 \\
Z Cha$^{*}$ & 0.0744992631(3)$^{\S}$ & 0.149(4) & 0.84(9) & 0.125(14) & 0.171(6) & RV & 26 \\
EX Hya & 0.068233843(1) & 0.13(2) & 0.78(3) & 0.10(2) & 0.150(10) & RV & 27 \\
U Gem & 0.17690617(6) & 0.35(5) & 1.20(5) & 0.42(4) & 0.456(14) & RV & 28,29 \\
DQ Her & 0.193620897 & 0.66(4) & 0.60(7) & 0.40(5) & 0.485(20) & RV & 30 \\
V347 Pup & 0.231936060(6) & 0.83(5) & 0.63(4) & 0.52(6) & 0.603(23) & RV & 31 \\
V348 Pup & 0.101838931(14) & 0.31(6) & 0.65(13) & 0.20(4) & 0.246(16) & RV & 32,33 \\
V603 Aql & 0.13820103(8) & 0.24(5) & 1.2(2) & 0.29(4) & 0.341(16) & RV & 34,35 \\
EM Cyg & 0.290909(4) & 0.77(4) & 1.00(12) & 0.77(8) & 0.797(28) & RV & 36,37 \\
AC Cnc & 0.30047747(4) & 1.02(4) & 0.76(3) & 0.77(5) & 0.827(18) & RV & 38 \\
V363 Aur & 0.32124187(8) & 1.17(7) & 0.90(6) & 1.06(11) & 0.97(4) & RV & 38 \\
WZ Sge & 0.0566878460(3) & 0.057(18) & 0.85(4) & 0.049(15) & 0.105(11) & RV,GR & 39,40,41 \\
VW Hyi & 0.074271038(14) & -- & 0.71(22) & -- & -- & GR & 42,43 \\
AM Her & 0.128927(2) & -- & 0.78(15) & -- & -- & SM & 44,45 \\
\hline
\end{tabular}
\caption[\textit{Continued.}]{\textit{Continued.}}
\end{table*}

\section{Journal of Observations}
\label{sec:journal}


\begin{table*}
\setlength{\tabcolsep}{5pt}
\begin{tabular}{lllcllclclcc}
\hline
\ \ \ \, Date & \ \ \ \ \ Object & \ \ Instrument & Filter(s) & \ \ \ \ \ \ \ \ $T_{\mathrm{mid}}$ & Cycle & Phase & \,$T_{\mathrm{exp}}$ & $N_{u'}$ & $N_{\mathrm{exp}}$ & Seeing & Flag \\
&& \ \ \ \ \, Setup && \ \ \ \ \ \, (MJD) & \, No. & Coverage & \ \,(s) &&& ($''$) & \\ \hline
2010 May 12 & CSS080623 & NTT+UCAM & $u'\,g'\,r'$ & 55329.23459(3)$^{h}$ & 0 & $-$0.110--0.177 & 3.301 & 3 & 504 & 1.2--1.4 & 2 \\
2010 May 17 & CSS080623 & NTT+UCAM & $u'\,g'\,r'$ & 55334.12012(3)$^{h}$ & 82 & $-$0.318--0.363 & 4.920 & 3 & 712 & 0.9--1.8 & 2 \\
2010 May 17 & CSS080623 & NTT+UCAM & $u'\,g'\,r'$ & 55334.17971(3)$^{h}$ & 83 & $-$0.244--0.451 & 3.923 & 4 & 909 & 1.1--1.4 & 2 \\
2010 May 17 & CSS080623 & NTT+UCAM & $u'\,g'\,r'$ & 55334.23925(3)$^{h}$ & 84 & $-$0.512--0.269 & 3.923 & 3 & 1021 & 1.2--1.6 & 2 \\
2010 Jun 07 & CSS080623 & NTT+UCAM & $u'\,g'\,r'$ & 55355.03231(3)$^{h}$ & 433 & $-$0.254--0.226 & 3.818 & 2 & 644 & 1.0--1.1 & 2 \\
2010 Jun 07 & CSS080623 & NTT+UCAM & $u'\,g'\,r'$ & 55355.15152(3)$^{h}$ & 435 & $-$0.318--0.250 & 3.818 & 2 & 764 & 1.0--1.8 & 2 \\
2011 May 27 & CSS080623 & NTT+UCAM & $u'\,g'\,r'$ & 55709.05056(3)$^{h}$ & 6375 & $-$0.100--0.141 & 2.890 & 2 & 427 & 0.9--1.1 & 2 \\
2011 May 30 & CSS080623 & NTT+UCAM & $u'\,g'\,r'$ & 55711.96995(3)$^{h}$ & 6424 & $-$0.192--0.146 & 3.941 & 3 & 440 & 1.1--1.3 & 2 \\
2011 May 30 & CSS080623 & NTT+UCAM & $u'\,g'\,r'$ & 55712.02952(3)$^{h}$ & 6425 & $-$0.479--0.136 & 3.941 & 3 & 798 & 1.1--1.5 & 2 \\
2011 May 30 & CSS080623 & NTT+UCAM & $u'\,g'\,r'$ & 55712.14867(3)$^{h}$ & 6427 & $-$0.505--0.279 & 3.941 & 3 & 1023 & 1.0--1.2 & 2 \\
\hline
2011 Jan 18 & CSS110113 & NTT+UCAM & $u'\,g'\,i'$ & 55580.12190(10)$^{h}$ & $-$5479 & $-$0.332--0.159 & 2.376 & 3 & 1168 & 0.9--1.0 & 8 \\
2012 Jan 14 & CSS110113 & WHT+UCAM & $u'\,g'\,r'$ & 55940.95783(3)$^{h}$ & $-$16 & $-$0.215--0.178 & 3.985 & 3 & 563 & 1.0--1.6 & 2 \\
2012 Jan 15 & CSS110113 & WHT+UCAM & $u'\,g'\,r'$ & 55942.01465(3)$^{h}$ & 0 & $-$0.167--0.175 & 4.980 & 4 & 393 & 1.2--1.8 & 2 \\
2012 Jan 16 & CSS110113 & WHT+UCAM & $u'\,g'\,r'$ & 55942.87330(3)$^{h}$ & 13 & $-$0.237--0.117 & 3.985 & 3 & 507 & 1.4--2.1 & 2 \\
2012 Sep 09 & CSS110113 & WHT+UCAM & $u'\,g'\,r'$ & 56180.12801(3)$^{h}$ & 3605 & $-$0.197--0.114 & 3.987 & 3 & 448 & 1.4--1.8 & 5 \\
2012 Oct 13 & CSS110113 & WHT+UCAM & $u'\,g'\,r'$ & 56214.07819(3)$^{h}$ & 4119 & $-$0.232--0.140 & 2.989 & 3 & 711 & 1.2--1.4 & 5 \\
2014 Jan 02 & CSS110113 & WHT+UCAM & $u'\,g'\,r'$ & 56659.92157(3)$^{h}$ & 10869 & $-$0.194--0.091 & 3.980 & 3 & 408 & 0.9--1.4 & 5 \\
2014 Jan 28 & CSS110113 & TNT+USPEC & \textit{KG5} & 56685.54929(3)$^{h}$ & 11257 & $-$0.030--0.174 & 9.352 & -- & 126 & 1.3--2.0 & 7 \\
2014 Jan 28 & CSS110113 & TNT+USPEC & \textit{KG5} & 56685.61535(3)$^{h}$ & 11258 & $-$0.186--0.185 & 9.352 & -- & 227 & 1.4--1.7 & 7 \\
2014 Jan 29 & CSS110113 & TNT+USPEC & $g'$ & 56686.54005(3)$^{h}$ & 11272 & $-$0.533--0.324 & 9.352 & -- & 523 & 1.4--3.0 & 7 \\
2014 Feb 01 & CSS110113 & TNT+USPEC & $g'$ & 56689.57840(3)$^{h}$ & 11318 & $-$0.106--0.358 & 8.958 & -- & 296 & 1.9--2.1 & 7 \\
2014 Mar 14 & CSS110113 & WHT+UCAM & $u'\,g'\,r'$ & 56730.86018(3)$^{h}$ & 11943 & $-$0.205--0.240 & 2.627 & 3 & 957 & 1.2--1.8 & 7 \\
\hline
2007 Jun 10 & CTCV 1300 & VLT+UCAM & $u'\,g'\,r'$ & 54262.09916(3)$^{h}$ & 0 & $-$0.270--0.193 & 1.002 & 4 & 3463 & 1.2--2.3 & 1 \\
2007 Jun 13 & CTCV 1300 & VLT+UCAM & $u'\,g'\,i'$ & 54265.12310(3)$^{h}$ & 34 & $-$0.261--0.144 & 1.952 & 3 & 1574 & 1.4--2.1 & 1 \\
2010 Jun 07 & CTCV 1300 & NTT+UCAM & $u'\,g'\,r'$ & 55355.00260(5)$^{h}$ & 12288 & $-$0.142--0.120 & 3.938 & 3 & 511 & 0.9--1.1 & 6 \\
2011 May 30 & CTCV 1300 & NTT+UCAM & $u'\,g'\,r'$ & 55712.18836(5)$^{h}$ & 16304 & $-$0.175--0.149 & 2.895 & 3 & 852 & 0.9--1.7 & 1 \\
\hline
2003 May 23 & GY Cnc & WHT+UCAM & $u'\,g'\,i'$ & 52782.93530(10)$^{b}$ & $-$17985 & $-$0.085--0.128 & 1.628 & 1 & 1945 & 0.9--1.1 & 8 \\
2012 Jan 11 & GY Cnc & WHT+UCAM & $u'\,g'\,r'$ & 55938.26366(5)$^{b}$ & 0 & $-$0.082--0.140 & 3.974 & 3 & 842 & 1.0--1.2 & 6 \\
2012 Jan 14 & GY Cnc & WHT+UCAM & $u'\,g'\,r'$ & 55941.24626(5)$^{b}$ & 17 & $-$0.077--0.128 & 3.077 & 2 & 1005 & 1.2--1.8 & 6 \\
2012 Jan 16 & GY Cnc & WHT+UCAM & $u'\,g'\,r'$ & 55943.00068(5)$^{b}$ & 27 & $-$0.066--0.115 & 2.480 & 3 & 1096 & 0.9--1.6 & 6 \\
2012 Jan 16 & GY Cnc & WHT+UCAM & $u'\,g'\,r'$ & 55943.17605(5)$^{b}$ & 28 & $-$0.168--0.112 & 2.480 & 3 & 1692 & 1.1--1.5 & 1 \\
2012 Jan 20 & GY Cnc & WHT+UCAM & $u'\,g'\,r'$ & 55947.21130(5)$^{b}$ & 51 & $-$0.109--0.093 & 3.879 & 3 & 784 & 2.1--2.7 & 6 \\
2013 Dec 30 & GY Cnc & WHT+UCAM & $u'\,g'\,r'$ & 56657.22682(5)$^{b}$ & 4098 & $-$0.180--0.115 & 3.974 & 3 & 1120 & 0.9--1.2 & 6 \\
2014 Jan 26 & GY Cnc & TNT+USPEC & \textit{KG5} & 56683.71865(5)$^{b}$ & 4249 & $-$0.112--0.098 & 1.273 & -- & 2462 & 1.0--1.6 & 1 \\
2015 Feb 27 & GY Cnc & TNT+USPEC & \textit{KG5} & 57080.56902(10)$^{b}$ & 6511 & $-$0.059--0.138 & 2.473 & -- & 1200 & 1.6--2.2 & 6 \\
2015 Dec 11 & GY Cnc & TNT+USPEC & \textit{KG5} & 57367.76812(10)$^{b}$ & 8148 & $-$0.146--0.213 & 3.967 & -- & 1370 & 1.3--1.6 & 6 \\
2016 Mar 13 & GY Cnc & TNT+USPEC & $g'$ & 57460.57718(10)$^{b}$ & 8677 & $-$0.111--0.123 & 3.926 & -- & 898 & 1.1--1.6 & 6 \\
2017 Feb 13 & GY Cnc & TNT+USPEC & $g'$ & 57797.60241(15)$^{b}$ & 10598 & $-$0.185--0.161 & 1.766 & -- & 2944 & 1.3--1.8 & 6 \\
\hline
2003 May 20 & DV UMa & WHT+UCAM & $u'\,g'\,i'$ & 52779.969152(20)$^{h}$ & $-$35 & $-$0.092--0.177 & 5.921 & 1 & 339 & 1.2--2.7 & 2 \\
2003 May 22 & DV UMa & WHT+UCAM & $u'\,g'\,i'$ & 52781.943747(20)$^{h}$ & $-$12 & $-$0.126--0.104 & 4.921 & 1 & 345 & 0.9--1.1 & 2 \\
2003 May 23 & DV UMa & WHT+UCAM & $u'\,g'\,i'$ & 52782.974025(20)$^{h}$ & 0 & $-$0.135--0.151 & 3.921 & 1 & 540 & 0.9--1.1 & 5 \\
2007 Oct 19 & DV UMa & WHT+UCAM & $u'\,g'\,r'$ & 54393.225867(10)$^{h}$ & 18756 & $-$0.172--0.180 & 2.754 & 2 & 940 & 1.2--2.4 & 2 \\
\end{tabular}
\caption[Journal of observations for systems modelled in this paper.]{\label{table:obsj}Journal of observations for systems modelled in this paper. The instrument setup column shows both the telescope and instrument (UCAM and USPEC refer to ULTRACAM and ULTRASPEC, respectively) used for each eclipse observation. $T_{\mathrm{mid}}$ represents the mid-eclipse time (taken to be that of the white dwarf, when visible), $T_{\mathrm{exp}}$ the exposure time and $N_{\mathrm{exp}}$ the number of exposures. $N_{u'}$ indicates the number of $u'$ band frames which were co-added on-chip to reduce the impact of readout noise. The final column is a flag for eclipse status: (1) modelled individually, (2) modelled as constituent of average eclipse, (3) usable for individual modelling but not chosen, (4) obtained post-modelling but usable, (5) clear bright spot features but not included in average eclipse due to significant change in disc radius/contribution, (6) lack of bright spot ingress/merged ingresses, (7) poor SNR, (8) in outburst.
\vspace{0.15cm}\\
$^{h}$Heliocentric time in HMJD(UTC), $^{b}$Barycentric time in BMJD(TDB).}
\end{table*}

\begin{table*}\ContinuedFloat
\setlength{\tabcolsep}{5pt}
\begin{tabular}{lllcllclclcc}
\hline
\ \ \ \, Date & \ \ \ Object & \ \ Instrument & Filter(s) & \ \ \ \ \ \ \ \ \ $T_{\mathrm{mid}}$ & Cycle & Phase & \,$T_{\mathrm{exp}}$ & $N_{u'}$ & $N_{\mathrm{exp}}$ & Seeing & Flag \\
&& \ \ \ \ \, Setup && \ \ \ \ \ \ \, (MJD) & \, No. & Coverage & \ \,(s) &&& ($''$) & \\ \hline
2014 Mar 30 & IY UMa & TNT+USPEC & \textit{KG5} & 56746.639516(20)$^{h}$ & 0 & $-$0.102--0.325 & 2.190 & -- & 1243 & 1.5--2.9 & 1 \\
2014 Mar 30 & IY UMa & TNT+USPEC & \textit{KG5} & 56746.713426(20)$^{h}$ & 1 & $-$0.060--0.201 & 2.190 & -- & 763 & 1.6--2.0 & 1 \\
2014 Mar 30 & IY UMa & TNT+USPEC & \textit{KG5} & 56746.787335(20)$^{h}$ & 2 & $-$0.066--0.249 & 2.190 & -- & 913 & 1.8--2.5 & 1 \\
2014 Nov 30 & IY UMa & TNT+USPEC & \textit{KG5} & 56991.94334(10)$^{h}$ & 3319 & $-$0.090--0.174 & 3.352 & -- & 502 & 1.6--1.9 & 8 \\
2015 Jan 03 & IY UMa & TNT+USPEC & $g'$ & 57025.94133(3)$^{h}$ & 3779 & $-$0.203--0.176 & 3.352 & -- & 713 & 1.1--1.4 & 1 \\
2015 Jan 06 & IY UMa & TNT+USPEC & $r'$ & 57028.89764(5)$^{h}$ & 3819 & $-$0.159--0.125 & 3.952 & -- & 458 & 1.2--1.5 & 3 \\
2015 Feb 23 & IY UMa & TNT+USPEC & $r'$ & 57076.86456(3)$^{h}$ & 4468 & $-$0.130--0.249 & 3.852 & -- & 632 & 1.9--2.3 & 1 \\
2016 Mar 11 & IY UMa & TNT+USPEC & $u'$ & 57458.67790(10)$^{h}$ & 9634 & $-$0.185--0.269 & 29.78 & -- & 99 & 1.7--2.3 & 6 \\
2016 Mar 13 & IY UMa & TNT+USPEC & $u'$ & 57460.67369(10)$^{h}$ & 9661 & $-$0.195--0.287 & 25.35 & -- & 122 & 1.5--2.0 & 3 \\
2016 Mar 13 & IY UMa & TNT+USPEC & $i'$ & 57460.74748(5)$^{h}$ & 9662 & $-$0.363--0.320 & 7.852 & -- & 655 & 1.1--1.5 & 1 \\
\hline
2010 Apr 27 & OY Car & NTT+UCAM & $u'\,g'\,r'$ & 55314.104056(8)$^{h}$ & $-$632 & $-$0.103--0.140 & 1.760 & 3 & 747 & 1.7--2.7 & 3 \\
2010 Jun 06 & OY Car & NTT+UCAM & $u'\,g'\,r'$ & 55353.996480(8)$^{h}$ & 0 & $-$0.119--0.170 & 1.424 & 3 & 3116 & 1.3--1.4 & 1 \\
2010 Jun 07 & OY Car & NTT+UCAM & $u'\,g'\,r'$ & 55355.069543(8)$^{h}$ & 17 & $-$0.293--0.249 & 1.369 & 3 & 2120 & 1.1--1.8 & 3 \\
2010 Nov 18 & OY Car & NTT+UCAM & $u'\,g'\,i'$ & 55519.310181(8)$^{h}$ & 2619 & $-$0.206--0.487 & 1.329 & 4 & 3894 & 1.3--2.7 & 1 \\
2010 Dec 17 & OY Car & NTT+UCAM & $u'\,g'\,i'$ & 55548.282678(8)$^{h}$ & 3078 & $-$0.189--0.126 & 2.814 & 2 & 606 & 0.8--1.0 & 1 \\
2011 May 24 & OY Car & NTT+UCAM & $u'\,g'\,r'$ & 55706.084989(8)$^{h}$ & 5578 & $-$0.368--0.180 & 1.329 & 1 & 2205 & 1.3--2.0 & 3 \\
2016 Nov 10 & OY Car & NTT+UCAM & $u'\,g'\,r'$ & 57703.294204(8)$^{h}$ & 37219 & $-$0.141--0.202 & 1.979 & 2 & 931 & 0.9--1.4 & 4 \\
\hline
2006 Mar 09 & SDSS 0901 & WHT+UCAM & $u'\,g'\,r'$ & 53803.906350(20)$^{h}$ & $-$27455 & $-$0.763--0.259 & 4.972 & 1 & 1374 & 1.2--2.0 & 2 \\
2006 Mar 10 & SDSS 0901 & WHT+UCAM & $u'\,g'\,r'$ & 53804.996665(20)$^{h}$ & $-$27441 & $-$0.135--0.092 & 4.972 & 1 & 307 & 1.1--1.4 & 2 \\
2006 Mar 10 & SDSS 0901 & WHT+UCAM & $u'\,g'\,r'$ & 53805.152456(20)$^{h}$ & $-$27439 & $-$0.258--0.178 & 4.972 & 1 & 590 & 1.2--1.7 & 2 \\
2010 Jan 07 & SDSS 0901 & WHT+UCAM & $u'\,g'\,r'$ & 55203.96452(5)$^{h}$ & $-$9478 & $-$0.289--0.137 & 1.677 & 4 & 1685 & 1.7--3.0 & 6 \\
2012 Jan 15 & SDSS 0901 & WHT+UCAM & $u'\,g'\,r'$ & 55942.116352(20)$^{h}$ & 0 & $-$0.350--0.152 & 4.480 & 3 & 752 & 0.8--1.0 & 2 \\
2012 Jan 15 & SDSS 0901 & WHT+UCAM & $u'\,g'\,r'$ & 55942.194233(20)$^{h}$ & 1 & $-$0.416--0.240 & 4.480 & 3 & 987 & 0.9--1.4 & 2 \\
2012 Jan 16 & SDSS 0901 & WHT+UCAM & $u'\,g'\,r'$ & 55942.973064(20)$^{h}$ & 11 & $-$0.132--0.146 & 4.480 & 3 & 417 & 1.1--1.6 & 2 \\
2012 Jan 16 & SDSS 0901 & WHT+UCAM & $u'\,g'\,r'$ & 55943.050921(20)$^{h}$ & 12 & $-$0.373--0.128 & 4.480 & 3 & 752 & 1.0--1.9 & 2 \\
2012 Jan 16 & SDSS 0901 & WHT+UCAM & $u'\,g'\,r'$ & 55943.206687(20)$^{h}$ & 14 & $-$0.125--0.179 & 4.480 & 3 & 456 & 1.1--1.7 & 2 \\
2012 Jan 16 & SDSS 0901 & WHT+UCAM & $u'\,g'\,r'$ & 55943.284578(20)$^{h}$ & 15 & $-$0.235--0.219 & 4.480 & 3 & 678 & 1.1--2.1 & 2 \\
\hline
2012 Jan 15 & SDSS 1006 & WHT+UCAM & $u'\,g'\,r'$ & 55942.05221(10)$^{h}$ & $-$3984 & $-$0.116--0.180 & 3.980 & 3 & 1187 & 0.8--1.7 & 6 \\
2014 Jan 25 & SDSS 1006 & TNT+USPEC & \textit{KG5} & 56682.72940(20)$^{h}$ & 0 & $-$0.114--0.129 & 3.352 & -- & 1165 & 1.6--2.3 & 1 \\
2014 Jan 25 & SDSS 1006 & TNT+USPEC & \textit{KG5} & 56682.91552(20)$^{h}$ & 1 & $-$0.190--0.127 & 3.352 & -- & 1514 & 1.6--2.0 & 6 \\ 
2014 Jan 26 & SDSS 1006 & TNT+USPEC & \textit{KG5} & 56683.84494(20)$^{h}$ & 6 & $-$0.127--0.155 & 3.352 & -- & 1342 & 1.2--1.4 & 3 \\
2014 Jan 28 & SDSS 1006 & TNT+USPEC & \textit{KG5} & 56685.88995(20)$^{h}$ & 17 & $-$0.171--0.131 & 3.352 & -- & 1446 & 1.2--1.4 & 1 \\ 
2014 Feb 02 & SDSS 1006 & TNT+USPEC & $g'$ & 56690.72366(20)$^{h}$ & 43 & $-$0.267--0.176 & 5.892 & -- & 1204 & 1.4--2.0 & 1 \\
2014 Feb 02 & SDSS 1006 & TNT+USPEC & $r'$ & 56690.90953(20)$^{h}$ & 44 & $-$0.143--0.117 & 3.352 & -- & 1244 & 1.3--2.1 & 1 \\ 
2015 Dec 06 & SDSS 1006 & TNT+USPEC & $g'$ & 57362.79981(20)$^{h}$ & 3658 & $-$0.144--0.138 & 4.946 & -- & 974 & 2.2--3.6 & 7 \\
2016 Mar 14 & SDSS 1006 & TNT+USPEC & $g'$ & 57461.70591(20)$^{h}$ & 4190 & $-$0.266--0.202 & 9.640 & -- & 783 & 1.0--3.1 & 7 \\ 
2017 Feb 20 & SDSS 1006 & TNT+USPEC & \textit{KG5} & 57804.71547(20)$^{h}$ & 6035 & $-$0.239--0.208 & 4.970 & -- & 1437 & 1.4--3.4 & 6 \\ 
2017 Feb 21 & SDSS 1006 & TNT+USPEC & $g'$ & 57805.64520(20)$^{h}$ & 6040 & $-$0.198--0.152 & 5.470 & -- & 1024 & 2.0--2.7 & 7 \\ 
\hline
2010 Jan 07 & SDSS 1152 & WHT+UCAM & $u'\,g'\,r'$ & 55204.101280(10)$^{h}$ & 0 & $-$0.840--0.137 & 3.800 & 3 & 1492 & 2.4--4.0 & 2 \\
2010 Jan 07 & SDSS 1152 & WHT+UCAM & $u'\,g'\,r'$ & 55204.169035(10)$^{h}$ & 1 & $-$0.274--0.116 & 3.800 & 3 & 600 & 1.2--2.6 & 2 \\
2010 Jan 07 & SDSS 1152 & WHT+UCAM & $u'\,g'\,r'$ & 55204.236772(10)$^{h}$ & 2 & $-$0.151--0.119 & 3.800 & 3 & 415 & 1.5--3.0 & 2 \\
2014 Mar 14 & SDSS 1152 & WHT+UCAM & $u'\,g'\,r'$ & 56730.908566(10)$^{h}$ & 22536 & $-$0.265--0.195 & 4.029 & 3 & 668 & 1.2--1.7 & 7 \\
2014 Mar 30 & SDSS 1152 & TNT+USPEC & \textit{KG5} & 56746.694264(10)$^{h}$ & 22769 & $-$0.385--0.195 & 5.352 & -- & 634 & 1.3--1.6 & 2 \\
2014 Mar 30 & SDSS 1152 & TNT+USPEC & \textit{KG5} & 56746.762006(10)$^{h}$ & 22770 & $-$0.322--0.259 & 5.352 & -- & 634 & 1.2--1.7 & 2 \\
2014 Mar 30 & SDSS 1152 & TNT+USPEC & \textit{KG5} & 56746.829759(10)$^{h}$ & 22771 & $-$0.290--0.285 & 5.352 & -- & 628 & 1.4--1.9 & 2 \\
\end{tabular}
\caption[\textit{Continued.}]{\textit{Continued.}}
\end{table*}

\begin{table*}\ContinuedFloat
\setlength{\tabcolsep}{5pt}
\begin{tabular}{lllcllclclcc}
\hline
\ \ \ \, Date & \ \, Object & \ \ Instrument & Filter(s) & \ \ \ \ \ \ \ \ \ $T_{\mathrm{mid}}$ & Cycle & Phase & \,$T_{\mathrm{exp}}$ & $N_{u'}$ & $N_{\mathrm{exp}}$ & Seeing & Flag \\
&& \ \ \ \ \, Setup && \ \ \ \ \ \ \, (MJD) & \, No. & Coverage & \ \,(s) &&& ($''$) & \\ \hline

2012 Apr 28 & SDSS 1057 & WHT+UCAM & $u'\,g'\,r'$ & 56046.002399(12)$^{h}$ & 0 & $-$0.581--0.149 & 4.021 & 3 & 981 & 1.2--2.7 & 2 \\
2012 Apr 29 & SDSS 1057 & WHT+UCAM & $u'\,g'\,r'$ & 56046.944270(12)$^{h}$ & 15 & $-$0.239--0.228 & 4.021 & 3 & 628 & 1.1--2.0 & 2 \\
2013 Dec 30 & SDSS 1057 & WHT+UCAM & $u'\,g'\,r'$ & 56657.28205(3)$^{h}$ & 9735 & $-$0.558--0.320 & 4.021 & 3 & 1178 & 1.0--1.6 & 7 \\
2014 Jan 25 & SDSS 1057 & TNT+USPEC & \textit{KG5} & 56682.775595(12)$^{h}$ & 10141 & $-$0.316--0.064 & 4.877 & -- & 422 & 1.4--2.7 & 7 \\
2014 Nov 28 & SDSS 1057 & TNT+USPEC & \textit{KG5} & 56989.82829(3)$^{h}$ & 15031 & $-$0.225--0.138 & 3.945 & -- & 498 & 1.3--2.5 & 7 \\
2014 Nov 29 & SDSS 1057 & TNT+USPEC & \textit{KG5} & 56990.89570(3)$^{h}$ & 15048 & $-$0.158--0.143 & 4.945 & -- & 331 & 0.9--1.4 & 7 \\
2015 Feb 24 & SDSS 1057 & TNT+USPEC & \textit{KG5} & 57077.862577(12)$^{h}$ & 16433 & $-$0.219--0.281 & 11.852 & -- & 230 & 1.4--2.1 & 2 \\
2015 Feb 25 & SDSS 1057 & TNT+USPEC & \textit{KG5} & 57078.867265(12)$^{h}$ & 16449 & $-$0.207--0.169 & 11.946 & -- & 172 & 2.0--2.4 & 2 \\
2015 Mar 01 & SDSS 1057 & TNT+USPEC & \textit{KG5} & 57082.885950(12)$^{h}$ & 16513 & $-$0.101--0.138 & 11.852 & -- & 111 & 1.4--1.8 & 2 \\
2015 Jun 21 & SDSS 1057 & WHT+UCAM & $u'\,g'\,r'$ & 57194.906824(12)$^{h}$ & 18297 & $-$0.390--0.182 & 4.021 & 3 & 769 & 1.2--2.1 & 2 \\
2015 Jun 22 & SDSS 1057 & WHT+UCAM & $u'\,g'\,r'$ & 57195.911476(12)$^{h}$ & 18313 & $-$0.170--0.171 & 4.021 & 3 & 460 & 1.2--2.3 & 2 \\
2015 Jun 23 & SDSS 1057 & WHT+UCAM & $u'\,g'\,r'$ & 57196.916157(12)$^{h}$ & 18329 & $-$0.179--0.130 & 4.021 & 3 & 416 & 1.1--2.0 & 5 \\
\hline
2004 May 17 & SDSS 1501 & WHT+UCAM & $u'\,g'\,r'$ & 53142.921635(10)$^{h}$ & $-$53411 & $-$0.198--0.218 & 6.115 & 1 & 335 & 0.9--1.4 & 1 \\
2006 Mar 04 & SDSS 1501 & WHT+UCAM & $u'\,g'\,r'$ & 53799.210838(10)$^{h}$ & $-$41865 & $-$0.663--0.165 & 4.971 & 1 & 813 & 1.4--2.4 & 6 \\
2006 Mar 05 & SDSS 1501 & WHT+UCAM & $u'\,g'\,r'$ & 53800.120302(10)$^{h}$ & $-$41849 & $-$0.845--0.195 & 5.971 & 1 & 853 & 2.1--3.9 & 6 \\
2006 Mar 07 & SDSS 1501 & WHT+UCAM & $u'\,g'\,r'$ & 53802.052900(10)$^{h}$ & $-$41815 & $-$0.294--0.192 & 4.971 & 1 & 480 & 1.0--1.5 & 6 \\
2006 Mar 07 & SDSS 1501 & WHT+UCAM & $u'\,g'\,r'$ & 53802.109742(10)$^{h}$ & $-$41814 & $-$0.316--0.217 & 4.971 & 1 & 525 & 1.1--1.4 & 6 \\
2006 Mar 07 & SDSS 1501 & WHT+UCAM & $u'\,g'\,r'$ & 53802.223433(10)$^{h}$ & $-$41812 & $-$0.189--0.214 & 4.971 & 1 & 397 & 1.0--1.3 & 6 \\
2006 Mar 08 & SDSS 1501 & WHT+UCAM & $u'\,g'\,r'$ & 53803.132876(10)$^{h}$ & $-$41796 & $-$0.175--0.143 & 4.971 & 1 & 315 & 1.1--1.6 & 6 \\
2006 Mar 08 & SDSS 1501 & WHT+UCAM & $u'\,g'\,r'$ & 53803.189718(10)$^{h}$ & $-$41795 & $-$0.057--0.141 & 4.971 & 1 & 197 & 1.1--1.5 & 6 \\
2006 Mar 08 & SDSS 1501 & WHT+UCAM & $u'\,g'\,r'$ & 53803.246575(10)$^{h}$ & $-$41794 & $-$0.088--0.139 & 4.971 & 1 & 227 & 0.9--1.1 & 6 \\
2010 Jan 07 & SDSS 1501 & WHT+UCAM & $u'\,g'\,r'$ & 55204.213149(15)$^{h}$ & $-$17147 & $-$0.217--0.120 & 3.800 & 3 & 435 & 1.4--3.9 & 7 \\
2010 Jan 07 & SDSS 1501 & WHT+UCAM & $u'\,g'\,r'$ & 55204.270013(15)$^{h}$ & $-$17146 & $-$0.119--0.129 & 3.800 & 3 & 321 & 1.2--2.7 & 7 \\
2012 Sep 08 & SDSS 1501 & WHT+UCAM & $u'\,g'\,r'$ & 56178.870508(10)$^{h}$ & 0 & $-$0.158--0.165 & 3.475 & 3 & 455 & 1.0--1.5 & 1 \\
\hline
2014 Aug 03 & SSS100615 & WHT+UCAM & $u'\,g'\,r'$ & 56873.023626(5)$^{h}$ & 0 & $-$0.764--0.157 & 3.005 & 4 & 1538 & 1.0--1.2 & 2 \\
2014 Aug 04 & SSS100615 & WHT+UCAM & $u'\,g'\,r'$ & 56874.021600(5)$^{h}$ & 17 & $-$0.084--0.162 & 3.005 & 3 & 416 & 1.0--1.1 & 2 \\
2014 Aug 05 & SSS100615 & WHT+UCAM & $u'\,g'\,i'$ & 56874.960873(5)$^{h}$ & 33 & $-$0.114--0.103 & 5.056 & 3 & 219 & 1.7--2.3 & 2 \\
\hline
2014 Jan 26 & SSS130413 & TNT+USPEC & \textit{KG5} & 56683.67392(5)$^{h}$ & 0 & $-$0.175--0.195 & 5.804 & -- & 362 & 1.6--2.2 & 3 \\
2014 Jan 26 & SSS130413 & TNT+USPEC & \textit{KG5} & 56683.73977(5)$^{h}$ & 1 & $-$0.085--0.141 & 5.804 & -- & 222 & 1.5--2.0 & 1 \\
2014 Jan 26 & SSS130413 & TNT+USPEC & \textit{KG5} & 56683.80551(5)$^{h}$ & 2 & $-$0.230--0.194 & 5.804 & -- & 415 & 1.3--1.5 & 1 \\
2014 Jan 27 & SSS130413 & TNT+USPEC & \textit{KG5} & 56684.72635(5)$^{h}$ & 16 & $-$0.602--0.349 & 5.804 & -- & 928 & 1.8--2.9 & 3 \\
2014 Jan 28 & SSS130413 & TNT+USPEC & \textit{KG5} & 56685.71280(5)$^{h}$ & 31 & $-$0.209--0.181 & 2.934 & -- & 752 & 1.4--1.7 & 3 \\
2014 Feb 01 & SSS130413 & TNT+USPEC & \textit{KG5} & 56689.65905(5)$^{h}$ & 91 & $-$0.291--0.191 & 2.934 & -- & 931 & 1.9--2.2 & 3 \\
2014 Feb 01 & SSS130413 & TNT+USPEC & \textit{KG5} & 56689.85631(5)$^{h}$ & 94 & $-$0.148--0.176 & 2.934 & -- & 627 & 1.5--2.0 & 1 \\
2014 Feb 02 & SSS130413 & TNT+USPEC & $g'$ & 56690.77702(5)$^{h}$ & 108 & $-$0.304--0.215 & 2.934 & -- & 998 & 1.5--1.7 & 1 \\
2014 Feb 02 & SSS130413 & TNT+USPEC & $r'$ & 56690.84278(5)$^{h}$ & 109 & $-$0.223--0.126 & 2.934 & -- & 673 & 1.3--1.5 & 1 \\
2014 Mar 23 & SSS130413 & TNT+USPEC & \textit{KG5} & 56739.64406(8)$^{h}$ & 851 & $-$0.177--0.149 & 2.934 & -- & 628 & 1.3--1.6 & 7 \\
2015 Jan 01 & SSS130413 & TNT+USPEC & $i'$ & 57023.76697(3)$^{h}$ & 5171 & $-$0.120--0.190 & 2.939 & -- & 595 & 1.0--1.4 & 7 \\
2016 Jan 29 & SSS130413 & TNT+USPEC & $u'$ & 57416.73848(5)$^{h}$ & 11146 & $-$0.388--0.161 & 9.252 & -- & 338 & 2.7--3.6 & 7 \\
2016 Jan 31 & SSS130413 & TNT+USPEC & $u'$ & 57418.71160(5)$^{h}$ & 11176 & $-$0.228--0.167 & 8.052 & -- & 280 & 2.0--2.7 & 1 \\
2017 Feb 12 & SSS130413 & TNT+USPEC & $r'$ & 57796.75351(10)$^{h}$ & 16924 & $-$0.656--0.244 & 8.052 & -- & 670 & 1.8--2.7 & 7 \\
2017 Feb 20 & SSS130413 & TNT+USPEC & $g'$ & 57804.77728(3)$^{h}$ & 17046 & $-$0.210--0.147 & 3.951 & -- & 511 & 2.1--3.1 & 4 \\
2017 Feb 21 & SSS130413 & TNT+USPEC & \textit{KG5} & 57805.69806(3)$^{h}$ & 17060 & $-$0.284--0.285 & 3.951 & -- & 815 & 1.4--2.7 & 4 \\
2017 Feb 24 & SSS130413 & TNT+USPEC & $r'$ & 57808.78922(3)$^{h}$ & 17107 & $-$0.305--0.186 & 3.952 & -- & 705 & 1.2--1.6 & 4 \\
\end{tabular}
\caption[\textit{Continued.}]{\textit{Continued.}}
\end{table*}

\begin{table*}\ContinuedFloat
\setlength{\tabcolsep}{5pt}
\begin{tabular}{lllcllclclcc}
\hline
\ \ \ \, Date & \ \ Object & \ \ Instrument & Filter(s) & \ \ \ \ \ \ \ \ \ $T_{\mathrm{mid}}$ & Cycle & Phase & \,$T_{\mathrm{exp}}$ & $N_{u'}$ & $N_{\mathrm{exp}}$ & Seeing & Flag \\
&& \ \ \ \ \, Setup && \ \ \ \ \ \ \, (MJD) & \, No. & Coverage & \ \,(s) &&& ($''$) & \\ \hline
2011 Aug 27 & V713 Cep & WHT+UCAM & $u'\,g'\,i'$ & 55801.180379(20)$^{h}$ & $-$4399 & $-$0.383--0.136 & 2.185 & 2 & 1737 & 1.2--1.6 & 6 \\
2012 Sep 06 & V713 Cep & WHT+UCAM & $u'\,g'\,r'$ & 56176.936374(20)$^{h}$ & 0 & $-$0.373--0.127 & 3.445 & 3 & 1065 & 1.0--1.2 & 3 \\
2012 Sep 07 & V713 Cep & WHT+UCAM & $u'\,g'\,r'$ & 56177.875986(20)$^{h}$ & 11 & $-$0.346--0.184 & 3.445 & 3 & 1126 & 1.0--1.2 & 1 \\
2012 Sep 09 & V713 Cep & WHT+UCAM & $u'\,g'\,r'$ & 56180.011450(20)$^{h}$  & 36 & $-$0.212--0.169 & 3.445 & 3 & 811 & 0.9--1.4 & 6 \\
2013 Jul 14 & V713 Cep & WHT+UCAM & $u'\,g'\,z'$ & 56488.201454(20)$^{h}$ & 3644 & $-$0.309--0.291 & 3.445 & 3 & 1277 & 1.0--1.2 & 3 \\
2013 Jul 15 & V713 Cep & WHT+UCAM & $u'\,g'\,i'$ & 56489.055641(20)$^{h}$ & 3654 & $-$0.124--0.162 & 3.445 & 3 & 609 & 1.1--1.3 & 3 \\
2013 Jul 15 & V713 Cep & WHT+UCAM & $u'\,g'\,i'$ & 56489.141063(20)$^{h}$ & 3655 & $-$0.105--0.111 & 3.445 & 3 & 460 & 0.9--4.5 & 1 \\
2013 Jul 15 & V713 Cep & WHT+UCAM & $u'\,g'\,i'$ & 56489.226484(20)$^{h}$ & 3656 & $-$0.429--0.110 & 3.445 & 3 & 1150 & 0.9--1.1 & 7 \\
2013 Jul 25 & V713 Cep & WHT+UCAM & $u'\,g'\,z'$ & 56499.220435(20)$^{h}$ & 3773 & $-$0.275--0.205 & 3.445 & 3 & 1022 & 1.1--1.3 & 6 \\
2013 Aug 04 & V713 Cep & WHT+UCAM & $u'\,g'\,r'$ & 56509.214395(20)$^{h}$ & 3890 & $-$0.323--0.308 & 3.445 & 3 & 1342 & 1.1--1.7 & 3 \\
2013 Aug 05 & V713 Cep & WHT+UCAM & $u'\,g'\,r'$ & 56510.154021(20)$^{h}$ & 3901 & $-$0.288--0.168 & 3.445 & 3 & 971 & 0.9--1.4 & 6 \\
2014 Aug 02 & V713 Cep & WHT+UCAM & $u'\,g'\,r'$ & 56872.07224(3)$^{h}$ & 8138 & $-$0.388--0.216 & 1.983 & 3 & 2224 & 1.4--3.0 & 6 \\
2014 Aug 10 & V713 Cep & WHT+UCAM & $u'\,g'\,i'$ & 56880.18702(3)$^{h}$ & 8233 & $-$0.182--0.155 & 3.445 & 3 & 719 & 1.1--2.4 & 7 \\
2015 Jun 24 & V713 Cep & WHT+UCAM & $u'\,g'\,r'$ & 57198.114652(20)$^{h}$ & 11955 & $-$0.213--0.070 & 3.445 & 3 & 603 & 0.8--1.0 & 6 \\
2015 Sep 17 & V713 Cep & WHT+UCAM & $u'\,g'\,r'$ & 57282.849778(20)$^{h}$ & 12947 & $-$0.047--0.118 & 4.985 & 3 & 245 & 1.1--1.5 & 6 \\
\hline
2005 May 07 & Z Cha & VLT+UCAM & $u'\,g'\,r'$ & 53498.011478(10)$^{h}$ & 0 & $-$0.147--0.142 & 0.476 & 1 & 4300 & 1.7--2.3 & 3 \\
2005 May 10 & Z Cha & VLT+UCAM & $u'\,g'\,r'$ & 53500.991449(10)$^{h}$ & 40 & $-$0.373--0.175 & 0.476 & 1 & 7632 & 1.3--4.8 & 6 \\
2005 May 12 & Z Cha & VLT+UCAM & $u'\,g'\,r'$ & 53503.002929(10)$^{h}$ & 67 & $-$0.071--0.113 & 0.476 & 1 & 10070 & 2.1--8.1 & 3 \\
2005 May 15 & Z Cha & VLT+UCAM & $u'\,g'\,r'$ & 53505.982900(10)$^{h}$ & 107 & $-$0.108--0.204 & 0.476 & 1 & 7007 & 1.7--2.7 & 3 \\
2005 May 17 & Z Cha & VLT+UCAM & $u'\,g'\,r'$ & 53507.994355(10)$^{h}$ & 134 & $-$0.586--0.176 & 0.476 & 1 & 9769 & 2.1--3.9 & 3 \\
2005 May 21 & Z Cha & VLT+UCAM & $u'\,g'\,r'$ & 53512.017323(10)$^{h}$ & 188 & $-$0.113--0.161 & 0.476 & 1 & 6674 & 1.8--3.6 & 3 \\
2010 Apr 26 & Z Cha & NTT+UCAM & $u'\,g'\,r'$ & 55313.03694(3)$^{h}$ & 24363 & $-$0.056--0.119 & 1.517 & 4 & 731 & 2.6--3.6 & 6 \\
2010 Jun 06 & Z Cha & NTT+UCAM & $u'\,g'\,r'$ & 55354.08603(10)$^{h}$ & 24914 & $-$0.337--0.090 & 1.331 & 3 & 2024 & 1.1--1.8 & 8 \\
2010 Nov 26 & Z Cha & NTT+UCAM & $u'\,g'\,i'$ & 55527.147891(20)$^{h}$ & 27237 & $-$0.063--0.097 & 1.381 & 3 & 1636 & 0.8--1.1 & 1 \\
2010 Dec 02 & Z Cha & NTT+UCAM & $u'\,g'\,r'$ & 55533.33134(10)$^{h}$ & 27320 & $-$0.153--0.265 & 1.324 & 3 & 1996 & 1.1--1.7 & 8 \\
2010 Dec 11 & Z Cha & NTT+UCAM & $u'\,g'\,i'$ & 55542.345728(20)$^{h}$ & 27441 & $-$0.169--0.171 & 1.331 & 3 & 1775 & 1.1--1.6 & 6 \\
2010 Dec 17 & Z Cha & NTT+UCAM & $u'\,g'\,i'$ & 55548.082204(20)$^{h}$ & 27518 & $-$0.136--0.128 & 2.874 & 2 & 589 & 1.1--2.0 & 1 \\
2010 Dec 17 & Z Cha & NTT+UCAM & $u'\,g'\,i'$ & 55548.305686(20)$^{h}$ & 27521 & $-$0.124--0.134 & 2.874 & 2 & 576 & 0.8--1.0 & 3 \\
2011 May 19 & Z Cha & NTT+UCAM & $u'\,g'\,r'$ & 55701.029233(20)$^{h}$ & 29571 & $-$0.295--0.202 & 2.855 & 2 & 1108 & 1.8--3.6 & 1 \\
\hline
\end{tabular}
\caption[\textit{Continued.}]{\textit{Continued.}}
\end{table*}  



\bsp	
\label{lastpage}
\end{document}